\documentclass[12pt]{article}
\pdfoutput=1
\usepackage{amsmath,amsfonts,graphicx,color,bbm,tikz}
\usepackage{comment}
\usepackage[nosort]{cite}
\usepackage{subfigure}
\usetikzlibrary{calc,positioning}
\usetikzlibrary{patterns,arrows,decorations.pathreplacing}
\usepackage{caption}
\usepackage{ulem}
\tikzset{>=stealth}
\usepackage{lipsum}
\usepackage{tabularx}
\usepackage{fullpage}
\usepackage{hyperref}

\definecolor{red}{RGB}{ 211, 0, 0}
\definecolor{blue}{RGB}{1, 62, 177}

\makeatletter\@addtoreset{equation}{section}\makeatother

\newcommand{\be}{\begin{equation}}
\newcommand{\ee}{\end{equation}}
\def\beq{\begin{equation}}
\def\eeq{\end{equation}}
\newcommand{\bea}{\begin{eqnarray}}
\newcommand{\eea}{\end{eqnarray}}

\newcommand{\bra}[1]{{\left< {#1} \right|}}
\newcommand{\ket}[1]{{\left| {#1} \right>}}

\renewcommand{\title}[1]{\vbox{\center\LARGE{#1}}\vspace{3mm}}
\renewcommand{\author}[1]{\vbox{\center{#1}}\vspace{3mm}}

\newcommand{\email}[1]{\vbox{\center\tt#1}\vspace{3mm}}

\begin{document}
\begin{titlepage}

\begin{center}
{\large {\bf Topological Quantum Computation \\ on Supersymmetric Spin Chains} }

\author{ Indrajit Jana,$^a$ Filippo Montorsi,$^{b}$
Pramod Padmanabhan,$^a$ and Diego Trancanelli$^{b,c,d}$}

{$^a${\it School of Basic Sciences,\\ Indian Institute of Technology, Bhubaneswar, India}}
\vskip0.1cm
{$^b${\it Dipartimento di Scienze Fisiche, Informatiche e Matematiche, \\
Universit\`a di Modena e Reggio Emilia, via Campi 213/A, 41125 Modena, Italy}}
\vskip0.1cm
{$^c${\it INFN Sezione di Bologna, via Irnerio 46, 40126 Bologna, Italy}}
\vskip0.1cm
{$^d${\it Institute of Physics, University of S\~ao Paulo, 05314-970 S\~ao Paulo, Brazil}}

\email{jana.indrajit, fillomonto, pramod23phys, dtrancan@gmail.com}

\vskip 3cm 

\end{center}


\abstract{
\noindent 
Quantum gates built out of braid group elements form the building blocks of topological quantum computation. They have been extensively studied in $SU(2)_k$ quantum group theories, a rich source of examples of non-Abelian anyons such as the Ising ($k=2$), Fibonacci ($k=3$) and Jones-Kauffman ($k=4$) anyons. We show that the fusion spaces of these anyonic systems can be precisely mapped to the product state zero modes of certain Nicolai-like supersymmetric spin chains. As a result, we can realize the braid group in terms of the product state zero modes of these supersymmetric systems. These operators kill all the other states in the Hilbert space, thus preventing the occurrence of errors while processing information, making them suitable for quantum computing.  
}

\end{titlepage}
\tableofcontents 

\section{Introduction}
\label{sec:Introduction}

One of the most promising approaches to achieve robust, fault-tolerant quantum computing is through the use of topology, with the so-called {\it topological quantum computing (TQC)}, see for example \cite{freedman2003topological,nayak2008non,wang2010topological, pachos2012introduction,Lahtinen_2017}. A preeminent role in this context is played by (2+1)-dimensional quasiparticle excitations known as {\it anyons}, which are thought to realize TQC because of their statistical interactions. These interactions are robust to local perturbations, due to their topological nature. More technically, anyons are representations of the {\it braid group} $\mathcal{B}_N$ and are characterized by  {\it fusion} and {\it braiding} properties, in which world-lines of these particles fuse together and intertwine. In topological quantum computers, information is encoded in the fusion spaces of anyons and quantum gates are implemented as elements of the appropriate braid group. Thus the premise for this construction makes them immune to errors that can occur while processing information. A general introduction to the applications of anyons in quantum computing can be found in the textbook \cite{pachos2012introduction} or in  \cite{Bonderson2007NonAbelianAA,Bonderson_2008,   panangaden2010categorical}.

Some of the most widely studied anyonic systems in the literature include the {\it Ising anyons}, the {\it Fibonacci anyons} and the {\it Jones-Kauffman anyons}. They are non-Abelian anyons associated to irreducible representations of the quantum group $SU(2)_k$ \cite{Biedenharn:1996vv,Majid1995FoundationsOQ}, appearing at levels $k=2,3$ and $k=4$, respectively. Among them, the Fibonacci anyons have found importance in quantum computing as they help generate a universal set of gates by braiding alone \cite{bonesteel2005braid,preskill1999lecture, field2018introduction, simon2006topological, rouabah2021compiling}.
On the other hand the level $k=4$ anyons can realize universal quantum computing by supplementing the braiding gates with non-topological ones \cite{https://doi.org/10.48550/arxiv.1501.02841, PhysRevA.92.012301}.

There are several physical realizations where these anyons are thought to exist, like Ryberg atoms, fractional quantum Hall systems and parafermionic models, among others \cite{djuric2017fibonacci,lesanovsky2012interacting,mong2017fibonacci,vaezi2014fibonacci,stoudenmire2015assembling}. While it is experimentally feasible to realize these systems, there have been theoretical challenges \cite{chandran2020absence}.
More formally, anyonic systems are also seen via Temperley-Lieb recoupling theories and spin networks \cite{kauffman1994temperley,kauffman2006spin,kauffman2007q,kauffman2008fibonacci,marzuoli2005computing, https://doi.org/10.48550/arxiv.1604.06429}, which we will use in this paper.

The scope of this paper is to approach the anyonic systems above from a novel prospective, the one of supersymmetry algebras realized on spin chains. Many lattice models have a continuum limit that possess supersymmetry, some examples being the tricritical Ising model and Ashkin-Teller models at special values of the coupling constants \cite{Friedan:1984rv,Qiu:1986if,Yang:1987bj,Yang:1987mf}. Moreover well studied spin chains like the $XXX$, and $XXZ$ models can be generated using dynamical lattice supercharges \cite{Yang2004NonlocalSS, Matsui2015DynamicalSO, Matsui2016SpinonEI}. All this has lead to a search for models with supercharges satisfying the supersymmetry algebra on one-dimensional lattices or models with explicit lattice supersymmetry. The first known example of such models is by Hermann Nicolai \cite{nicolai1976supersymmetry}. Then there is also a class of lattice models called the $\mathcal{M}_k$ models, see  \cite{fendley2003lattice,fendley2003lattice1, hagendorf2013spin,fokkema2017m,Hagendorf_2012}.
In general these models have an extensive ground state degeneracy, with exponentially growing numbers of states being the common trend more often than not \cite{huijse2010supersymmetric,katsura2017characterization,la2018ground,sannomiya2017supersymmetry,Moriya2018SupersymmetryBF,Fendley_2011}.
Some of these supersymmetric spin chains have also shown to exhibit localization effects similar to {\it many body localized} systems \cite{Padmanabhan:2017ekk,moriya2018ergodicity} and these supercharges have been used to obtain solutions of the Yang-Baxter equation and its generalization \cite{padmanabhan2019quantum}. 

In this paper we consider three models similar to the Nicolai supersymmetric spin chain and we study their zero modes (ground states) in detail. This allows us to establish a connection between these zero mode states and the fusion spaces of anyons and, consequently, topological quantum computation. In this regard  \cite{Fokkema_2015} considers defects in the $\mathcal{M}_k$ lattice supersymmetric models that have fusion rules of the Ising and Fibonacci anyons. And also in this line there has been an earlier attempt to connect supersymmetry and quantum computation \cite{https://doi.org/10.48550/arxiv.2011.01239}, however this is quite different from what we intend to do here. 

More specifically, we construct three different supercharges on a spin chain, where each chain site is endowed with a Hilbert space which is either $\mathbb{C}^2$ or $\mathbb{C}^2 \otimes \mathbb{C}^2$. The spectrum inherits a $\mathbb{Z}_2$-grading from supersymmetry and is divided in a bosonic and fermionic sector. We study the zero modes of these systems, which have entangled and non-entangled (product) states. The product states display a very specific scaling with the number of spin sites, which can be precisely mapped to the dimension of the fusion spaces of the Fibonacci anyons (when the Hilbert space is $\mathbb{C}^2$) and of the Jones-Kauffman and Ising anyons (when the Hilbert space is $\mathbb{C}^2 \otimes \mathbb{C}^2$). Equipped with this correspondence, we go on with expressing the braid group generators acting on those anyonic fusion spaces in terms of the supersymmetric zero modes of the spin chain. 

This shows that we can mimic TQC via anyons on these spin chains without actually realizing non-Abelian anyons. This is especially useful given the fact that it is difficult to find actual non-Abelian anyons in fractional quantum Hall systems. Another advantage in using supersymmetric spin chains seems to arise from the fact that the quantum information now needs to be stored in product states which are essentially classical. This could reduce errors during the encoding process. Furthermore, as explained in detail in \ref{subsec:SUSYqc}, we find that the braid generators are supersymmetric and hence the quantum gates built using these braid generators cannot access parts of the supersymmetric Hilbert space that are not the product zero modes, thus reducing errors while implementing the quantum gates as well.   

This paper is organized as follows. In general, we try to keep things as self-contained as possible and for this reason we begin with a brief review of anyons and the construction of their fusion spaces in Sec. \ref{sec:Anyons}. The mathematical treatment of this subject can get rather abstract, so we keep the language of categories to a minimum and list exactly those properties that are useful for the goals of this work. We provide the examples of $SU(2)_k$ anyons in the same section and also count the dimensions of the fusion basis of the $k=2,3,4$ cases. We then proceed to construct three kinds of supersymmetric spin chains in Sec. \ref{sec:SUSYspinchains}. There we also study in detail the product zero modes of the three systems. The main correspondence between the two spaces - the space of product supersymmetric zero modes and the fusion basis of the Fibonacci, the Jones-Kauffman and the Ising anyons - is then established in Sec. \ref{sec:SUSYanyoncorrespondence}. Using this correspondence the braid groups on the space of supersymmetric zero modes are constructed in Sec. \ref{sec:Zeromodesbraidgroup} after reviewing the construction of braid groups on anyon fusion spaces in Sec. \ref{sec:Fusionspacebraidgroup}. In Sec. \ref{subsec:stability} we discuss the important issue of the stability of the product zero modes to deformations of the SUSY Hamiltonians. This explains the robustness properties of topological quantum computation on SUSY spin chains considered here. Before we conclude we look at how the correspondence can be extended to the case where the local Hilbert space is not $\mathbb{C}^2$, but more generally $\mathbb{C}^d$ in Sec. \ref{sec:SISbraidgroup}. This requires the use of the supersymmetric systems based on inverse semigroups developed in \cite{Padmanabhan:2017ekk}. 
We end with an outlook in Sec. \ref{sec:Outlook} and keep some of the more tangential details in appendices.

\section{Review of anyons and their fusion spaces}
\label{sec:Anyons}

To keep this article self-contained we briefly review anyons and their fusion spaces, closely following \cite{Bonderson_2008,pachos2012introduction}. 

Since the discovery of anyons as quasiparticle excitations in two-dimensional systems, an elaborate mathematical approach based on category theory has been developed to study their properties. In this perspective, anyonic systems are thought of as {\it unitary braided fusion categories}, described by a set $\mathcal{C}$ of anyon species 
\bea
\mathcal{C}= \{a,b,c,\ldots\}.
\eea
Different anyonic particles $\phi_a, \phi_b, \ldots$ are labelled by the species to which they belong and obey a set of {\it fusion rules}
\begin{equation}
\label{eq:fusion}
\phi_a\times \phi_b =\sum\limits_{c\in\mathcal{C}}~N^c_{ab}\, \phi_c.
\end{equation}
Here $N^c_{ab}$ are non-negative integers denoting the multiplicity of a particular fusion channel, namely the number of different ways in which anyons of species $a$ and $b$ can combine to form an anyon of species $c$. 
The fusion rules in \eqref{eq:fusion} are both commutative and associative, implying
\begin{equation}
N^c_{ab} = N^c_{ba},\qquad \sum\limits_{e\in\mathcal{C}}~N^e_{ab}N^f_{ec} = \sum\limits_{e\in\mathcal{C}}~N^f_{ae}N^e_{bc},
\end{equation}
respectively. Among the anyonic species there is the 1, which is unique and is called the {\it vacuum}. It is defined by the fact that its fusion with any other species gives back that species itself, that is $N^c_{a1}=N^c_{1a}=\delta_{ac}$.  An important distinction is between Abelian anyons, for which $N^c_{ab}=1$ for only one value of $c$ and $N^{c'}_{ab}=0$ for all $c'\neq c$, and non-Abelian anyons, for which there is at least one $a$ and $b$ such that there are multiple fusion channels $c$ with $N^c_{ab}\neq 0$. We will mostly be interested in examples of non-Abelian anyons with no multiplicity, and henceforth we only consider the case in which $N^c_{ab}=0,1$. Every anyon $\phi_a$ has an anti-particle $\phi_{\bar{a}}$ (which can possibly be $\phi_a$ itself), such that they fuse to the vacuum in precisely one way, that is $N^1_{a \bar{a}}=1$. Besides annihilating into the vacuum, they may also annihilate into other species.

To every fusion channel one can associate a vector space $V^c_{ab}$, called {\it fusion space}, with $\dim(V^c_{ab})=N^c_{ab}$ and basis vectors $\ket{a,b;c}$. 
The dual to the fusion space is the {\it splitting space} $V^{ab}_c$ spanned by $\bra{a,b;c}$. Physically, splitting can be thought of as the time reversed process of fusion. Together they define the inner product, which we take to be the canonical one. 
\begin{equation}\label{eq:innerproduct}
    \langle a,b;i_1|a,b;i_2\rangle = \delta(i_1, i_2),
\end{equation}
where $i_1$ and $i_2$ represent two different anyon species corresponding to different fusion channels of the anyons $a$ and $b$.

\begin{figure}[h]
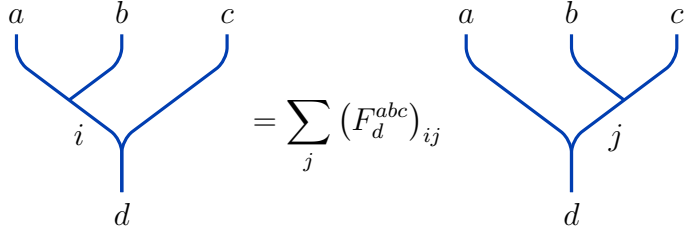

		\centering
		\begin{align*}
		\tikz[baseline=-7ex, scale = 0.7]{
			\def\a{0}
			\def\b{0}
			\draw[very thick, blue, rounded corners=5] (\a, \b) -- (\a, \b-0.5) -- (2+\a, \b-2) -- (2+\a, \b -3);
			\draw[very thick, blue, rounded corners=5] (\a + 2, \b) -- (\a + 2, \b-0.5) -- (1+\a, \b-1.25);
			\draw[very thick, blue, rounded corners=5] (\a + 4, \b) -- (\a + 4, \b-0.5) -- (2+\a, \b-2) -- (2+\a, \b -3); 
			\node[anchor=south] at (\a, \b) {$a$};
			\node[anchor=south] at (\a + 2, \b) {$b$};
			\node[anchor=south] at (\a + 4, \b) {$c$};
			\node[anchor=north east] at (\a + 1.5, \b - 1.5) {$i$};
			\node[anchor=north] at (\a + 2, \b - 3) {$d$};
		} = \sum_{j}\left(F_{d}^{abc}\right)_{ij}
		\tikz[baseline=-7ex, scale = 0.7]{
			\def\a{0}
			\def\b{0}
			\draw[very thick, blue, rounded corners=5] (\a, \b) -- (\a, \b-0.5) -- (2+\a, \b-2) -- (2+\a, \b -3);
			\draw[very thick, blue, rounded corners=5] (\a + 2, \b) -- (\a + 2, \b-0.5) -- (3+\a, \b-1.25);
			\draw[very thick, blue, rounded corners=5] (\a + 4, \b) -- (\a + 4, \b-0.5) -- (2+\a, \b-2) -- (2+\a, \b -3); 
			\node[anchor=south] at (\a, \b) {$a$};
			\node[anchor=south] at (\a + 2, \b) {$b$};
			\node[anchor=south] at (\a + 4, \b) {$c$};
			\node[anchor=north west] at (\a + 2.5, \b - 1.5) {$j$};
			\node[anchor=north] at (\a + 2, \b - 3) {$d$};
		}
		\end{align*}
	\caption{Definition of the $F$-matrix as a basis change in fusion space. Graphically, this corresponds to sliding the $b$ line from $a$ to $c$.} 
 \label{fig:Fmove}
	\end{figure}

One can generalize the vector space $V^c_{ab}$ ($V^{ab}_c$) to the case where more particles fuse (split) into (from) $d$. For example, $V^d_{abc}$ has three initial particles of species $a,b,c$ and a final particle of species $d$. This space can be obtained in two ways, according to the order in which the anyons fuse, either as $(ab)c$ or as $a(bc)$. These two processes are unitarily related by so-called {\it $F$-matrices}, as shown in Fig. \ref{fig:Fmove}.
 

In terms of basis kets, this is encoded in the following relation
\begin{equation}
\ket{ab;i}\ket{ic;d} = \sum_j\left(F^{abc}_d\right)_{ij}\ket{bc;j}\ket{aj;d}.
\end{equation}
Later on we will  refrain from mentioning the anyons that are fusing (splitting) when it is understood from the context. In such situations, $\ket{ab;i}\ket{ic;d}\equiv \ket{i}$ and $\ket{bc;j}\ket{aj;d}\equiv \ket{j}$.
The $F$-matrices can be thought of as sliding the anyon line $b$ from $a$ to $c$, see Fig. \ref{fig:Fmove}, justifying the nomenclature of {\it $F$-move}. As this operation also resembles the associativity condition in algebra, this is sometimes also denoted as the {\it associativity move}. 

\begin{figure}[h!]
	\centering
	\begin{tikzpicture}[scale=0.4]
		\def\c{4}
		\def\d{-2}
		\def\s{3}
		\def\t{13}
		\def\r{18}
		\def\reduce{20}
		\draw[thick] ({\c+\s*cos(0+\r)}, {\d+\s*sin(0+\r)}) -- ({\c+\s*cos(72+\r)}, {\d+\s*sin(72+\r)}) -- ({\c+\s*cos(2*72+\r)}, {\d+\s*sin(2*72+\r)}) -- ({\c+\s*cos(3*72+\r)}, {\d+\s*sin(3*72+\r)}) -- ({\c+\s*cos(4*72+\r)}, {\d+\s*sin(4*72+\r)}) -- ({\c+\s*cos(\r)}, {\d+\s*sin(\r)});
		\def\a{-14.27}
		\def\b{4.64}
		\draw[very thick, blue, rounded corners=5] (\a, \b) -- (\a, \b-0.5) -- (3+\a, \b-2.75) -- (3+\a, \b - 3.75);
		\draw[very thick, blue, rounded corners=5] (\a + 2, \b) -- (\a + 2, \b-0.5) -- (1+\a, \b-1.25);
		\draw[very thick, blue, rounded corners=5] (\a + 4, \b) -- (\a + 4, \b-0.5) -- (2+\a, \b-2);
		\draw[very thick, blue, rounded corners=5] (\a + 6, \b) -- (\a + 6, \b-0.5) -- (3+\a, \b-2.75) -- (3+\a, \b - 3.75);  
		\node[anchor=south] at (\a, \b) {$a$};
		\node[anchor=south] at (\a + 2, \b) {$b$};
		\node[anchor=south] at (\a + 4, \b) {$c$};
		\node[anchor=south] at (\a + 6, \b) {$d$};
		\node[anchor=north east] at (\a + 1.75, \b - 1.5) {$i_{1}$};
		\node[anchor=north east] at (\a + 2.75, \b - 2.25) {$i_{2}$};
		\node[anchor=north] at (\a + 3, \b - 3.75) {$e$};
		\draw[very thick, red,  ->] ({\c+\t*cos(2*72+\r - \reduce)}, {\d+\t*sin(2*72+\r - \reduce)}) -- ({\c+\t*cos(72+\r + \reduce)}, {\d+\t*sin(72+\r + \reduce)});
		\node[anchor=south east] at ({\c+\t*cos(1.5*72+\r)}, {\d+\t*sin(1.5*72+\r)}) {\Large $F$};
		\def\a{0}
		\def\b{15}
		\draw[very thick, blue, rounded corners=5] (\a, \b) -- (\a, \b-0.5) -- (3+\a, \b-2.75) -- (3+\a, \b - 3.75);
		\draw[very thick, blue, rounded corners=5] (\a + 2, \b) -- (\a + 2, \b-0.5) -- (1+\a, \b-1.25);
		\draw[very thick, blue, rounded corners=5] (\a + 4, \b) -- (\a + 4, \b-0.5) -- (5+\a, \b-1.25);
		\draw[very thick, blue, rounded corners=5] (\a + 6, \b) -- (\a + 6, \b-0.5) -- (3+\a, \b-2.75) -- (3+\a, \b - 3.75);  
		\node[anchor=south] at (\a, \b) {$a$};
		\node[anchor=south] at (\a + 2, \b) {$b$};
		\node[anchor=south] at (\a + 4, \b) {$c$};
		\node[anchor=south] at (\a + 6, \b) {$d$};
		\node[anchor=north east] at (\a + 2.5, \b - 2) {$i_{1}$};
		\node[anchor=north west] at (\a + 3.75, \b - 2) {$j_{2}$};
		\node[anchor=north] at (\a + 3, \b - 3.75) {$e$};
		\draw[very thick, red,  ->] ({\c+\t*cos(72+\r - \reduce)}, {\d+\t*sin(72+\r - \reduce)}) -- ({\c+\t*cos(0+\r + \reduce)}, {\d+\t*sin(0+\r + \reduce)});
		\node[anchor=south west] at ({\c+\t*cos(0.5*72+\r)}, {\d+\t*sin(0.5*72+\r)}) {\Large $F$};
		\def\a{14.27}
		\def\b{4.64}
		\draw[very thick, blue, rounded corners=5] (\a, \b) -- (\a, \b-0.5) -- (3+\a, \b-2.75) -- (3+\a, \b - 3.75);
		\draw[very thick, blue, rounded corners=5] (\a + 2, \b) -- (\a + 2, \b-0.5) -- (4+\a, \b-2);
		\draw[very thick, blue, rounded corners=5] (\a + 4, \b) -- (\a + 4, \b-0.5) -- (5+\a, \b-1.25);
		\draw[very thick, blue, rounded corners=5] (\a + 6, \b) -- (\a + 6, \b-0.5) -- (3+\a, \b-2.75) -- (3+\a, \b - 3.75);  
		\node[anchor=south] at (\a, \b) {$a$};
		\node[anchor=south] at (\a + 2, \b) {$b$};
		\node[anchor=south] at (\a + 4, \b) {$c$};
		\node[anchor=south] at (\a + 6, \b) {$d$};
		\node[anchor=north west] at (\a + 3.5, \b - 2.25) {$j_{1}$};
		\node[anchor=north west] at (\a + 4.5, \b - 1.5) {$j_{2}$};
		\node[anchor=north] at (\a + 3, \b - 3.75) {$e$};
		\draw[very thick, red,  ->] ({\c+\t*cos(2*72+\r + \reduce)}, {\d+\t*sin(2*72+\r + \reduce)}) -- ({\c+\t*cos(3*72+\r - \reduce)}, {\d+\t*sin(3*72+\r - \reduce)});
		\node[anchor=north east] at ({\c+\t*cos(2.5*72+\r)}, {\d+\t*sin(2.5*72+\r)}) {\Large $F$};
		\def\a{-8.82}
		\def\b{-12.14}
		\draw[very thick, blue, rounded corners=5] (\a, \b) -- (\a, \b-0.5) -- (3+\a, \b-2.75) -- (3+\a, \b - 3.75);
		\draw[very thick, blue, rounded corners=5] (\a + 2, \b) -- (\a + 2, \b-0.5) -- (3+\a, \b-1.25);
		\draw[very thick, blue, rounded corners=5] (\a + 4, \b) -- (\a + 4, \b-0.5) -- (2+\a, \b-2);
		\draw[very thick, blue, rounded corners=5] (\a + 6, \b) -- (\a + 6, \b-0.5) -- (3+\a, \b-2.75) -- (3+\a, \b - 3.75);  
		\node[anchor=south] at (\a, \b) {$a$};
		\node[anchor=south] at (\a + 2, \b) {$b$};
		\node[anchor=south] at (\a + 4, \b) {$c$};
		\node[anchor=south] at (\a + 6, \b) {$d$};
		\node[anchor=north east] at (\a + 2.75, \b - 2.25) {$i_{2}$};
		\node[anchor=north west] at (\a + 2.25, \b - 1.25) {$k_{1}$};
		\node[anchor=north] at (\a + 3, \b - 3.75) {$e$};
		\draw[very thick, red,  ->] ({\c+\t*cos(3*72+\r + \reduce)}, {\d+\t*sin(3*72+\r + \reduce)}) -- ({\c+\t*cos(4*72+\r - \reduce)}, {\d+\t*sin(4*72+\r - \reduce)});
		\node[anchor=north] at ({\c+\t*cos(3.5*72+\r)}, {\d+\t*sin(3.5*72+\r)}) {\Large $F$};
		\def\a{8.82}
		\def\b{-12.14}
		\draw[very thick, blue, rounded corners=5] (\a, \b) -- (\a, \b-0.5) -- (3+\a, \b-2.75) -- (3+\a, \b - 3.75);
		\draw[very thick, blue, rounded corners=5] (\a + 2, \b) -- (\a + 2, \b-0.5) -- (4+\a, \b-2);
		\draw[very thick, blue, rounded corners=5] (\a + 4, \b) -- (\a + 4, \b-0.5) -- (3+\a, \b-1.25);
		\draw[very thick, blue, rounded corners=5] (\a + 6, \b) -- (\a + 6, \b-0.5) -- (3+\a, \b-2.75) -- (3+\a, \b - 3.75);  
		\node[anchor=south] at (\a, \b) {$a$};
		\node[anchor=south] at (\a + 2, \b) {$b$};
		\node[anchor=south] at (\a + 4, \b) {$c$};
		\node[anchor=south] at (\a + 6, \b) {$d$};
		\node[anchor=north west] at (\a + 3.5, \b - 2.25) {$j_{1}$};
		\node[anchor=south west] at (\a + 3.5, \b - 2) {$k_{1}$};
		\node[anchor=north] at (\a + 3, \b - 3.75) {$e$};
		\draw[very thick, red,  ->] ({\c+\t*cos(4*72+\r + \reduce)}, {\d+\t*sin(4*72+\r + \reduce)}) -- ({\c+\t*cos(5*72+\r - \reduce)}, {\d+\t*sin(5*72+\r - \reduce)});
		\node[anchor=north west] at ({\c+\t*cos(4.5*72+\r)}, {\d+\t*sin(4.5*72+\r)}) {\Large $F$};
	\end{tikzpicture}
	\caption{One can go from left to right either with the two $F$-moves in the upper part of the diagram or with the three $F$-moves in the lower part. Equating these two ways to fuse four anyons gives the pentagon identity in \eqref{pentid}.}
	\label{fig:pentagon}
\end{figure}
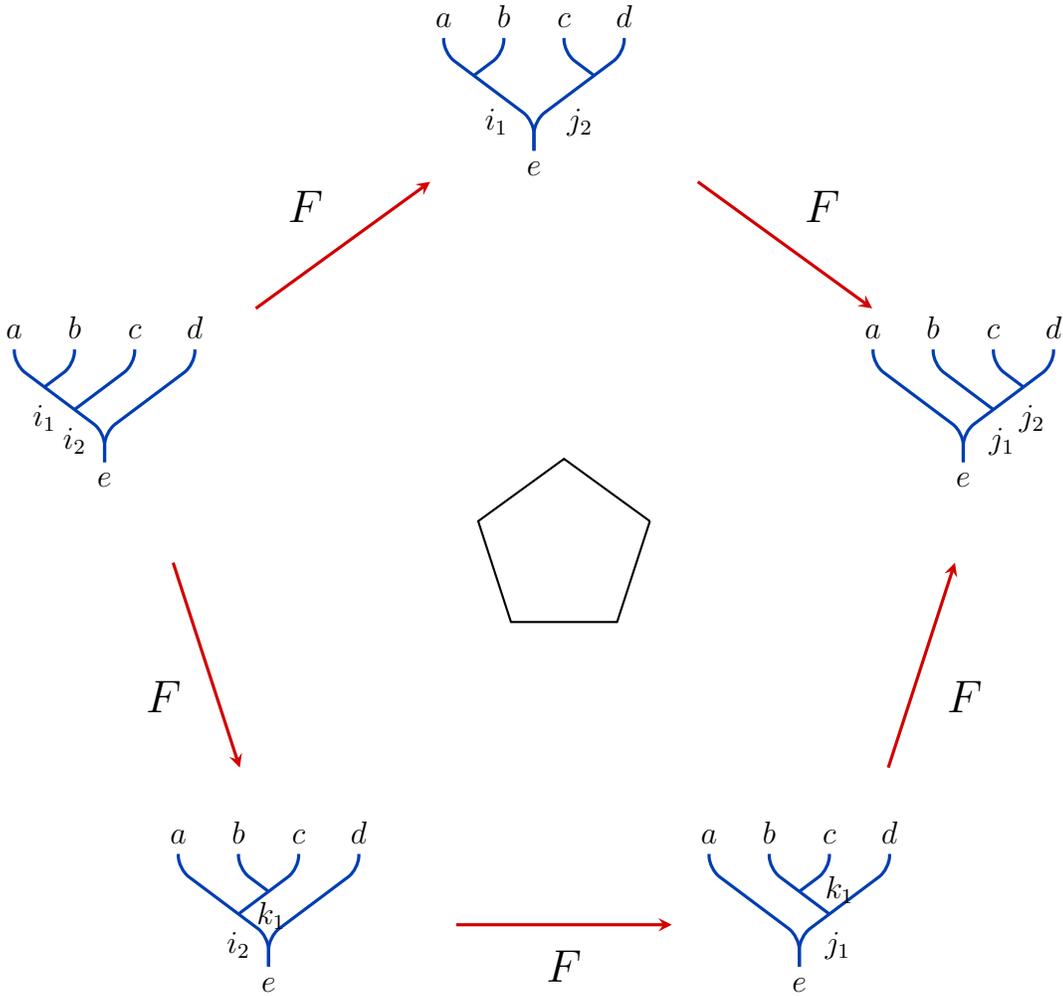

One can also consider a general fusion space $V^{a_1,\ldots, a_m}_{a_1',\ldots, a_n'}$, with arbitrarily many in and out particles. This can be decomposed into tensor products of the elementary fusion (splitting) spaces $V^c_{ab}$ ($V^{ab}_c$). The order of fusion (splitting) of the anyons should not matter and this is guaranteed to be the case if the $F$-matrices satisfy the so-called {\it pentagon identity}, schematized in Fig. \ref{fig:pentagon} for the $V^{abcd}_e$ case.  

From the figure, one can read off that the pentagon identity for this process is
\begin{equation}
\label{pentid}
\sum\limits_{j_1, j_2}~\left(F^{abj_2}_e\right)_{i_1j_1}\left(F^{i_1cd}_e\right)_{i_2j_2} = \sum\limits_{k_1, j_1, j_2}~\left(F^{bcd}_{j_1}\right)_{k_1 j_2}\left(F^{ak_1d}_e\right)_{i_2j_1}\left(F^{abc}_{i_2}\right)_{i_1k_1}.
\end{equation}
Fusing or splitting with the vacuum does not change the state of the anyon. This physical condition imposes that the corresponding $F^{abc}_d$ is trivial, i.e. $F^{abc}_d=1$ \cite{Bonderson2007NonAbelianAA}, when either of $a$, $b$ or $c$ is 1. Note that this is not true when anyon $d$ is the vacuum and we will see such cases soon. 

\begin{figure}[h]
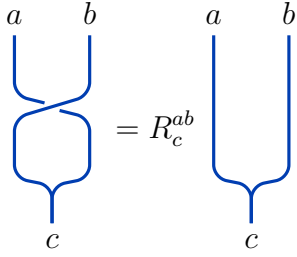

	\centering
	\begin{align*}
		\tikz[baseline=3.5ex]{
			\def\a{0}
			\def\b{0}
			\draw[very thick, blue, rounded corners=5] (0.5+\a, 2+\b) -- (0.5+\a, 1.2+\b) -- (\a - 0.5, 0.9+\b) -- (-0.5+\a, \b + 0.1) -- (\a, \b) -- (\a, \b - 0.5);
			\draw[very thick, blue, rounded corners=5] (-0.5+\a, 2+\b) -- (-0.5+\a, 1.2+\b) -- (\a-0.1, 1.1+\b);
			\draw[very thick, blue, rounded corners=5] (0.1+\a, 1+\b) -- (\a + 0.5, 0.9+\b) -- (0.5+\a, \b + 0.1) -- (\a, \b) -- (\a, \b - 0.5);
			\node[anchor=south] at (\a - 0.5, 2 + \b) {$a$};
			\node[anchor=south] at (\a + 0.5, 2 + \b) {$b$};
			\node[anchor=north] at (\a, \b - 0.5) {$c$};
		} = R_{c}^{ab}
		\tikz[baseline=3.5ex]{
			\def\a{0}
			\def\b{0}
			\draw[very thick, blue, rounded corners=5] (0.5+\a, 2+\b) -- (0.5+\a, \b + 0.1) -- (\a, \b) -- (\a, \b - 0.5);
			\draw[very thick, blue, rounded corners=5] (-0.5+\a, 2+\b) -- (-0.5+\a, \b + 0.1) -- (\a, \b) -- (\a, \b - 0.5);
			\node[anchor=south] at (\a - 0.5, 2 + \b) {$a$};
			\node[anchor=south] at (\a + 0.5, 2 + \b) {$b$};
			\node[anchor=north] at (\a, \b - 0.5) {$c$};
		} 
	\end{align*}
	\caption{The $R$-move untwists fusing anyon lines. 
		Note that the anyon line $b$ crosses over $a$. An undercrossing would correspond to  using $R^{-1}$ instead of $R$.}
	\label{fig:Rmove}
\end{figure}

Another operation that can be performed on anyonic lines is braiding, with the so-called {\it R-move}. This is done by the unitary $R$-matrix as defined in Fig. \ref{fig:Rmove}.


In equations this is
\begin{equation}
\ket{b,a;c} = R^{ab}_c\ket{a,b;c},
\end{equation}
for every $c$ in the fusion channel of $a$ and $b$. 
This `disentangling' of the anyon lines is essential for finding the braid group on these fusion spaces, as we shall see. 

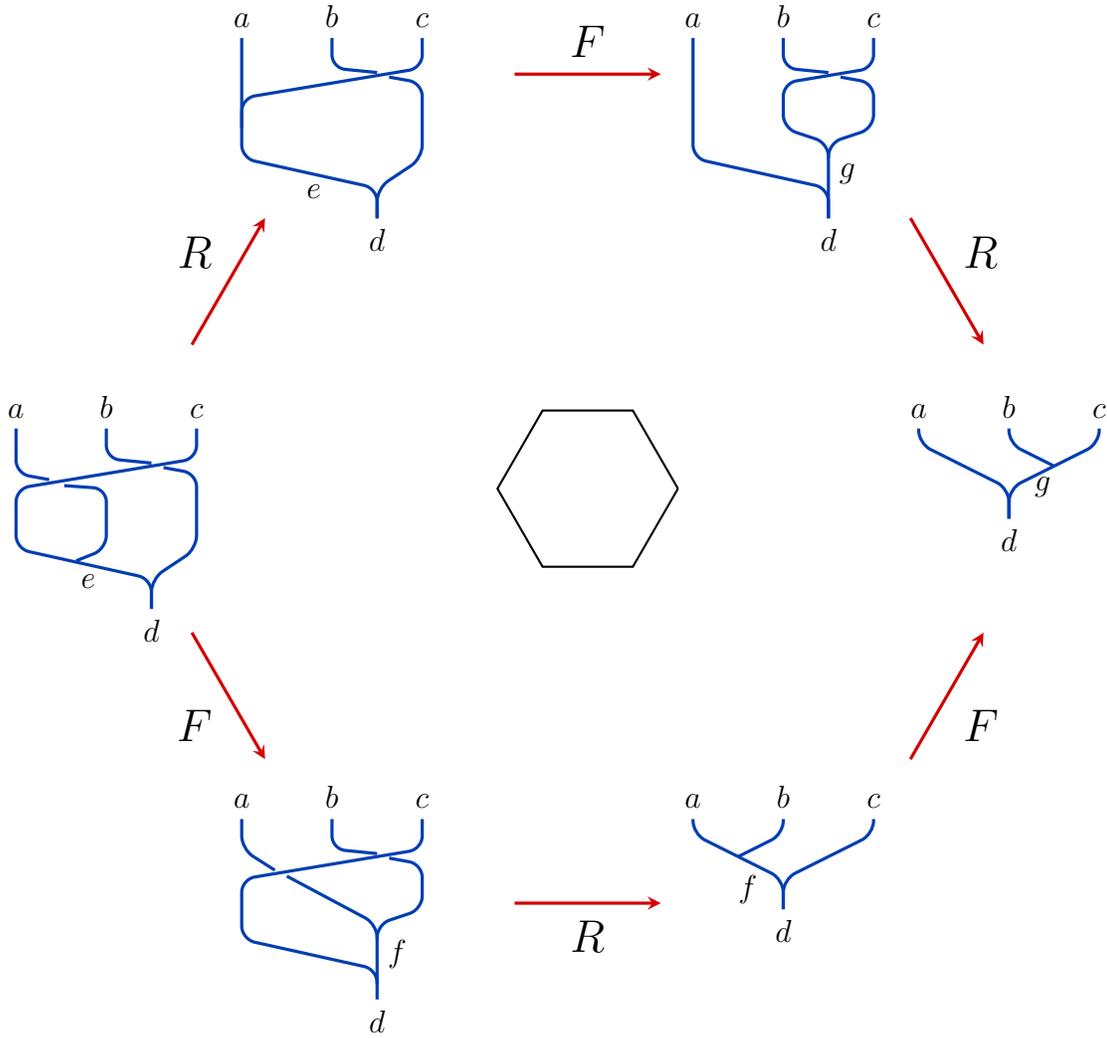
\begin{figure}[h!]
	\centering
	\begin{tikzpicture}[scale=0.4]
		\def\c{4}
		\def\d{-2}
		\def\s{3}
		\def\t{14}
		\def\r{0}
		\def\reduce{20}
		\draw[thick] ({\c+\s*cos(0+\r)}, {\d+\s*sin(0+\r)}) -- ({\c+\s*cos(60+\r)}, {\d+\s*sin(60+\r)}) -- ({\c+\s*cos(2*60+\r)}, {\d+\s*sin(2*60+\r)}) -- ({\c+\s*cos(3*60+\r)}, {\d+\s*sin(3*60+\r)}) -- ({\c+\s*cos(4*60+\r)}, {\d+\s*sin(4*60+\r)}) -- ({\c+\s*cos(5*60+\r)}, {\d+\s*sin(5*60+\r)}) -- ({\c+\s*cos(0+\r)}, {\d+\s*sin(0+\r)});
		\def\a{-15}
		\def\b{0}
		\draw[very thick, blue, rounded corners=5] (\a, \b) -- (\a, \b - 1.5) -- (\a + 1.1, \b - 1.7);
		\draw[very thick, blue, rounded corners=5] (\a + 1.6, \b - 1.9) -- (\a + 3, \b - 2) -- (\a + 3, \b - 4) -- (\a + 2, \b - 4.4);
		\draw[very thick, blue, rounded corners=5] (\a + 3, \b) -- (\a + 3, \b-1) -- (\a + 4.5, \b-1.15);
		\draw[very thick, blue, rounded corners=5] (\a + 4.9, \b-1.3) -- (\a + 6, \b-1.5) -- (\a + 6, \b - 4) -- (\a + 4.5, \b - 5) -- (\a + 4.5, \b - 6);
		\draw[very thick, blue, rounded corners=5] (\a + 6, \b) -- (\a + 6, \b-1) -- (\a, \b-2) -- (\a, \b - 4) -- (\a + 4.5, \b - 5) -- (\a + 4.5, \b - 6); 
		\node[anchor=south] at (\a, \b) {$a$};
		\node[anchor=south] at (\a + 3, \b) {$b$};
		\node[anchor=south] at (\a + 6, \b) {$c$};
		\node[anchor=north east] at (\a + 3, \b - 4.5) {$e$};
		\node[anchor=north] at (\a + 4.5, \b - 6) {$d$};
		\draw[very thick, red,  ->] ({\c+\t*cos(3*60+\r - \reduce)}, {\d+\t*sin(3*60+\r - \reduce)}) -- ({\c+\t*cos(2*60+\r + \reduce)}, {\d+\t*sin(2*60+\r + \reduce)});
		\node[anchor=south east] at ({\c+\t*cos(2.5*60+\r)}, {\d+\t*sin(2.5*60+\r)}) {\Large $R$};
		\def\a{-7.5}
		\def\b{12.99}
		\draw[very thick, blue, rounded corners=5] (\a, \b) -- (\a, \b - 4) -- (\a + 4.5, \b - 5) -- (\a + 4.5, \b - 6);
		\draw[very thick, blue, rounded corners=5] (\a + 3, \b) -- (\a + 3, \b-1) -- (\a + 4.5, \b-1.15);
		\draw[very thick, blue, rounded corners=5] (\a + 4.9, \b-1.3) -- (\a + 6, \b-1.5) -- (\a + 6, \b - 4) -- (\a + 4.5, \b - 5) -- (\a + 4.5, \b - 6);
		\draw[very thick, blue, rounded corners=5] (\a + 6, \b) -- (\a + 6, \b-1) -- (\a, \b-2) -- (\a, \b - 3); 
		\node[anchor=south] at (\a, \b) {$a$};
		\node[anchor=south] at (\a + 3, \b) {$b$};
		\node[anchor=south] at (\a + 6, \b) {$c$};
		\node[anchor=north east] at (\a + 3, \b - 4.5) {$e$};
		\node[anchor=north] at (\a + 4.5, \b - 6) {$d$};
		\draw[very thick, red,  ->] ({\c+\t*cos(2*60+\r - \reduce)}, {\d+\t*sin(2*60+\r - \reduce)}) -- ({\c+\t*cos(60+\r + \reduce)}, {\d+\t*sin(60+\r + \reduce)});
		\node[anchor=south] at ({\c+\t*cos(1.5*60+\r)}, {\d+\t*sin(1.5*60+\r)}) {\Large $F$};
		\def\a{7.5}
		\def\b{12.99}
		\draw[very thick, blue, rounded corners=5] (\a, \b) -- (\a, \b - 4) -- (\a + 4.5, \b - 5) -- (\a + 4.5, \b - 6);
		\draw[very thick, blue, rounded corners=5] (\a + 3, \b) -- (\a + 3, \b-1) -- (\a + 4.5, \b-1.15);
		\draw[very thick, blue, rounded corners=5] (\a + 4.9, \b-1.3) -- (\a + 6, \b-1.5) -- (\a + 6, \b - 3) -- (\a + 4.5, \b - 3.5) -- (\a + 4.5, \b - 6);
		\draw[very thick, blue, rounded corners=5] (\a + 6, \b) -- (\a + 6, \b-1) -- (\a + 3, \b-1.5) -- (\a + 3, \b - 3) -- (\a + 4.5, \b - 3.5) -- (\a + 4.5, \b - 4); 
		\node[anchor=south] at (\a, \b) {$a$};
		\node[anchor=south] at (\a + 3, \b) {$b$};
		\node[anchor=south] at (\a + 6, \b) {$c$};
		\node[anchor=west] at (\a + 4.5, \b - 4.5) {$g$};
		\node[anchor=north] at (\a + 4.5, \b - 6) {$d$};
		\draw[very thick, red,  ->] ({\c+\t*cos(60+\r - \reduce)}, {\d+\t*sin(60+\r - \reduce)}) -- ({\c+\t*cos(0+\r + \reduce)}, {\d+\t*sin(0+\r + \reduce)});
		\node[anchor=south west] at ({\c+\t*cos(0.5*60+\r)}, {\d+\t*sin(0.5*60+\r)}) {\Large $R$};
		\def\a{15}
		\def\b{0}
		\draw[very thick, blue, rounded corners=5] (\a, \b) -- (\a, \b - 0.5) -- (\a + 3, \b - 2) -- (\a + 3, \b - 3);
		\draw[very thick, blue, rounded corners=5] (\a + 3, \b) -- (\a + 3, \b-0.5) -- (\a + 4.5, \b-1.25);
		\draw[very thick, blue, rounded corners=5] (\a + 6, \b) -- (\a + 6, \b - 0.5) -- (\a + 3, \b - 2) -- (\a + 3, \b - 3); 
		\node[anchor=south] at (\a, \b) {$a$};
		\node[anchor=south] at (\a + 3, \b) {$b$};
		\node[anchor=south] at (\a + 6, \b) {$c$};
		\node[anchor=north west] at (\a + 3.5, \b - 1.25) {$g$};
		\node[anchor=north] at (\a + 3, \b - 3) {$d$};
		\draw[very thick, red,  ->] ({\c+\t*cos(3*60+\r + \reduce)}, {\d+\t*sin(3*60+\r + \reduce)}) -- ({\c+\t*cos(4*60+\r - \reduce)}, {\d+\t*sin(4*60+\r - \reduce)});
		\node[anchor=north east] at ({\c+\t*cos(3.5*60+\r)}, {\d+\t*sin(3.5*60+\r)}) {\Large $F$};
		\def\a{-7.5}
		\def\b{-12.99}
		\draw[very thick, blue, rounded corners=5] (\a, \b) -- (\a, \b - 1) -- (\a + 1.1, \b - 1.7);
		\draw[very thick, blue, rounded corners=5] (\a + 1.5, \b - 1.9) -- (\a + 4.5, \b - 3.5) -- (\a + 4.5, \b - 4);
		\draw[very thick, blue, rounded corners=5] (\a + 3, \b) -- (\a + 3, \b-1) -- (\a + 4.5, \b-1.15);
		\draw[very thick, blue, rounded corners=5] (\a + 4.9, \b-1.3) -- (\a + 6, \b-1.5) -- (\a + 6, \b - 3) -- (\a + 4.5, \b - 3.5) -- (\a + 4.5, \b - 5.5);
		\draw[very thick, blue, rounded corners=5] (\a + 6, \b) -- (\a + 6, \b-1) -- (\a, \b-2) -- (\a, \b - 4) -- (\a + 4.5, \b - 5) -- (\a + 4.5, \b - 6); 
		\node[anchor=south] at (\a, \b) {$a$};
		\node[anchor=south] at (\a + 3, \b) {$b$};
		\node[anchor=south] at (\a + 6, \b) {$c$};
		\node[anchor=west] at (\a + 4.5, \b - 4.5) {$f$};
		\node[anchor=north] at (\a + 4.5, \b - 6) {$d$};
		\draw[very thick, red,  ->] ({\c+\t*cos(4*60+\r + \reduce)}, {\d+\t*sin(4*60+\r + \reduce)}) -- ({\c+\t*cos(5*60+\r - \reduce)}, {\d+\t*sin(5*60+\r - \reduce)});
		\node[anchor=north] at ({\c+\t*cos(4.5*60+\r)}, {\d+\t*sin(4.5*60+\r)}) {\Large $R$};
		\def\a{7.5}
		\def\b{-12.99}
		\draw[very thick, blue, rounded corners=5] (\a, \b) -- (\a, \b - 0.5) -- (\a + 3, \b - 2) -- (\a + 3, \b - 3);
		\draw[very thick, blue, rounded corners=5] (\a + 3, \b) -- (\a + 3, \b-0.5) -- (\a + 1.5, \b-1.25);
		\draw[very thick, blue, rounded corners=5] (\a + 6, \b) -- (\a + 6, \b - 0.5) -- (\a + 3, \b - 2) -- (\a + 3, \b - 3); 
		\node[anchor=south] at (\a, \b) {$a$};
		\node[anchor=south] at (\a + 3, \b) {$b$};
		\node[anchor=south] at (\a + 6, \b) {$c$};
		\node[anchor=north east] at (\a + 2.5, \b - 1.5) {$f$};
		\node[anchor=north] at (\a + 3, \b - 3) {$d$};
		\draw[very thick, red,  ->] ({\c+\t*cos(5*60+\r + \reduce)}, {\d+\t*sin(5*60+\r + \reduce)}) -- ({\c+\t*cos(0+\r - \reduce)}, {\d+\t*sin(0+\r - \reduce)});
		\node[anchor=north west] at ({\c+\t*cos(5.5*60+\r)}, {\d+\t*sin(5.5*60+\r)}) {\Large $F$};
	\end{tikzpicture}
	\caption{The compatibility between the $F$- and $R$-matrices is encoded in the diagram above, in which one can go from left to right along the upper or lower path. The equivalence of the two procedures leads to the hexagon identity \eqref{hexagonid}.}
	\label{fig:hexagon}
\end{figure}

While fusing anyons we can braid some of them in the process. This can be done in several different orders and for them to be consistent we require the $F$- and $R$-matrices to satisfy the so-called {\it hexagon identity}, see Fig. \ref{fig:hexagon}. This reads 

\begin{equation}
\label{hexagonid}
\sum\limits_g~R^{ac}_e\left(F^{acb}_d\right)_{eg}R^{bc}_g = \sum\limits_{f,g}~\left(F^{cab}_d\right)_{ef}R^{cf}_d\left(F^{abc}_d\right)_{fg}.
\end{equation}
This identity guarantees the compatibility between the $F$- and $R$-matrices.
This relation will become quite ubiquitous and crucial for the construction of the braid group on fusion spaces.

\begin{figure}[h!]
	\centering
	\begin{tikzpicture}[scale=0.4]
		\def\c{4}
		\def\d{-2}
		\def\s{3}
		\def\t{14}
		\def\r{0}
		\def\reduce{20}
		\draw[thick] ({\c+\s*cos(0+\r)}, {\d+\s*sin(0+\r)}) -- ({\c+\s*cos(60+\r)}, {\d+\s*sin(60+\r)}) -- ({\c+\s*cos(2*60+\r)}, {\d+\s*sin(2*60+\r)}) -- ({\c+\s*cos(3*60+\r)}, {\d+\s*sin(3*60+\r)}) -- ({\c+\s*cos(4*60+\r)}, {\d+\s*sin(4*60+\r)}) -- ({\c+\s*cos(5*60+\r)}, {\d+\s*sin(5*60+\r)}) -- ({\c+\s*cos(0+\r)}, {\d+\s*sin(0+\r)});
		\def\a{-15}
		\def\b{0}
		\draw[very thick, blue, rounded corners=5] (\a, \b) -- (\a, \b - 1.5) -- (\a + 3, \b - 2) -- (\a + 3, \b - 4) -- (\a + 2, \b - 4.4);
		\draw[very thick, blue, rounded corners=5] (\a + 3, \b) -- (\a + 3, \b-1) -- (\a + 6, \b-1.5) -- (\a + 6, \b - 4) -- (\a + 4.5, \b - 5) -- (\a + 4.5, \b - 6);
		\draw[very thick, blue, rounded corners=5] (\a + 6, \b) -- (\a + 6, \b - 1) -- (\a + 4.7, \b - 1.2);
		\draw[very thick, blue, rounded corners=5] (\a + 4.4, \b - 1.4) -- (\a + 1.8, \b-1.7);
		\draw[very thick, blue, rounded corners=5] (\a + 1.3, \b-1.9) -- (\a, \b-2) -- (\a, \b - 4) -- (\a + 4.5, \b - 5) -- (\a + 4.5, \b - 6); 
		\node[anchor=south] at (\a, \b) {$a$};
		\node[anchor=south] at (\a + 3, \b) {$b$};
		\node[anchor=south] at (\a + 6, \b) {$c$};
		\node[anchor=north east] at (\a + 3, \b - 4.5) {$e$};
		\node[anchor=north] at (\a + 4.5, \b - 6) {$d$};
		\draw[very thick, red,  ->] ({\c+\t*cos(3*60+\r - \reduce)}, {\d+\t*sin(3*60+\r - \reduce)}) -- ({\c+\t*cos(2*60+\r + \reduce)}, {\d+\t*sin(2*60+\r + \reduce)});
		\node[anchor=south east] at ({\c+\t*cos(2.5*60+\r)}, {\d+\t*sin(2.5*60+\r)}) {\Large $R^{-1}$};
		\def\a{-7.5}
		\def\b{12.99}
		\draw[very thick, blue, rounded corners=5] (\a, \b) -- (\a, \b - 4) -- (\a + 4.5, \b - 5) -- (\a + 4.5, \b - 6);
		\draw[very thick, blue, rounded corners=5] (\a + 3, \b) -- (\a + 3, \b-1) -- (\a + 6, \b-1.5) -- (\a + 6, \b - 4) -- (\a + 4.5, \b - 5) -- (\a + 4.5, \b - 6);
		\draw[very thick, blue, rounded corners=5] (\a + 6, \b) -- (\a + 6, \b - 1) -- (\a + 4.9, \b - 1.2);
		\draw[very thick, blue, rounded corners=5] (\a + 4.3, \b-1.4) -- (\a, \b-2) -- (\a, \b - 3); 
		\node[anchor=south] at (\a, \b) {$a$};
		\node[anchor=south] at (\a + 3, \b) {$b$};
		\node[anchor=south] at (\a + 6, \b) {$c$};
		\node[anchor=north east] at (\a + 3, \b - 4.5) {$e$};
		\node[anchor=north] at (\a + 4.5, \b - 6) {$d$};
		\draw[very thick, red,  ->] ({\c+\t*cos(2*60+\r - \reduce)}, {\d+\t*sin(2*60+\r - \reduce)}) -- ({\c+\t*cos(60+\r + \reduce)}, {\d+\t*sin(60+\r + \reduce)});
		\node[anchor=south] at ({\c+\t*cos(1.5*60+\r)}, {\d+\t*sin(1.5*60+\r)}) {\Large $F$};
		\def\a{7.5}
		\def\b{12.99}
		\draw[very thick, blue, rounded corners=5] (\a, \b) -- (\a, \b - 4) -- (\a + 4.5, \b - 5) -- (\a + 4.5, \b - 6);
		\draw[very thick, blue, rounded corners=5] (\a + 3, \b) -- (\a + 3, \b-1) -- (\a + 6, \b-1.5) -- (\a + 6, \b - 3) -- (\a + 4.5, \b - 3.5) -- (\a + 4.5, \b - 6);
		\draw[very thick, blue, rounded corners=5] (\a + 6, \b) -- (\a + 6, \b - 1) -- (\a + 4.9, \b - 1.2);
		\draw[very thick, blue, rounded corners=5] (\a + 4.3, \b-1.4) -- (\a + 3, \b-1.5) -- (\a + 3, \b - 3) -- (\a + 4.5, \b - 3.5) -- (\a + 4.5, \b - 4); 
		\node[anchor=south] at (\a, \b) {$a$};
		\node[anchor=south] at (\a + 3, \b) {$b$};
		\node[anchor=south] at (\a + 6, \b) {$c$};
		\node[anchor=west] at (\a + 4.5, \b - 4.5) {$g$};
		\node[anchor=north] at (\a + 4.5, \b - 6) {$d$};
		\draw[very thick, red,  ->] ({\c+\t*cos(60+\r - \reduce)}, {\d+\t*sin(60+\r - \reduce)}) -- ({\c+\t*cos(0+\r + \reduce)}, {\d+\t*sin(0+\r + \reduce)});
		\node[anchor=south west] at ({\c+\t*cos(0.5*60+\r)}, {\d+\t*sin(0.5*60+\r)}) {\Large $R^{-1}$};
		\def\a{15}
		\def\b{0}
		\draw[very thick, blue, rounded corners=5] (\a, \b) -- (\a, \b - 0.5) -- (\a + 3, \b - 2) -- (\a + 3, \b - 3);
		\draw[very thick, blue, rounded corners=5] (\a + 3, \b) -- (\a + 3, \b-0.5) -- (\a + 4.5, \b-1.25);
		\draw[very thick, blue, rounded corners=5] (\a + 6, \b) -- (\a + 6, \b - 0.5) -- (\a + 3, \b - 2) -- (\a + 3, \b - 3); 
		\node[anchor=south] at (\a, \b) {$a$};
		\node[anchor=south] at (\a + 3, \b) {$b$};
		\node[anchor=south] at (\a + 6, \b) {$c$};
		\node[anchor=north west] at (\a + 3.5, \b - 1.25) {$g$};
		\node[anchor=north] at (\a + 3, \b - 3) {$d$};
		\draw[very thick, red,  ->] ({\c+\t*cos(3*60+\r + \reduce)}, {\d+\t*sin(3*60+\r + \reduce)}) -- ({\c+\t*cos(4*60+\r - \reduce)}, {\d+\t*sin(4*60+\r - \reduce)});
		\node[anchor=north east] at ({\c+\t*cos(3.5*60+\r)}, {\d+\t*sin(3.5*60+\r)}) {\Large $F$};
		\def\a{-7.5}
		\def\b{-12.99}
		\draw[very thick, blue, rounded corners=5] (\a, \b) -- (\a, \b - 1) -- (\a + 4.5, \b - 3.5) -- (\a + 4.5, \b - 4);
		\draw[very thick, blue, rounded corners=5] (\a + 3, \b) -- (\a + 3, \b-1) -- (\a + 6, \b-1.5) -- (\a + 6, \b - 3) -- (\a + 4.5, \b - 3.5) -- (\a + 4.5, \b - 5.5);
		\draw[very thick, blue, rounded corners=5] (\a + 6, \b) -- (\a + 6, \b - 1) -- (\a + 4.7, \b - 1.2);
		\draw[very thick, blue, rounded corners=5] (\a + 4.4, \b - 1.4) -- (\a + 1.8, \b-1.7);
		\draw[very thick, blue, rounded corners=5] (\a + 1.3, \b-1.9) -- (\a, \b-2) -- (\a, \b - 4) -- (\a + 4.5, \b - 5) -- (\a + 4.5, \b - 6); 
		\node[anchor=south] at (\a, \b) {$a$};
		\node[anchor=south] at (\a + 3, \b) {$b$};
		\node[anchor=south] at (\a + 6, \b) {$c$};
		\node[anchor=west] at (\a + 4.5, \b - 4.5) {$f$};
		\node[anchor=north] at (\a + 4.5, \b - 6) {$d$};
		\draw[very thick, red,  ->] ({\c+\t*cos(4*60+\r + \reduce)}, {\d+\t*sin(4*60+\r + \reduce)}) -- ({\c+\t*cos(5*60+\r - \reduce)}, {\d+\t*sin(5*60+\r - \reduce)});
		\node[anchor=north] at ({\c+\t*cos(4.5*60+\r)}, {\d+\t*sin(4.5*60+\r)}) {\Large $R^{-1}$};
		\def\a{7.5}
		\def\b{-12.99}
		\draw[very thick, blue, rounded corners=5] (\a, \b) -- (\a, \b - 0.5) -- (\a + 3, \b - 2) -- (\a + 3, \b - 3);
		\draw[very thick, blue, rounded corners=5] (\a + 3, \b) -- (\a + 3, \b-0.5) -- (\a + 1.5, \b-1.25);
		\draw[very thick, blue, rounded corners=5] (\a + 6, \b) -- (\a + 6, \b - 0.5) -- (\a + 3, \b - 2) -- (\a + 3, \b - 3); 
		\node[anchor=south] at (\a, \b) {$a$};
		\node[anchor=south] at (\a + 3, \b) {$b$};
		\node[anchor=south] at (\a + 6, \b) {$c$};
		\node[anchor=north east] at (\a + 2.5, \b - 1.5) {$f$};
		\node[anchor=north] at (\a + 3, \b - 3) {$d$};
		\draw[very thick, red,  ->] ({\c+\t*cos(5*60+\r + \reduce)}, {\d+\t*sin(5*60+\r + \reduce)}) -- ({\c+\t*cos(0+\r - \reduce)}, {\d+\t*sin(0+\r - \reduce)});
		\node[anchor=north west] at ({\c+\t*cos(5.5*60+\r)}, {\d+\t*sin(5.5*60+\r)}) {\Large $F$};
	\end{tikzpicture}
	\caption{The same idea applied to the $F$- and $R^{-1}$-matrices leads to the alternate hexagon identity \eqref{hexagonidalt}.}
	\label{fig:hexagon2}
\end{figure}
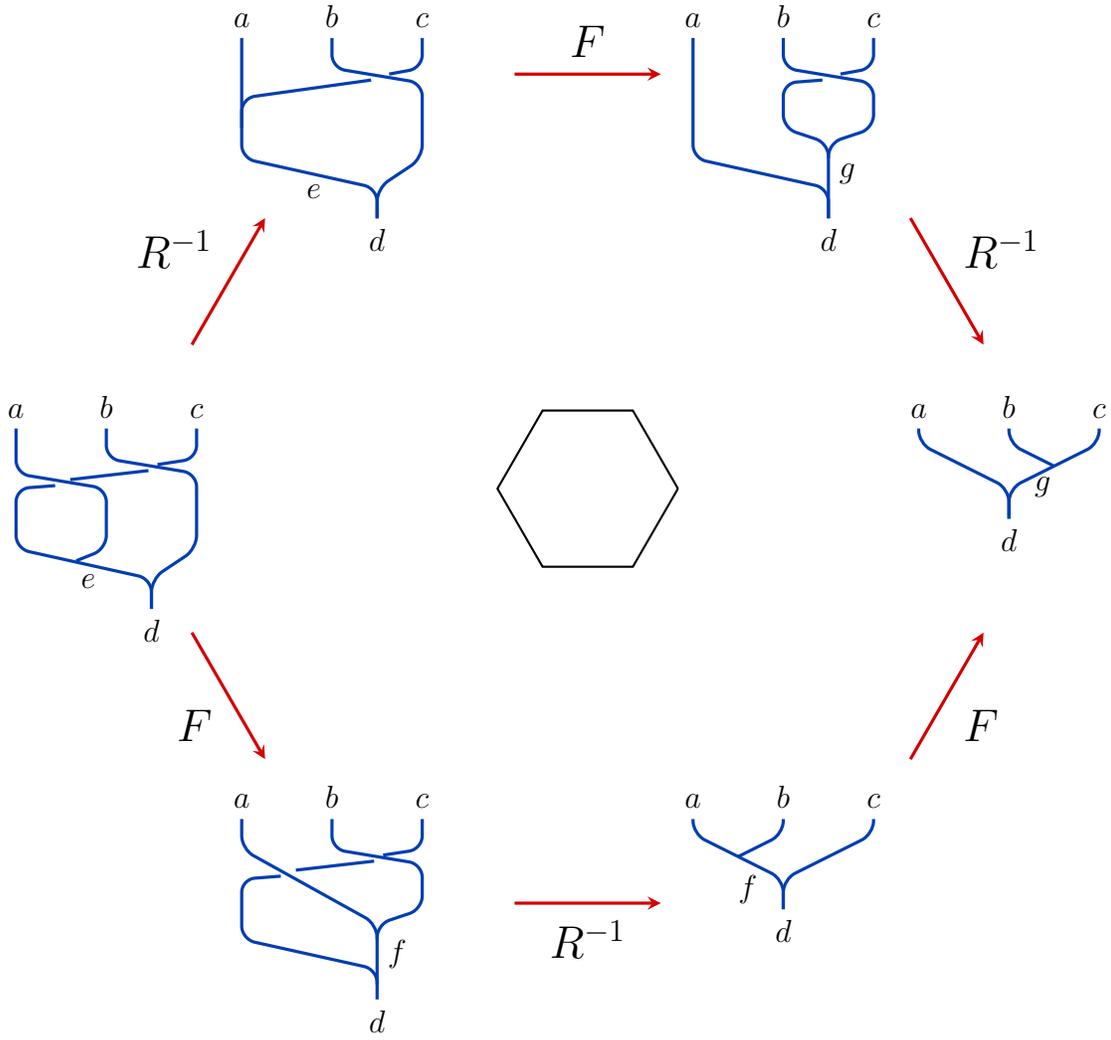

An alternative hexagon identity is obtained when the $c$ line in Fig. \ref{fig:hexagon} is wound below the $a$ and $b$  lines, as shown in Fig. \ref{fig:hexagon2}.

This leads to the equation
\begin{equation}
\label{hexagonidalt}
\sum\limits_g~\left(R^{-1}\right)^{ac}_e\left(F^{acb}_d\right)_{eg}\left(R^{-1}\right)^{bc}_g = \sum\limits_{f,g}~\left(F^{cab}_d\right)_{ef}\left(R^{-1}\right)^{cf}_d\left(F^{abc}_d\right)_{fg}.
\end{equation}

There are many more quantities that can be defined on fusion spaces and we refer the interested reader to \cite{Bonderson_2008}. However, for our purposes of finding the braid group, the quantities defined above and their relations will suffice. Solving pentagon and hexagon identities is enough to define anyonic systems, as stated by the MacLane coherence theorem \cite{maclane}. This is however very difficult to do in general. For this reason we shall only work with specific examples, based on the $SU(2)_k$ quantum group. 

\subsection{$SU(2)_k$ anyonic systems}
\label{subsec:SU2kanyons}

The quantum group $SU(2)_k$ is a rich source of examples of anyonic systems \cite{pachos2012introduction}. At level $k$, its irreducible representations (IRR's) are labelled by $\mathcal{C}=\{0, \frac{1}{2},1, \ldots, \frac{k}{2}\}$, which can be interpreted as different anyonic species. Analogously to what is done for $SU(2)$, one can combine IRR's (`add angular momenta') to form other representations. The {\it fusion rule} to be followed to this scope is
\begin{equation}
j_1\times j_2 = |j_1-j_2| \oplus |j_1-j_2|+1 \oplus \cdots \oplus \textrm{min}~\{j_1+j_2, k-\left(j_1+j_2\right)\}.
\end{equation}
We now illustrate this for specific values of $k$.

\paragraph{$k=2$ - Ising anyons}
The first system we consider is provided by $SU(2)_2$. This has only three possible IRR's labelled by $\{0,\frac{1}{2},1\}$, with fusion table given in Table \ref{tab:k=2fusion}.
\begin{table}[h!]
  \begin{center}
    \begin{tabular}{|c|ccc|} 
     \hline
     $\times$ & 0 & $\frac{1}{2}$ & 1\\
     \hline
      0 & 0 & $\frac{1}{2}$ & 1\\
      $\frac{1}{2}$ & $\frac{1}{2}$ & $0+1$ & $\frac{1}{2}$\\
      1 & 1 & $\frac{1}{2}$ & 0\\
\hline
\end{tabular}
    \caption{Fusion rules at level $k=2$.}
    \label{tab:k=2fusion}
  \end{center}
\end{table}
Identifying 0 with the vacuum, $\frac{1}{2}$ with the anyonic species $\sigma$ and 1 with the anyonic species $\psi$, one obtains the fusion rules
\begin{eqnarray}
1\times a & = & a\times 1 = a,\qquad a\in\{1, \sigma, \psi\} \nonumber \\
\sigma\times\psi & = & \psi\times\sigma = \sigma,\qquad \psi\times\psi=1, \qquad 
\sigma\times\sigma = 1 + \psi,
\end{eqnarray}
which are precisely the fusion rules of the {\it Ising anyons}. 

\paragraph{$k=3$ - Fibonacci anyons}
At level $k=3$ the fusion table is given by Table \ref{tab:k=3fusion}.
\begin{table}[h!]
  \begin{center}
    \begin{tabular}{|c|cccc|} 
     \hline
     $\times$ & 0 & $\frac{1}{2}$ & 1 & $\frac{3}{2}$\\
     \hline
      0 & 0 & $\frac{1}{2}$ & 1 & $\frac{3}{2}$\\
      $\frac{1}{2}$ & $\frac{1}{2}$ & $0+1$ & $\frac{1}{2}$ + $\frac{3}{2}$ & 1\\
      1 & 1 & $\frac{1}{2}$ + $\frac{3}{2}$ & 0 + 1 & $\frac{1}{2}$\\
      $\frac{3}{2}$ &  $\frac{3}{2}$ &  1  & $\frac{1}{2}$ & 0 \\
\hline
\end{tabular}
    \caption{Fusion rules at level $k=3$.}
    \label{tab:k=3fusion}
  \end{center}
\end{table}
Focusing only on the integer representations, upon identifying 0 with the vacuum and 1 with the anyonic species $\tau$, one obtains the fusion rules
\begin{equation}
1\times\tau = \tau\times 1 = \tau,\qquad \tau\times\tau = 1 +\tau
\end{equation}
of the {\it Fibonacci anyons}. Note that we have not included the half-integral IRR's in the definition of the Fibonacci anyons. As we shall see in Sec. \ref{subsec:basisfusionspace}, the fusion of the $\tau$'s result in a fusion basis whose dimension scales as the Fibonacci number. This is the origin of the nomenclature for these anyons. However this scaling property does not hold when we fuse the half-integral anyons and hence the half-integral anyons are not considered in the definition of the Fibonacci anyons.


\paragraph{$k=4$ - Jones-Kauffman anyons}
At level $k=4$ the IRR's are labelled by $\{0, \frac{1}{2}, 1, \frac{3}{2}, 2\}$. The fusion table of the integer IRR's of $SU(2)_4$ is reported in Table \ref{tab:k=4fusion}.
\begin{table}[h!]
  \begin{center}
    \begin{tabular}{|c|ccc|} 
     \hline
     $\times$ & 0 & 1 & 2\\
     \hline
      0 & 0 & 1 & 2\\
      1 & 1 & $0+1+2$ & 1\\
      2 & 2 & 1 & 0\\
\hline
\end{tabular}
    \caption{Fusion rules at level $k=4$ for the integer IRR's.}
    \label{tab:k=4fusion}
  \end{center}
\end{table}
Identifying 0 with the vacuum, 1 with the anyonic species $\tau$ and 2 with the anyonic species $\mu$ gives the fusion rules (omitting the trivial ones involving fusing with $1$)
\begin{eqnarray}
\tau\times\tau = 1 + \tau + \mu, \qquad
\mu\times\tau  =  \tau\times\mu = \tau, \qquad
\mu\times\mu  = 1
\end{eqnarray}
of the {\it Jones-Kauffman anyons}. For the same reason as in the definition of the Fibonacci anyons the half-integral IRR's are not included in the defintion of the Jones-Kauffman anyons.

\subsection{Basis of the fusion spaces}
\label{subsec:basisfusionspace}
\begin{figure}[h]
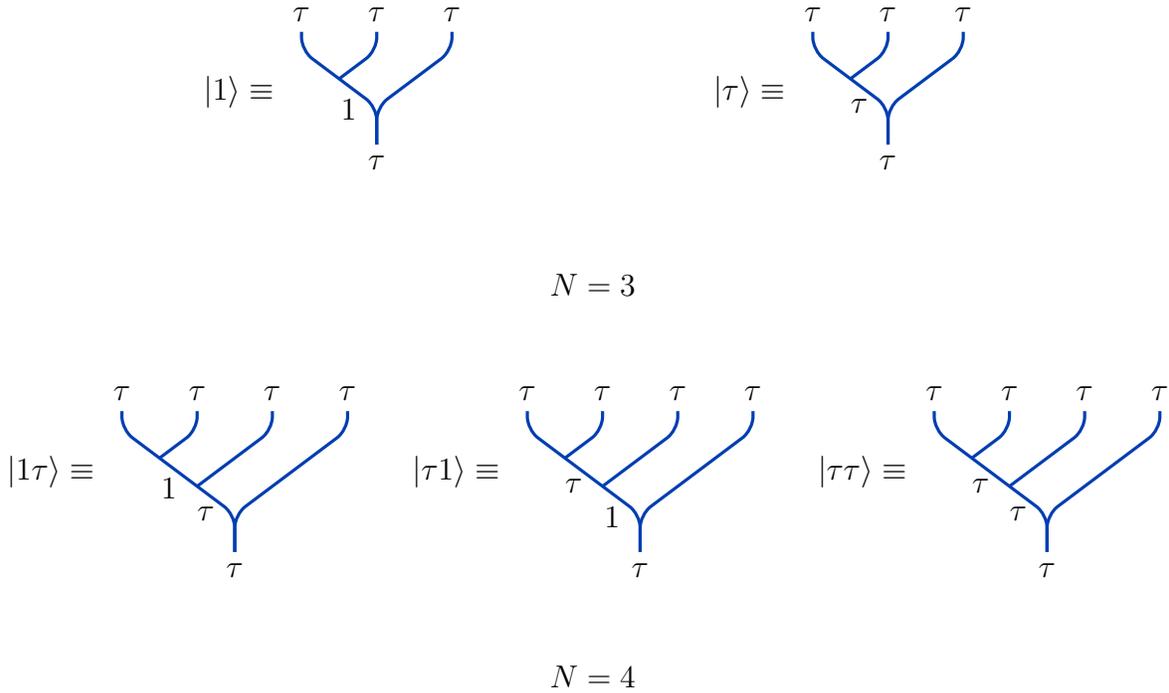

	\centering
	\begin{align*}
		|1 \rangle &\equiv
		\tikz[scale=0.5, baseline=-5ex]{
			\def\a{0}
			\def\b{0}
			\draw[very thick, blue, rounded corners=5] (\a, \b) -- (\a, \b-0.5) -- (2+\a, \b-2) -- (2+\a, \b - 3);
			\draw[very thick, blue, rounded corners=5] (\a + 2, \b) -- (\a + 2, \b-0.5) -- (1+\a, \b-1.25);
			\draw[very thick, blue, rounded corners=5] (\a + 4, \b) -- (\a + 4, \b-0.5) -- (2+\a, \b-2) -- (2+\a, \b - 3); 
			\node[anchor=south] at (\a, \b) {$\tau$};
			\node[anchor=south] at (\a + 2, \b) {$\tau$};
			\node[anchor=south] at (\a + 4, \b) {$\tau$};
			\node[anchor=north east] at (\a + 1.75, \b - 1.5) {$1$};
			\node[anchor=north] at (\a + 2, \b - 3) {$\tau$};
		}&
		|\tau \rangle &\equiv
		\tikz[scale=0.5, baseline=-5ex]{
			\def\a{0}
			\def\b{0}
			\draw[very thick, blue, rounded corners=5] (\a, \b) -- (\a, \b-0.5) -- (2+\a, \b-2) -- (2+\a, \b - 3);
			\draw[very thick, blue, rounded corners=5] (\a + 2, \b) -- (\a + 2, \b-0.5) -- (1+\a, \b-1.25);
			\draw[very thick, blue, rounded corners=5] (\a + 4, \b) -- (\a + 4, \b-0.5) -- (2+\a, \b-2) -- (2+\a, \b - 3); 
			\node[anchor=south] at (\a, \b) {$\tau$};
			\node[anchor=south] at (\a + 2, \b) {$\tau$};
			\node[anchor=south] at (\a + 4, \b) {$\tau$};
			\node[anchor=north east] at (\a + 1.75, \b - 1.5) {$\tau$};
			\node[anchor=north] at (\a + 2, \b - 3) {$\tau$};
		}
	\end{align*}

	\begin{align*}
		N = 3
	\end{align*}
	\begin{align*}
		|1\tau \rangle &\equiv \tikz[scale=0.5, baseline=-5ex]{
		\def\a{0}
		\def\b{0}
		\draw[very thick, blue, rounded corners=5] (\a, \b) -- (\a, \b-0.5) -- (3+\a, \b-2.75) -- (3+\a, \b - 3.75);
		\draw[very thick, blue, rounded corners=5] (\a + 2, \b) -- (\a + 2, \b-0.5) -- (1+\a, \b-1.25);
		\draw[very thick, blue, rounded corners=5] (\a + 4, \b) -- (\a + 4, \b-0.5) -- (2+\a, \b-2);
		\draw[very thick, blue, rounded corners=5] (\a + 6, \b) -- (\a + 6, \b-0.5) -- (3+\a, \b-2.75) -- (3+\a, \b - 3.75);  
		\node[anchor=south] at (\a, \b) {$\tau$};
		\node[anchor=south] at (\a + 2, \b) {$\tau$};
		\node[anchor=south] at (\a + 4, \b) {$\tau$};
		\node[anchor=south] at (\a + 6, \b) {$\tau$};
		\node[anchor=north east] at (\a + 1.75, \b - 1.5) {$1$};
		\node[anchor=north east] at (\a + 2.75, \b - 2.25) {$\tau$};
		\node[anchor=north] at (\a + 3, \b - 3.75) {$\tau$};
		}&
	|\tau 1 \rangle &\equiv \tikz[scale=0.5, baseline=-5ex]{
		\def\a{0}
		\def\b{0}
		\draw[very thick, blue, rounded corners=5] (\a, \b) -- (\a, \b-0.5) -- (3+\a, \b-2.75) -- (3+\a, \b - 3.75);
		\draw[very thick, blue, rounded corners=5] (\a + 2, \b) -- (\a + 2, \b-0.5) -- (1+\a, \b-1.25);
		\draw[very thick, blue, rounded corners=5] (\a + 4, \b) -- (\a + 4, \b-0.5) -- (2+\a, \b-2);
		\draw[very thick, blue, rounded corners=5] (\a + 6, \b) -- (\a + 6, \b-0.5) -- (3+\a, \b-2.75) -- (3+\a, \b - 3.75);  
		\node[anchor=south] at (\a, \b) {$\tau$};
		\node[anchor=south] at (\a + 2, \b) {$\tau$};
		\node[anchor=south] at (\a + 4, \b) {$\tau$};
		\node[anchor=south] at (\a + 6, \b) {$\tau$};
		\node[anchor=north east] at (\a + 1.75, \b - 1.5) {$\tau$};
		\node[anchor=north east] at (\a + 2.75, \b - 2.25) {$1$};
		\node[anchor=north] at (\a + 3, \b - 3.75) {$\tau$};
	}&
	|\tau\tau \rangle &\equiv \tikz[scale=0.5, baseline=-5ex]{
		\def\a{0}
		\def\b{0}
		\draw[very thick, blue, rounded corners=5] (\a, \b) -- (\a, \b-0.5) -- (3+\a, \b-2.75) -- (3+\a, \b - 3.75);
		\draw[very thick, blue, rounded corners=5] (\a + 2, \b) -- (\a + 2, \b-0.5) -- (1+\a, \b-1.25);
		\draw[very thick, blue, rounded corners=5] (\a + 4, \b) -- (\a + 4, \b-0.5) -- (2+\a, \b-2);
		\draw[very thick, blue, rounded corners=5] (\a + 6, \b) -- (\a + 6, \b-0.5) -- (3+\a, \b-2.75) -- (3+\a, \b - 3.75);  
		\node[anchor=south] at (\a, \b) {$\tau$};
		\node[anchor=south] at (\a + 2, \b) {$\tau$};
		\node[anchor=south] at (\a + 4, \b) {$\tau$};
		\node[anchor=south] at (\a + 6, \b) {$\tau$};
		\node[anchor=north east] at (\a + 1.75, \b - 1.5) {$\tau$};
		\node[anchor=north east] at (\a + 2.75, \b - 2.25) {$\tau$};
		\node[anchor=north] at (\a + 3, \b - 3.75) {$\tau$};
	}
	\end{align*}
	\begin{align*}
		N=4
	\end{align*}
\caption{Fusion basis resulting from the fusion of $N=3$ and $N=4$ Fibonacci anyons.}
\label{fig:Fibfusionbasis}
\end{figure}

We now proceed to writing down explicitly the basis of different fusion spaces, starting with the case of the Fibonacci anyons.
Consider $V^\tau_{\tau^{\otimes N}}$, the space in which $N$ Fibonacci anyons of species $\tau$ fuse to give another $\tau$. This space is spanned by the possible outcomes in the fusion channels of $N$ $\tau$'s. For example, the $N=3$ and $4$ cases are shown in Fig. \ref{fig:Fibfusionbasis}.

For $N=3$ there are two elements in the basis, given by $\{\ket{1}, \ket{\tau}\}$, while for $N=4$ there are three elements, given by $\{\ket{1\tau}, \ket{\tau 1}, \ket{\tau\tau}\}$. This shows that the basis states spanning this fusion space are sequences made of 1's and $\tau$'s. It is easy to see that it is impossible to obtain consecutive $1$'s in these sequences. As an example, for $N=5$ the allowed sequences are $1\tau 1, 1\tau\tau, \tau 1 \tau, \tau\tau 1, \tau\tau\tau$. We can summarize the allowed and forbidden configurations in Table \ref{tab:afFibonacci}.
\begin{table}[ht!]
  \begin{center}
    \begin{tabular}{|c|c|} 
     \hline
     Forbidden & Allowed \\
     \hline
      11 & $1\tau$, $\tau 1$, $\tau\tau$ \\
      \hline
\end{tabular}
    \caption{The forbidden and allowed configurations in the basis states of Fibonacci anyons.}
    \label{tab:afFibonacci}
  \end{center}
\end{table}

The allowed sequences are nothing else than the {\it Fibonacci sequences}, as we now show. Let $f(N)$ be the number of basis states with $N$ anyons. From the above, we see that $f(3)=2, f(4)=3, f(5) = 5$. For general $N$, we may count $f(N)$ recursively by the following algorithm. First, we decompose $f(N)$ into two categories: let $f_{1}(N)$ and $f_{\tau}(N)$ be the number of sequences starting with $1$ and $\tau$, respectively, so that
\begin{align}\label{eqn:decomposition of f to 1 and tau}
    f(N)=f_{1}(N)+f_{\tau}(N).
\end{align}
However, if a sequence starts with $1$, then the next state must be a  $\tau$. Therefore, the number of fusion sequences of length $N$ starting with $1$ is the same as the number of fusion sequences of length $N-1$ starting with $\tau$. In other words, $f_{1}(N)=f_{\tau}(N-1)$ and \eqref{eqn:decomposition of f to 1 and tau} can be now expressed as 
\begin{align}\label{eqn:decomposition of f to tau}
    f(N)=f_{\tau}(N-1)+f_{\tau}(N).
\end{align}
Now we may count $f_{\tau}(l)$ in the following way. Let us take an admissible fusion sequence $s$ of length $l-1$. Now, if we attach a $\tau$ at the beginning of $s$, then  $\tau s$ will become an admissible fusion sequence of length $l+1$ starting with $\tau$. Therefore $f_{\tau}(l+1)=f(l).$
Combining this with \eqref{eqn:decomposition of f to tau}, we conclude
\begin{align}\label{eqn:fibonacci anyon formula}
    f(N)=f(N-2)+f(N-1),
\end{align}
which is the definition of the Fibonacci sequence.

In a similar manner, one can construct the fusion basis for $N$ Jones-Kauffman anyons of species $\tau$, by identifying the forbidden and allowed configurations that make up the sequence. These are listed in Table \ref{tab:afJK} and some examples of fusion basis states are shown in Fig. \ref{fig:JKfusionbasis}.
\begin{table}[h!]
  \begin{center}
    \begin{tabular}{|c|c|} 
     \hline
     Forbidden & Allowed \\
     \hline
      11, $\mu\mu$ & $1\tau$, $\tau 1$, $\tau\tau$ \\
      $1\mu$, $\mu 1$ & $\tau\mu$, $\mu\tau$ \\
      \hline
\end{tabular}
    \caption{The forbidden and allowed configurations in the Jones-Kauffman sequences.}
    \label{tab:afJK}
  \end{center}
\end{table}

\begin{figure}[h]
	\centering
	\begin{align*}
		|1 \rangle &\equiv
		\tikz[scale=0.5, baseline=-5ex]{
			\def\a{0}
			\def\b{0}
			\draw[very thick, blue, rounded corners=5] (\a, \b) -- (\a, \b-0.5) -- (2+\a, \b-2) -- (2+\a, \b - 3);
			\draw[very thick, blue, rounded corners=5] (\a + 2, \b) -- (\a + 2, \b-0.5) -- (1+\a, \b-1.25);
			\draw[very thick, blue, rounded corners=5] (\a + 4, \b) -- (\a + 4, \b-0.5) -- (2+\a, \b-2) -- (2+\a, \b - 3); 
			\node[anchor=south] at (\a, \b) {$\tau$};
			\node[anchor=south] at (\a + 2, \b) {$\tau$};
			\node[anchor=south] at (\a + 4, \b) {$\tau$};
			\node[anchor=north east] at (\a + 1.75, \b - 1.5) {$1$};
			\node[anchor=north] at (\a + 2, \b - 3) {$\tau$};
		}&
		|\tau \rangle &\equiv
		\tikz[scale=0.5, baseline=-5ex]{
			\def\a{0}
			\def\b{0}
			\draw[very thick, blue, rounded corners=5] (\a, \b) -- (\a, \b-0.5) -- (2+\a, \b-2) -- (2+\a, \b - 3);
			\draw[very thick, blue, rounded corners=5] (\a + 2, \b) -- (\a + 2, \b-0.5) -- (1+\a, \b-1.25);
			\draw[very thick, blue, rounded corners=5] (\a + 4, \b) -- (\a + 4, \b-0.5) -- (2+\a, \b-2) -- (2+\a, \b - 3); 
			\node[anchor=south] at (\a, \b) {$\tau$};
			\node[anchor=south] at (\a + 2, \b) {$\tau$};
			\node[anchor=south] at (\a + 4, \b) {$\tau$};
			\node[anchor=north east] at (\a + 1.75, \b - 1.5) {$\tau$};
			\node[anchor=north] at (\a + 2, \b - 3) {$\tau$};
		}&
		|\mu \rangle &\equiv
		\tikz[scale=0.5, baseline=-5ex]{
			\def\a{0}
			\def\b{0}
			\draw[very thick, blue, rounded corners=5] (\a, \b) -- (\a, \b-0.5) -- (2+\a, \b-2) -- (2+\a, \b - 3);
			\draw[very thick, blue, rounded corners=5] (\a + 2, \b) -- (\a + 2, \b-0.5) -- (1+\a, \b-1.25);
			\draw[very thick, blue, rounded corners=5] (\a + 4, \b) -- (\a + 4, \b-0.5) -- (2+\a, \b-2) -- (2+\a, \b - 3); 
			\node[anchor=south] at (\a, \b) {$\tau$};
			\node[anchor=south] at (\a + 2, \b) {$\tau$};
			\node[anchor=south] at (\a + 4, \b) {$\tau$};
			\node[anchor=north east] at (\a + 1.75, \b - 1.5) {$\mu$};
			\node[anchor=north] at (\a + 2, \b - 3) {$\tau$};
		}
	\end{align*}
	
	\begin{align*}
		N= 3
	\end{align*}
	\begin{align*}
		|1\tau \rangle &\equiv \tikz[scale=0.35, baseline=-5ex]{
			\def\a{0}
			\def\b{0}
			\draw[very thick, blue, rounded corners=5] (\a, \b) -- (\a, \b-0.5) -- (3+\a, \b-2.75) -- (3+\a, \b - 3.75);
			\draw[very thick, blue, rounded corners=5] (\a + 2, \b) -- (\a + 2, \b-0.5) -- (1+\a, \b-1.25);
			\draw[very thick, blue, rounded corners=5] (\a + 4, \b) -- (\a + 4, \b-0.5) -- (2+\a, \b-2);
			\draw[very thick, blue, rounded corners=5] (\a + 6, \b) -- (\a + 6, \b-0.5) -- (3+\a, \b-2.75) -- (3+\a, \b - 3.75);  
			\node[anchor=south] at (\a, \b) {$\tau$};
			\node[anchor=south] at (\a + 2, \b) {$\tau$};
			\node[anchor=south] at (\a + 4, \b) {$\tau$};
			\node[anchor=south] at (\a + 6, \b) {$\tau$};
			\node[anchor=north east] at (\a + 1.75, \b - 1.5) {$1$};
			\node[anchor=north east] at (\a + 2.75, \b - 2.25) {$\tau$};
			\node[anchor=north] at (\a + 3, \b - 3.75) {$\tau$};
		}&
		|\tau 1 \rangle &\equiv \tikz[scale=0.35, baseline=-5ex]{
			\def\a{0}
			\def\b{0}
			\draw[very thick, blue, rounded corners=5] (\a, \b) -- (\a, \b-0.5) -- (3+\a, \b-2.75) -- (3+\a, \b - 3.75);
			\draw[very thick, blue, rounded corners=5] (\a + 2, \b) -- (\a + 2, \b-0.5) -- (1+\a, \b-1.25);
			\draw[very thick, blue, rounded corners=5] (\a + 4, \b) -- (\a + 4, \b-0.5) -- (2+\a, \b-2);
			\draw[very thick, blue, rounded corners=5] (\a + 6, \b) -- (\a + 6, \b-0.5) -- (3+\a, \b-2.75) -- (3+\a, \b - 3.75);  
			\node[anchor=south] at (\a, \b) {$\tau$};
			\node[anchor=south] at (\a + 2, \b) {$\tau$};
			\node[anchor=south] at (\a + 4, \b) {$\tau$};
			\node[anchor=south] at (\a + 6, \b) {$\tau$};
			\node[anchor=north east] at (\a + 1.75, \b - 1.5) {$\tau$};
			\node[anchor=north east] at (\a + 2.75, \b - 2.25) {$1$};
			\node[anchor=north] at (\a + 3, \b - 3.75) {$\tau$};
		}&
		|\tau\mu \rangle &\equiv \tikz[scale=0.35, baseline=-5ex]{
			\def\a{0}
			\def\b{0}
			\draw[very thick, blue, rounded corners=5] (\a, \b) -- (\a, \b-0.5) -- (3+\a, \b-2.75) -- (3+\a, \b - 3.75);
			\draw[very thick, blue, rounded corners=5] (\a + 2, \b) -- (\a + 2, \b-0.5) -- (1+\a, \b-1.25);
			\draw[very thick, blue, rounded corners=5] (\a + 4, \b) -- (\a + 4, \b-0.5) -- (2+\a, \b-2);
			\draw[very thick, blue, rounded corners=5] (\a + 6, \b) -- (\a + 6, \b-0.5) -- (3+\a, \b-2.75) -- (3+\a, \b - 3.75);  
			\node[anchor=south] at (\a, \b) {$\tau$};
			\node[anchor=south] at (\a + 2, \b) {$\tau$};
			\node[anchor=south] at (\a + 4, \b) {$\tau$};
			\node[anchor=south] at (\a + 6, \b) {$\tau$};
			\node[anchor=north east] at (\a + 1.75, \b - 1.5) {$\tau$};
			\node[anchor=north east] at (\a + 2.75, \b - 2.25) {$\mu$};
			\node[anchor=north] at (\a + 3, \b - 3.75) {$\tau$};
		}
	\end{align*}
	\begin{align*}
		|\mu\tau \rangle &\equiv \tikz[scale=0.35, baseline=-5ex]{
			\def\a{0}
			\def\b{0}
			\draw[very thick, blue, rounded corners=5] (\a, \b) -- (\a, \b-0.5) -- (3+\a, \b-2.75) -- (3+\a, \b - 3.75);
			\draw[very thick, blue, rounded corners=5] (\a + 2, \b) -- (\a + 2, \b-0.5) -- (1+\a, \b-1.25);
			\draw[very thick, blue, rounded corners=5] (\a + 4, \b) -- (\a + 4, \b-0.5) -- (2+\a, \b-2);
			\draw[very thick, blue, rounded corners=5] (\a + 6, \b) -- (\a + 6, \b-0.5) -- (3+\a, \b-2.75) -- (3+\a, \b - 3.75);  
			\node[anchor=south] at (\a, \b) {$\tau$};
			\node[anchor=south] at (\a + 2, \b) {$\tau$};
			\node[anchor=south] at (\a + 4, \b) {$\tau$};
			\node[anchor=south] at (\a + 6, \b) {$\tau$};
			\node[anchor=north east] at (\a + 1.75, \b - 1.5) {$\mu$};
			\node[anchor=north east] at (\a + 2.75, \b - 2.25) {$\tau$};
			\node[anchor=north] at (\a + 3, \b - 3.75) {$\tau$};
		}&
		|\tau\tau \rangle &\equiv \tikz[scale=0.35, baseline=-5ex]{
			\def\a{0}
			\def\b{0}
			\draw[very thick, blue, rounded corners=5] (\a, \b) -- (\a, \b-0.5) -- (3+\a, \b-2.75) -- (3+\a, \b - 3.75);
			\draw[very thick, blue, rounded corners=5] (\a + 2, \b) -- (\a + 2, \b-0.5) -- (1+\a, \b-1.25);
			\draw[very thick, blue, rounded corners=5] (\a + 4, \b) -- (\a + 4, \b-0.5) -- (2+\a, \b-2);
			\draw[very thick, blue, rounded corners=5] (\a + 6, \b) -- (\a + 6, \b-0.5) -- (3+\a, \b-2.75) -- (3+\a, \b - 3.75);  
			\node[anchor=south] at (\a, \b) {$\tau$};
			\node[anchor=south] at (\a + 2, \b) {$\tau$};
			\node[anchor=south] at (\a + 4, \b) {$\tau$};
			\node[anchor=south] at (\a + 6, \b) {$\tau$};
			\node[anchor=north east] at (\a + 1.75, \b - 1.5) {$\tau$};
			\node[anchor=north east] at (\a + 2.75, \b - 2.25) {$\tau$};
			\node[anchor=north] at (\a + 3, \b - 3.75) {$\tau$};
		}
	\end{align*}
	\begin{align*}
		N=4
	\end{align*}
	\caption{Fusion basis resulting from the fusion of $N=3$ and $N=4$ Jones-Kauffman anyons.}
	\label{fig:JKfusionbasis}
\end{figure}


As done above for the Fibonacci case, we can now count the basis states for $N$ Jones-Kauffman anyons, which we denote by $j(N)$. From the Fig. \ref{fig:JKfusionbasis}, we see that $j(3)=3, j(4)=5$. In general, we may write $j(N)=j_\tau(N)+j_{1\mu}(N)$, where $j_{\tau}(N)$ is the number of basis states starting with $\tau$ and $j_{1\mu}(N)$ is the number of basis states starting with $1$ or $\mu$. Now, if a sequence starts with $1$ or $\mu$, then the next state has to be $\tau$. Therefore $j_{1\mu}(N)=2j_{\tau}(N-1)$ and
\bea
j(N)=j_{\tau}(N)+2j_{\tau}(N-1).
\eea
 It is then enough to compute $j_{\tau}(N)$. However, $j_{\tau}(N)=j(N-1)$, because by simply inserting $\tau$ in the first position of fusion sequences with $N-1$ anyons, one obtains the fusion sequences for $N$ anyons starting with $\tau$ .
%
%
One finally has
\begin{align}\label{eqn:JK_anyon_counting_equation}
	j(N) = j(N-1)+2j(N-2).
\end{align}
Table \ref{tab:JK_anyons} gives an illustration of the above algorithm.
\begin{table}[h!]
    \centering
    	\begin{tabular}{|c|m{0.12\textwidth}|m{0.12\textwidth}|m{0.12\textwidth}|m{0.1\textwidth}|m{0.1\textwidth}|m{0.1\textwidth}|c|c|}
		\hline
		$N$ & $j_{1\mu}(N)$ & $l=2$ & $l=3$ & $l=4$ & $l=5$ & $l=6$ & $j_{\tau}(N)$ & $j(N)$\\\hline
		3 & $|1\rangle, |\mu\rangle $ & $|\tau\rangle $ &&&&& 1 &3\\\hline
		4 
		& $|1\tau\rangle, |\mu\tau\rangle$ 
		& $|\tau 1\rangle, |\tau\mu\rangle$ 
		& $|\tau\tau\rangle$ & &&& 3& 5\\\hline
		5 
		& $|1\tau\tau\rangle, |\mu\tau \tau\rangle $ $|1\tau\mu\rangle, |\mu\tau\mu\rangle $ $|1\tau 1\rangle, |\mu\tau 1\rangle$ 
		& $|\tau 1 \tau\rangle, |\tau \mu\tau\rangle$ 
		&$|\tau\tau 1\rangle, |\tau\tau \mu\rangle$
		& $|\tau\tau\tau\rangle $&&& 5 & 11 \\ \hline
		6 & $10$ &
		$|\tau 1\tau\tau\rangle$ $ |\tau 1\tau 1\rangle$ $ |\tau 1\tau \mu\rangle$ 
		$|\tau \mu\tau\tau\rangle$ $ |\tau \mu\tau 1\rangle$ $ |\tau \mu\tau \mu\rangle$
		& $|\tau\tau 1 \tau\rangle, $ $ |\tau\tau \mu\tau\rangle$ &$|\tau\tau\tau 1\rangle $ $ |\tau\tau\tau \mu\rangle$& $|\tau\tau\tau\tau\rangle$&& 11 & 21\\
		\hline
		7 & $22$ & $10$
		&
		$|\tau\tau 1\tau\tau\rangle$ $ |\tau\tau 1\tau 1\rangle$ $ |\tau\tau 1\tau \mu\rangle$ 
		$|\tau\tau \mu\tau\tau\rangle$ $ |\tau\tau \mu\tau 1\rangle $ $ |\tau\tau \mu\tau \mu\rangle$
		& $|\tau\tau\tau 1 \tau\rangle$ $ |\tau\tau\tau \mu\tau\rangle$ &$|\tau\tau\tau\tau 1\rangle$ $ |\tau\tau\tau\tau \mu\rangle$
		& $|\tau\tau\tau\tau\tau\rangle$& 21 & 43\\
		\hline
		8 & $42$ & $22$ & $10$
		&
		$|\tau\tau\tau 1\tau\tau\rangle$ $ |\tau\tau\tau 1\tau 1\rangle$ $ |\tau\tau\tau 1\tau \mu\rangle$ 
		$|\tau\tau\tau \mu\tau\tau\rangle$ $ |\tau\tau\tau \mu\tau 1\rangle$ $ |\tau\tau\tau \mu\tau \mu\rangle$
		& $|\tau\tau\tau\tau 1 \tau\rangle $ $ |\tau\tau\tau\tau \mu\tau\rangle$ &$|\tau\tau\tau\tau\tau 1\rangle $ $ |\tau\tau\tau\tau\tau \mu\rangle$
		& 43 & 85\\
		\hline
	\end{tabular}
    \caption{Number of basis states in the Jones-Kauffman anyons space. Here $l$ denotes the first location where the sequence of $\tau$ is broken by $1$ or $\mu$.}
    \label{tab:JK_anyons}
\end{table}

Finally, we look at the fusion basis obtained by fusing $N$ Ising anyons $\sigma$. In this case one needs to consider the odd and even $N$ cases separately, as the outcomes are fixed in each case respectively. In the odd case the outcome can only be another $\sigma$, whereas in the even case it can either be a $1$ or a $\psi$. Unlike the $k=3$ and $k=4$ cases, the fusion space for both odd and even $N$ has a tensor product structure. This is easily seen by inspection, for example in the odd case the basis states of the fusion space take the form, 
\bea
\ket{1/\psi,~\sigma,~1/\psi,~\sigma,~\cdots, \sigma},
\eea
and in the even case they look like
\bea
\ket{1/\psi,~\sigma,~1/\psi,~\sigma,~\cdots, \sigma, 1/\psi}.
\eea
The forbidden and allowed configurations of these basis states are given in Table \ref{tab:afIsing}.
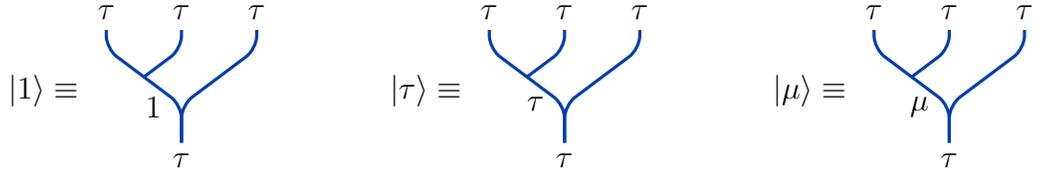
\begin{table}[h!]
  \begin{center}
    \begin{tabular}{|c|c|} 
     \hline
     Forbidden & Allowed \\
     \hline
      11, $\sigma\sigma$, $\psi\psi$ & $1\sigma$, $\sigma 1$,  \\
      $1\psi$, $\psi 1$ & $\psi\sigma$, $\sigma\psi$ \\
      \hline
\end{tabular}
    \caption{The forbidden and allowed configurations in the Ising sequences.}
    \label{tab:afIsing}
  \end{center}
\end{table}

The dimensions of the fusion basis can be computed by looking at the form of the basis states. We arrive at 
\begin{equation}
    \textrm{dim}~V^\sigma_{\sigma^{\otimes N}} = 2^{\frac{N-1}{2}},\qquad \textrm{dim}~V^{1/\psi}_{\sigma^{\otimes N}} = 2^{\frac{N-2}{2}},
\end{equation}
for the odd and even cases, respectively. This formula also confirms the tensor product structure of these fusion spaces. 

\section{Nicolai-like supersymmetric spin chains}
\label{sec:SUSYspinchains}

The scope of this article is to reproduce the anyonic fusion spaces reviewed above from spin chains endowed with a certain supersymmetric grading of their spectra.\footnote{Supersymmetry is usually seen as an extension of the Poincar\'e symmetry in relativistic systems. In this article, we simply think of supersymmetry as a way to enforce a $\mathbb{Z}_2$-grading on a spin chain.}  Specifically, we are interested in 0+1-dimensional spin chains with $N$ sites and a $\mathcal{N}=2$ SUSY algebra generated by {\it supercharges} $Q$ and $Q^\dag$, satisfying
\begin{equation}
\label{susyalgebra}
Q^2=(Q^\dag)^2=0, \qquad \{Q, Q^\dag\}\equiv Q Q^\dag+Q^\dag Q= H,
\end{equation}
where $H$ is the Hamiltonian of the system. This is supersymmetric, as it commutes with both $Q$ and $Q^\dag$. It follows from the supersymmetry algebra that $H = \left(Q + Q^\dag\right)^2$, making the spectrum of the system non-negative. The resulting Hilbert space is now divided into two parts, which we dub `bosonic' and the `fermionic' sectors, the projectors\footnote{They are projectors as the supercharges follow the additional identity $QQ^\dag Q= Q$. This is easily verified for all the supercharges considered in this paper.} to which are $B=QQ^\dag$ and $G=Q^\dag Q$, respectively. If supersymmetry is spontaneously broken , then
\bea
Q\ket{0}\neq 0, \qquad \bra{0} H \ket{0}>0,
\eea
where $\ket{0}$ is the vacuum state. For later purposes, it is convenient to define the operator $W=\left(-1\right)^G$, known as the {\it Witten index}. This operator serves as an order parameter for spontaneous supersymmetry breaking and reduces to $W=1-2G$, which follows from the relation $G^2=G$. 

In order to define the global supercharges $Q$ and $Q^\dag$ on the whole spin chain, we start by introducing local supercharges $q_i$ on each site, with $i=1,\ldots, N$. The supercharges on different sites commute with each other. To obtain anticommuting supercharges we define
\begin{equation}
\theta_j = \prod\limits_{k<j}~\left(1-2G_k\right)q_j.
\end{equation}
It is an easy check that $\{\theta_i, \theta_j\}=0$, when $qq^\dag q=q$ and $q^\dag qq^\dag = q^\dag$. 
Using these local supercharges as basic ingredients, we can now construct global supercharges $Q$  \footnote{This operator is indeed a supercharge as each term in the sum squares to zero and as neighboring terms anticommute due to $\{\theta_i, \theta_j\}=0$.} on the whole spin chain obeying the supersymmetry algebra \eqref{susyalgebra} as follows
\begin{equation}\label{eq:Q1}
Q = \sum\limits_{j=1}^{N-2}~\theta_j\theta_{j+1}\theta_{j+2}.
\end{equation}

This construction for various representations of the supercharge was  presented in \cite{Padmanabhan:2017ekk}, to which the reader is referred for more details. A deformed version of this supercharge with periodic and open boundary conditions was studied in \cite{sannomiya2017supersymmetry} and is denoted as the $\mathbb{Z}_2$ Nicolai model.\footnote{The Nicolai model on an open chain with odd number of sites is defined by the supercharge,
\begin{equation}
    Q_{Nic} = \sum\limits_{j=1}^{\frac{N-1}{2}}~\theta_{2j-1}\theta_{2j}^\dag\theta_{2j+1}.
\end{equation}
Notice that unlike the supercharge in \eqref{eq:Q1} there are fewer terms as consecutive terms skip two sites. Due to this difference we call the models defined by \eqref{eq:Q1} as Nicolai-like models.} 

At this point we endow the spin chain with a Hilbert space. We are going to be interested in two different choices, which we dub ${\cal H}_F$ and ${\cal H}_{JK}\equiv\cal{H}_I$ and which will lead to three different models. The former will realize a Fibonacci anyonic system, whereas the latter will realize the Jones-Kauffman anyonic system and the Ising anyon system.

As first choice we take
\bea
\mathcal{H}_F = \bigotimes\limits_{j=1}^N~\mathbb{C}^2_j,
\eea
where $\mathbb{C}^2_j$ is a complex two-dimensional vector space placed on every site. It is spanned by the vectors $\ket{b}=\left(\begin{array}{c} 1 \\ 0\end{array}\right)$ and $\ket{f}=\left(\begin{array}{c} 0 \\ 1\end{array}\right)$, such that $q\ket{f}=\ket{b}$, $q\ket{b}=0$ and $q^\dag\ket{f}=0$, $q^\dag\ket{b}=\ket{f}$. It follows from these expressions that 
\begin{eqnarray}\label{eq:localq1}
q  & = & \left(\begin{array}{cc} 0 & 1\\ 0 & 0 \end{array}\right), \qquad q^\dag  =  \left(\begin{array}{cc} 0 & 0\\ 1 & 0 \end{array}\right), \qquad
qq^\dag  =  \left(\begin{array}{cc} 1 & 0\\ 0 & 0 \end{array}\right),\qquad q^\dag q = \left(\begin{array}{cc} 0 & 0\\ 0 & 1 \end{array}\right).
\end{eqnarray}

This system has a global $b\leftrightarrow f$ symmetry given by the operator
\begin{equation}\label{eq:phsymmetry}
P = \prod\limits_{j=1}^N~\left(\theta_j + \theta^\dag_j\right).
\end{equation}
We have
\begin{equation}
\left(\theta_j + \theta^\dag_j\right)\theta_j = \theta^\dag_j\left(\theta_j + \theta^\dag_j\right) ,
\end{equation}
for each site and hence $P$ is a global symmetry of the Hamiltonian. This is often denoted as particle-hole symmetry in the literature.

To construct the other two supercharges we use the Hilbert space
\bea
\mathcal{H}_{JK}\equiv\mathcal{H}_I = \bigotimes\limits_{j=1}^N~\left[\mathbb{C}^2\otimes\mathbb{C}^2\right]_j,
\eea
spanned by
\begin{equation}
\left\{\ket{b_1}= \left(\begin{array}{c} b \\ 0\end{array}\right), \ket{b_2}=\left(\begin{array}{c} 0 \\ b\end{array}\right), \ket{f_1} = \left(\begin{array}{c} f \\ 0\end{array}\right), \ket{f_2}=\left(\begin{array}{c} 0 \\ f\end{array}\right),  \right\},
\end{equation}
where $b$ and $f$ are shorthand for $\left(\begin{array}{c} 1 \\ 0\end{array}\right)$ and $\left(\begin{array}{c} 0 \\ 1\end{array}\right)$, respectively. 

To build the fusion space of the Jones-Kauffman anyons the global charge in \eqref{eq:Q1} is built out of the local supercharge,
\begin{equation}\label{eq:Qjk}
    1_{2\times 2}\otimes \theta = \left(\begin{array}{cc}\theta & 0 \\ 0 & \theta\end{array}\right),
\end{equation}
on every site. Here $\theta$ is built using the supercharge given in \eqref{eq:localq1} and $1_{2\times 2}$ is the two-dimensional identity matrix. 

The fusion basis of Ising anyons is built on the same Hilbert space with the global supercharge,
\begin{eqnarray}\label{eq:QIsing}
    Q & = & \sum\limits_{j=1}^{N-2}~\left(1_{2\times 2}\otimes G\right)_j\left(1_{2\times 2}\otimes \theta\right)_{j+1}\left(1_{2\times 2}\otimes G\right)_{j+2} + \left(1_{2\times 2}\otimes B\right)_j\left(1_{2\times 2}\otimes \theta^\dag\right)_{j+1}\left(1_{2\times 2}\otimes B\right)_{j+2}, \nonumber \\
\end{eqnarray}
Both these systems continue to enjoy the global particle-hole symmetry of \eqref{eq:phsymmetry} with appropriately constructed supercharges $\theta$.

\subsection{Product zero modes}
\label{sec:SUSYzeromodes}
The next step in our analysis is to write down the zero modes of the supersymmetric systems introduced above, namely the states that satisfy
\begin{equation}
Q\ket{z}=Q^\dag\ket{z} =0.
\end{equation}
In general, the supersymmetric spin chains governed by the supercharge in \eqref{eq:Q1}
have both non-entangled (or product) and entangled zero modes. Here we count the number of product zero modes. This will allow to establish the correspondence with the basis vectors of the anyonic fusion spaces studied earlier. 

 
We start with the first case, based on ${\cal H}_F$. For the supercharge $Q$ built out of the local supercharges in \eqref{eq:localq1}, it is easy to realize that the product zero modes occur when the configuration on the chain made of local $\ket{f}$'s and $\ket{b}$'s do not contain the sequences $\ket{f_jf_{j+1}f_{j+2}}$ or $\ket{b_jb_{j+1}b_{j+2}}$ on three consecutive sites. 

The counting then goes as follows. Let $f_{P}(N)$ be the number of product zero modes on $N$ sites. For $N=1, 2, 3$, we have the results in Table \ref{tab:simple_product_ground}, from which one obtains $f_{P}(1)=2, f_{P}(2)=4, f_{P}(3) = 6$.
 \begin{table}
     \centering
     \begin{tabular}{|c|c|}
        \hline
		$N$ & Product ground states\\
		\hline
		1 & $|b_{1}\rangle, |f_{1}\rangle$\\
		2 & $|b_{1}b_{2}\rangle, |b_{1}f_{2}\rangle, |f_{1}b_{2}\rangle, |f_{1}f_{2}\rangle$\\
		3 & $|b_{1}b_{2}f_{3}\rangle, |b_{1}f_{2}b_{3}\rangle, |b_{1}f_{2}f_{3}\rangle,|f_{1}b_{2}b_{3}\rangle, |f_{1}b_{2}f_{3}\rangle, |f_{1}f_{2}b_{3}\rangle$\\\hline
	\end{tabular}
     \caption{Product ground states}
     \label{tab:simple_product_ground}
 \end{table}
	
 In general, we may find the expression of $f_{P}(N)$ recursively with the following procedure: 
	\begin{enumerate}
		\item Start the sequence with $|b \rangle$ (or $| f\rangle$).
		\item Let $k$ be the first site which is occupied by $|f\rangle$ (or $|b\rangle$). For example, $|b_{1}b_{2}f_{3}\cdots\rangle$ and $|b_{1}b_{2}b_{3}f_{4}\cdots\rangle$ correspond to $k=3$ and $k=4$ respectively.
		\item If $k=2,3$, then we may treat it as a new sequence of length $N-k+1$ starting with $|f_{k}\rangle$ (or $|b_{k}\rangle$). Discard the product state if $k\geq 4$.
	\end{enumerate}
Thus we conclude that
\begin{align}\label{eqn:product_ground_states_fibo}
    f_{P}(N)=f_{P}(N-1)+f_{P}(N-2).
\end{align}
Consequently, $f_{P}(N)$ is a Fibonacci sequence with initial conditions $f_{P}(1)=2, f_{P}(2)=4$.

As for the choice ${\cal H}_{JK}$, when the global supercharges are constructed out of $1_{2\times 2}\otimes \theta$ on each site, the product zero mode configurations exclude more sequences. More precisely, the product states are built out of configurations made of $\ket{f_1}$, $\ket{f_2}$ and $\ket{b_1}$, $\ket{b_2}$ and we have a four-dimensional vector space on each site. The excluded sequences in this case are given by $\ket{(f_{i_1})_j(f_{i_2})_{j+1}(f_{i_3})_{j+2}}$ and $\ket{(b_{i_1})_j(b_{i_2})_{j+1}(b_{i_3})_{j+2}}$, with $i_1$, $i_2$ and $i_3$ taking values 1 or 2 and $j$ being the site index as before.

We may also count the number of product zero modes using the similar algorithm as above. Let $j_{P}(N)$ be the number of product zero modes on $N$ sites. From inspection we see that $j_{P}(1)=4, j_{P}(2)=16, j_{P}(3)=48$. In general, we may find the product zero modes of $\mathcal{H}_{JK}$ from those of $\mathcal{H}_{F}$ as follows. Let $|b_{1}f_{2}\cdots f_{n}\rangle$ be a product zero modes of $\mathcal{H}_{F}$. Then by taking $i_{k}=1, 2$, we may obtain $|(b_{i_{1}})_{1}(f_{i_{2}})_{2}\cdots (f_{i_{n}})_{n}\rangle$ as a product ground state of $\mathcal{H}_{JK}$. Moreover, all the product zero modes of $\mathcal{H}_{JK}$ can be obtained via this correspondence. Thus, we conclude that
\begin{align}\label{eqn: jk zero modes is 2^n fibonacci}
	j_{P}(N)=2^{N}f_{P}(N).
\end{align}
In addition, we deduce the following recursion relation
\begin{align}\label{eqn: recursion relation of j(N) zero modes}
	j_{P}(N)=2[j_{P}(N-1)+2j_{P}(N-2)],
\end{align}
which is similar to the counting of the JK anyon fusion basis, \eqref{eqn:JK_anyon_counting_equation}.

The supercharge in \eqref{eq:QIsing} on $\cal{H}_I$ is responsible the product zero modes that exclude the sequences
$$ \ket{(f_{i_1})_j(f_{i_2})_{j+1}(f_{i_3})_{j+2}},\qquad  \ket{(b_{i_1})_j(b_{i_2})_{j+1}(b_{i_3})_{j+2}},$$ and 
$$ \ket{(f_{i_1})_j(b_{i_2})_{j+1}(f_{i_3})_{j+2}},\qquad  \ket{(b_{i_1})_j(f_{i_2})_{j+1}(b_{i_3})_{j+2}},$$
with $i_1, i_2, i_3=1, 2$. The number of such zero modes can be easily counted by noting that the only allowed sequences with these four exclusions are either 
$$ ffbbffbb\cdots\quad \textrm{or}\quad bbffbbff\cdots. $$  
However on each of the $f$ and $b$ labels we can add the index 1 or 2 making the total $2^{N+1}$ for a $N$-site chain. 

\section{Anyonic fusion space from zero modes}
\label{sec:SUSYanyoncorrespondence}

We are now in the position to conjecture a correspondence between the product zero modes of the two supersymmetric systems considered in the previous section and the basis of the anyon fusion spaces of the level $k=3$ and $k=4$ anyons of the $SU(2)_k$ quantum group. The main evidence pointing to such correspondence is the observation that the number of product zero modes is proportional to the number of allowed configurations used to build the basis states of the anyonic fusion space. This is due to the fact that the same kind of sequences are excluded in the construction of the product zero modes on the chain and in the allowed configurations of the basis of the anyonic fusion spaces. As a consequence, one can identify each anyon with a pair of sites on the supersymmetric chain.

Here we note that this correspondence is not really a function but a dictionary between the two spaces. More precisely, given a sequence of anyons of a fusion basis element, we can use this dictionary to find two product zero modes of the SUSY spin chain. The reason for getting two zero modes is the global $b\longleftrightarrow f$ symmetry of the SUSY spin chains. The zero mode we get is determined by the type of local supersymmetric state the first anyon in the sequence gets mapped to. These statements will become clear when we see them through the three examples considered in this paper. 

\subsection{Fibonacci anyons and the chain with ${\cal H}_F$}
\label{subsec:Fib2Q1}

We start by considering the Fibonacci anyons, which as we have seen in Section \ref{subsec:SU2kanyons} come in two species: 1 and $\tau$. We make the following identifications
\begin{eqnarray}\label{eq:Fib2q1z}
1 & \longrightarrow & bb ,\,  ff \nonumber \\
\tau & \longrightarrow & bf,\, fb
\end{eqnarray}
between the two anyonic species and the sequences of $b$'s and $f$'s in the product zero modes of \eqref{eq:localq1}. Note how the global $b\leftrightarrow f$ symmetry doubles this space, as there are two configurations for each anyon that are related by this symmetry. We find that the number of zero modes on an $N$ site chain is
\begin{equation}
\tilde{F}(N) = 2\, \textrm{dim}\left(V^\tau_{\otimes \tau^{N+1}}\right)=2F(N+1),
\end{equation}
with the factor 2 due to the global $b\leftrightarrow f$ symmetry and $F(N)$ the $N$-th Fibonacci number.

As an example consider the case of four $\tau$'s fusing. As seen before, this space is spanned by the vectors $\{1\tau, \tau 1, \tau\tau\}$. By the correspondence in \eqref{eq:Fib2q1z}, one has
\begin{eqnarray}
1\tau & \longrightarrow & bbf,\, ffb, \nonumber \\
\tau 1 & \longrightarrow & bff,\, fbb, \nonumber \\
\tau\tau & \longrightarrow & bfb,\, fbf,
\end{eqnarray}
which are precisely the six zero modes we would expect on a chain with three sites. The correspondence for $N=5$ is illustrated in Table \ref{tab:N5Fib2Q1z}.
\begin{table}[h!]
  \begin{center}
    \begin{tabular}{|c|c|} 
     \hline
     $N=5$ Fibonacci basis & Product zero modes of $N=4$ SUSY chain \\
     \hline
     $\ket{1\tau\tau}$ & $\ket{bbfb}$,\, $\ket{ffbf}$ \\
     $\ket{1\tau 1}$ & $\ket{bbff}$,\, $\ket{ffbb}$ \\
     $\ket{\tau\tau 1}$ & $\ket{bfbb}$,\, $\ket{fbff}$ \\
     $\ket{\tau 1\tau}$ & $\ket{bffb}$,\, $\ket{fbbf}$ \\
     $\ket{\tau\tau\tau}$ & $\ket{bfbf}$,\, $\ket{fbfb}$ \\
      \hline
\end{tabular}
    \caption{The correspondence for $N=5$.}
    \label{tab:N5Fib2Q1z}
  \end{center}
\end{table}

We can write down more general supercharges with product zero modes corresponding to the basis states of the Fibonacci fusion spaces. Consider global supercharges built out of the local supercharges
\begin{equation}
q = \left(\begin{array}{cc} 0 & 1_{m\times m}\\ 0 & 0 \end{array}\right),\qquad q^\dag = \left(\begin{array}{cc} 0 & 0\\ 1_{m\times m} & 0 \end{array}\right).
\end{equation}
The local fermions and bosons are given by 
\begin{equation}\label{eq:bifi}
\ket{f_i} = \left(\begin{array}{ccccc} 0 & \cdots & \underbrace{f}_i & \cdots & 0\end{array}\right)^T,\qquad \ket{b_i} = \left(\begin{array}{ccccc} 0 & \cdots & \underbrace{b}_i & \cdots & 0\end{array}\right)^T, \qquad i=1,\ldots, m.
\end{equation}
Here $f=\left(\begin{array}{c}0 \\ 1\end{array}\right)$ and $b=\left(\begin{array}{c}1 \\ 0\end{array}\right)$, as done before.
The correspondence then becomes
\begin{eqnarray}\label{eq:Fib2q1zgeneral}
1 & \longrightarrow & b_ib_j,\, f_if_j \nonumber \\
\tau & \longrightarrow & b_if_j,\, f_ib_j,\qquad i, j=1, \ldots, m.
\end{eqnarray}
The number of zero modes in this case is given by 
\begin{equation}
\tilde{F}(N) = 2m^2~\textrm{dim}\left(V^\tau_{\otimes \tau^{N+1}}\right)=2m^2F(N+1).
\end{equation}

\subsection{Jones-Kauffman anyons and the chain with ${\cal H}_{JK}$}
\label{subsec:JK2Q2}
We now move on to the Jones-Kauffman anyons. The identifications
\begin{eqnarray}\label{eq:JKzIdentification}
1 & \longrightarrow & b_1b_1, b_2b_2/f_1f_1, f_2f_2, \nonumber \\
\mu & \longrightarrow & b_1b_2, b_2b_1/f_1f_2, f_2f_1, \nonumber \\
\tau & \longrightarrow & b_1f_1, b_1f_2, b_2f_1, b_2f_2/f_1b_1, f_1b_2, f_2b_1, f_2b_2,
\end{eqnarray}
with $f_1$, $f_2$ and $b_1$, $b_2$ given by \eqref{eq:bifi} lead to the desired correspondence. Here too we have a doubling of the space due to the global $b\leftrightarrow f$ symmetry of \eqref{eq:phsymmetry}. It is easy to see that $11$, $1\mu$, $\mu 1$ and $\mu\mu$ result in forbidden sequences in the supersymmetric space that do not lead to product zero modes. Every other pair, namely $1\tau$, $\tau 1$, $\tau\tau$, $\tau\mu$ and $\mu\tau$ that are used to construct the basis of the Jones-Kauffman anyon fusion space, lead instead to sequences that are part of the product zero modes in the supersymmetric system. 

More generally, one can consider supersymmetric systems built out of the local supercharge
\begin{equation}
1_{m\times m}\otimes \theta,
\end{equation}
with $\theta$ built out of the $q$ given by \eqref{eq:localq1}. The product zero modes of this system corresponds to the Jones-Kauffman anyons via
\begin{eqnarray}\label{eq:JKzIdentificationGeneral}
1 & \longrightarrow & b_jb_j/f_jf_j, \qquad j=1,\ldots, m, \nonumber \\
\mu & \longrightarrow & b_jb_k/f_jf_k, \qquad j\neq k=1,\ldots, m, \nonumber \\
\tau & \longrightarrow & b_jf_k/f_jb_k, \qquad j\neq k=1,\ldots, m,
\end{eqnarray}
where $f_j$ and $b_j$ are given in \eqref{eq:bifi}. 

\subsection{Ising anyons and the chain with ${\cal H}_I$}
\label{subsec:IsingQ3}

To establish the correspondence between the fusion basis of Ising anyons to the zero modes of the supersymmetric system defined by \eqref{eq:QIsing}, we use exactly the same identification as in \eqref{eq:JKzIdentification}, with $\mu$ replaced by $\psi$ and $\tau$ replaced by $\sigma$. Then we see that the forbidden configurations in Table \ref{tab:afIsing} ($11$, $\sigma\sigma$, $\psi\psi$, $1\psi$, $\psi 1$) are precisely the product zero modes of this supersymmetric system. 

We can obtain a more general correspondence between the fusion space of Ising anyons and the spin chain constructed by replacing $1_{2\times 2}$ by $1_{m\times m}$ in the supercharge in \eqref{eq:QIsing}.
The identification for this purpose is precisely the same as the one shown in \eqref{eq:JKzIdentificationGeneral}.

\section{Braid group on anyonic fusion spaces}
\label{sec:Fusionspacebraidgroup}
Again, in the spirit of keeping this article self-contained, we briefly review the braid group and its action on anyonic fusion spaces. This will serve later as the basis for the formulation of the braid group in terms of the zero modes of the supersymmetric spin chain.

\begin{figure}[h]
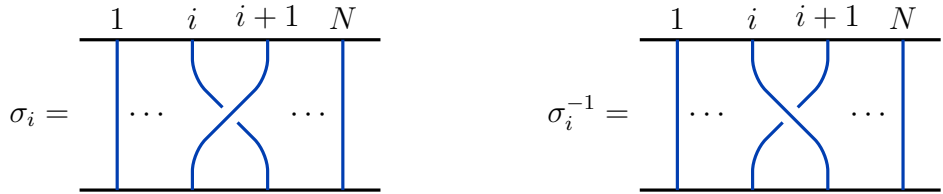

	\centering
	\begin{align*}
		\sigma_{i} &= \tikz[scale=1, baseline=-6ex]{
			\def\a{0}
			\def\b{0}
			\node[anchor=south] at (\a, \b) {$1$};
			\node[anchor=south] at (\a  + 1, \b) {$i$};
			\node[anchor=south] at (\a + 2, \b) {$i+1$};
			\node[anchor=south] at (\a + 3, \b) {$N$};
			\draw[very thick] (\a-0.5, \b) -- (\a + 3.5, \b);
			\draw[very thick] (\a - 0.5, \b -2) -- (\a + 3.5, \b - 2);
			\draw[very thick, blue] (\a, \b) -- (\a, \b - 2);
			\node[anchor=west] at (\a, \b - 1) {$\cdots$};
			\draw[very thick, blue, rounded corners=7] (\a + 1, \b) -- (\a + 1, \b - 0.5) -- (\a + 1.4, \b - 0.9);
			\draw[very thick, blue, rounded corners=7] (\a + 1.6, \b - 1.1) -- (\a + 2, \b - 1.5) -- (\a + 2, \b - 2);
			\draw[very thick, blue, rounded corners=7] (\a + 2, \b) -- (\a + 2, \b - 0.5) -- (\a + 1, \b - 1.5) -- (\a + 1, \b - 2);
			\node[anchor=east] at (\a + 3, \b - 1) {$\cdots$};
			\draw[very thick, blue] (\a + 3, \b) -- (\a + 3, \b - 2);
		}&
		\sigma_{i}^{-1} &= \tikz[scale=1, baseline=-6ex]{
			\def\a{0}
			\def\b{0}
			\node[anchor=south] at (\a, \b) {$1$};
			\node[anchor=south] at (\a  + 1, \b) {$i$};
			\node[anchor=south] at (\a + 2, \b) {$i+1$};
			\node[anchor=south] at (\a + 3, \b) {$N$};
			\draw[very thick] (\a-0.5, \b) -- (\a + 3.5, \b);
			\draw[very thick] (\a - 0.5, \b -2) -- (\a + 3.5, \b - 2);
			\draw[very thick, blue] (\a, \b) -- (\a, \b - 2);
			\node[anchor=west] at (\a, \b - 1) {$\cdots$};
			\draw[very thick, blue, rounded corners=7] (\a + 1, \b) -- (\a + 1, \b - 0.5) -- (\a + 2, \b - 1.5) -- (\a + 2, \b - 2);
			\draw[very thick, blue, rounded corners=7] (\a + 2, \b) -- (\a + 2, \b - 0.5) -- (\a + 1.6, \b - 0.9);
			\draw[very thick, blue, rounded corners=7] (\a + 1.4, \b - 1.1) -- (\a + 1, \b - 1.5) -- (\a + 1, \b - 2);
			\node[anchor=east] at (\a + 3, \b - 1) {$\cdots$};
			\draw[very thick, blue] (\a + 3, \b) -- (\a + 3, \b - 2);
		}
	\end{align*}
	\caption{Pictorial realization of the $N$-strand braid generators, $\sigma_i$ and $\sigma_i^{-1}$.}
	\label{fig:BNgenerators}
\end{figure}

The braid group on $N$ strands, $\mathcal{B}_N$, is generated by $\sigma_i$ ($i=1,\ldots, N-1$) obeying braid and far commutativity relations
\begin{eqnarray}
\sigma_i\sigma_{i+1}\sigma_i & = & \sigma_{i+1}\sigma_i\sigma_{i+1},\qquad i=1,\ldots, N-2, \nonumber \\
\sigma_i\sigma_j & = & \sigma_j\sigma_i, \qquad |i-j|>1.
\end{eqnarray}
They can be represented pictorially as in Fig. \ref{fig:BNgenerators}.

\begin{figure}[h]
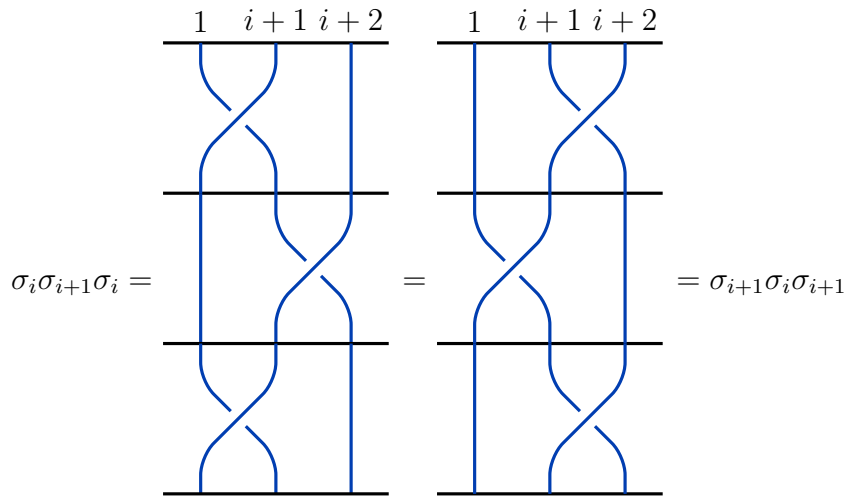

	\begin{align*}
		\sigma_{i}\sigma_{i+1}\sigma_{i} = \tikz[scale=1, baseline=-18ex]{
			\def\a{0}
			\def\b{0}
			\node[anchor=south] at (\a  + 1, \b) {$1$};
			\node[anchor=south] at (\a  + 2, \b) {$i+1$};
			\node[anchor=south] at (\a + 3, \b) {$i+2$};
			\draw[very thick] (\a+0.5, \b) -- (\a + 3.5, \b);
			\draw[very thick] (\a + 0.5, \b -2) -- (\a + 3.5, \b - 2);
			\draw[very thick, blue, rounded corners=7] (\a + 1, \b) -- (\a + 1, \b - 0.5) -- (\a + 1.4, \b - 0.9);
			\draw[very thick, blue, rounded corners=7] (\a + 1.6, \b - 1.1) -- (\a + 2, \b - 1.5) -- (\a + 2, \b - 2);
			\draw[very thick, blue, rounded corners=7] (\a + 2, \b) -- (\a + 2, \b - 0.5) -- (\a + 1, \b - 1.5) -- (\a + 1, \b - 2);
			\draw[very thick, blue] (\a + 3, \b) -- (\a + 3, \b - 2);
			\def\a{1}
			\def\b{-2}
			\draw[very thick, blue] (\a, \b) -- (\a, \b - 2);
			\draw[very thick, blue, rounded corners=7] (\a + 1, \b) -- (\a + 1, \b - 0.5) -- (\a + 1.4, \b - 0.9);
			\draw[very thick, blue, rounded corners=7] (\a + 1.6, \b - 1.1) -- (\a + 2, \b - 1.5) -- (\a + 2, \b - 2);
			\draw[very thick, blue, rounded corners=7] (\a + 2, \b) -- (\a + 2, \b - 0.5) -- (\a + 1, \b - 1.5) -- (\a + 1, \b - 2);
			\draw[very thick] (\a - 0.5, \b -2) -- (\a + 2.5, \b - 2);
			\def\a{0}
			\def\b{-4}
			\draw[very thick, blue, rounded corners=7] (\a + 1, \b) -- (\a + 1, \b - 0.5) -- (\a + 1.4, \b - 0.9);
			\draw[very thick, blue, rounded corners=7] (\a + 1.6, \b - 1.1) -- (\a + 2, \b - 1.5) -- (\a + 2, \b - 2);
			\draw[very thick, blue, rounded corners=7] (\a + 2, \b) -- (\a + 2, \b - 0.5) -- (\a + 1, \b - 1.5) -- (\a + 1, \b - 2);
			\draw[very thick, blue] (\a + 3, \b) -- (\a + 3, \b - 2);
			\draw[very thick] (\a + 0.5, \b - 2) -- (\a + 3.5, \b - 2);
		}
		= \tikz[scale=1, baseline=-18ex]{
			\def\a{1}
			\def\b{0}
			\node[anchor=south] at (\a, \b) {$1$};
			\node[anchor=south] at (\a  + 1, \b) {$i+1$};
			\node[anchor=south] at (\a + 2, \b) {$i+2$};
			\draw[very thick] (\a-0.5, \b) -- (\a + 2.5, \b);
			\draw[very thick] (\a - 0.5, \b -2) -- (\a + 2.5, \b - 2);
			\draw[very thick, blue] (\a, \b) -- (\a, \b - 2);
			\draw[very thick, blue, rounded corners=7] (\a + 1, \b) -- (\a + 1, \b - 0.5) -- (\a + 1.4, \b - 0.9);
			\draw[very thick, blue, rounded corners=7] (\a + 1.6, \b - 1.1) -- (\a + 2, \b - 1.5) -- (\a + 2, \b - 2);
			\draw[very thick, blue, rounded corners=7] (\a + 2, \b) -- (\a + 2, \b - 0.5) -- (\a + 1, \b - 1.5) -- (\a + 1, \b - 2);
			\def\a{0}
			\def\b{-2}
			\draw[very thick] (\a + 0.5, \b -2) -- (\a + 3.5, \b - 2);
			\draw[very thick, blue, rounded corners=7] (\a + 1, \b) -- (\a + 1, \b - 0.5) -- (\a + 1.4, \b - 0.9);
			\draw[very thick, blue, rounded corners=7] (\a + 1.6, \b - 1.1) -- (\a + 2, \b - 1.5) -- (\a + 2, \b - 2);
			\draw[very thick, blue, rounded corners=7] (\a + 2, \b) -- (\a + 2, \b - 0.5) -- (\a + 1, \b - 1.5) -- (\a + 1, \b - 2);
			\draw[very thick, blue] (\a + 3, \b) -- (\a + 3, \b - 2);
			\def\a{1}
			\def\b{-4}
			\draw[very thick] (\a - 0.5, \b -2) -- (\a + 2.5, \b - 2);
			\draw[very thick, blue, rounded corners=7] (\a + 1, \b) -- (\a + 1, \b - 0.5) -- (\a + 1.4, \b - 0.9);
			\draw[very thick, blue, rounded corners=7] (\a + 1.6, \b - 1.1) -- (\a + 2, \b - 1.5) -- (\a + 2, \b - 2);
			\draw[very thick, blue, rounded corners=7] (\a + 2, \b) -- (\a + 2, \b - 0.5) -- (\a + 1, \b - 1.5) -- (\a + 1, \b - 2);
			\draw[very thick, blue] (\a, \b) -- (\a, \b - 2);
		}  = \sigma_{i+1}\sigma_{i}\sigma_{i+1}
	\end{align*}
	\caption{Pictorial realization of the braid relation.}
	\label{fig:BNmultiplication}
\end{figure}

Multiplication in $\mathcal{B}_N$ is obtained by stacking the generators one above the other. For example, the braid relation is proved pictorially in Fig. \ref{fig:BNmultiplication}.



Representations of $\mathcal{B}_N$ are obtained via the homomorphism
\bea
 \mathcal{B}_N \mapsto GL(m),
\eea
where $m$ is the dimension of the representation.
In general, one can obtain both {\it local} and {\it non-local} representations. By non-local representations we mean that the representation space does not have a tensor product structure like those of the local representations. More often than not, the representations on anyonic fusion spaces are non-local. 

On the other hand for local representations $\sigma_i$ acts on a tensor product of vector spaces, $V\otimes V\otimes\cdots\otimes V$. In this case the generator $\sigma_i$ acts non-trivially only on strands $i$ and $i+1$. For example, when $N=3$ one has that $\sigma_1=\sigma\otimes I$ and $\sigma_2=I\otimes\sigma$, with $I$ being the identity operator on $V$. In such cases the far commutativity is automatically satisfied and the braid relation reduces to the {\it Yang-Baxter equation (YBE)}
\begin{equation} 
\left(\sigma\otimes I\right)\left(I\otimes\sigma\right)\left(\sigma\otimes I\right) = \left(I\otimes\sigma\right)\left(\sigma\otimes I\right)\left(I\otimes\sigma\right),
\end{equation}
on $V\otimes V\otimes V$. Thus solving the YBE amounts to solving for the matrix $\sigma$ which helps us build all the generators of $\mathcal{B}_N$. However this is not the case for non-local representations where we need to construct the generators of $\mathcal{B}_N$ separately for each $N$. This is indeed the case for the Fibonacci example as we shall see next. 

\subsection{Braid group on the Fibonacci anyon fusion space}
\label{subsec:braidgroup}

We will now find the representation of $\mathcal{B}_N$ on the Fibonacci anyon fusion space following \cite{kauffmanlomonaco}. This space does not have a tensor product structure,  hence the resulting representations are non-local and are derived from the {\it Temperley-Lieb-Jones algebra}. They are known as the Jones representations and result in the {\it Jones polynomials}, a set of knot/link invariants. 

The $F$-, $R$-moves and the pictorial representation of the $\mathcal{B}_N$ generators guide us in finding the desired representations. The procedure goes as follows:
\begin{enumerate}
    \item The anyon lines in the basis vectors of the fusion space are taken as the strands on which the braid group generators act. For an arbitrary case the action of $\sigma_{i, i+1}$ is shown in Fig. \ref{fig:sigmaiactionfusion}.\footnote{We pick a convention where the left strand crosses behind the neighboring right strand (See Fig. \ref{fig:BNgenerators}) The construction is equally valid for the opposite convention using the alternate hexagon identities, \ref{hexagonidalt}.}
\begin{figure}[h]
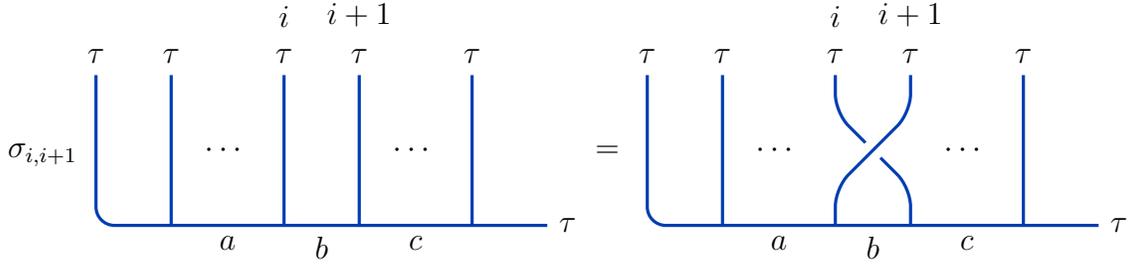

	\centering
	\begin{align*}
		\sigma_{i, i+1}\tikz[baseline=5ex, scale=1]{
			\def\a{0}
			\def\b{0}
			\node[anchor=south] at (\a + 2.5, \b + 2.5) {$i$};
			\node[anchor=south] at (\a + 3.5, \b + 2.5) {$i+1$};
			\draw[very thick, blue, rounded corners=7] (\a, \b+2) node[anchor=south, black] {$\tau$} -- (\a, \b) -- (\a + 6, \b) node[anchor=west, black] {$\tau$};
			\draw[very thick, blue] (\a+1, \b) -- (\a+1, \b+2) node[anchor=south, black] {$\tau$};
			\node[anchor=west] at (\a+1.3, \b+1) {$\cdots$};
			\draw[very thick, blue] (\a+2.5, \b) -- (\a+2.5, \b+2) node[anchor=south, black] {$\tau$};
			\draw[very thick, blue] (\a+3.5, \b) -- (\a+3.5, \b+2) node[anchor=south, black] {$\tau$};
			\node[anchor=west] at (\a+3.8, \b+1) {$\cdots$};
			\draw[very thick, blue] (\a+5, \b) -- (\a+5, \b+2) node[anchor=south, black] {$\tau$};
			\node[anchor=north] at (\a + 1.75, \b) {$a$};
			\node[anchor=north] at (\a + 3, \b) {$b$};
			\node[anchor=north] at (\a + 4.25, \b) {$c$};
		}=
		\tikz[scale=1, baseline=5ex]{
			\def\a{0}
			\def\b{0}
			\node[anchor=south] at (\a + 2.5, \b + 2.5) {$i$};
			\node[anchor=south] at (\a + 3.5, \b + 2.5) {$i+1$};
			\draw[very thick, blue, rounded corners=7] (\a, \b+2) node[anchor=south, black] {$\tau$} -- (\a, \b) -- (\a + 6, \b) node[anchor=west, black] {$\tau$};
			\draw[very thick, blue] (\a+1, \b) -- (\a+1, \b+2) node[anchor=south, black] {$\tau$};
			\node[anchor=west] at (\a+1.3, \b+1) {$\cdots$};
			\def\a{1.5}
			\def\b{2}
			\draw[very thick, blue, rounded corners=7] (\a + 1, \b) node[anchor=south, black] {$\tau$} -- (\a + 1, \b - 0.5) -- (\a + 1.4, \b - 0.9);
			\draw[very thick, blue, rounded corners=7] (\a + 1.6, \b - 1.1) -- (\a + 2, \b - 1.5) -- (\a + 2, \b - 2);
			\draw[very thick, blue, rounded corners=7] (\a + 2, \b) node[anchor=south, black] {$\tau$} -- (\a + 2, \b - 0.5) -- (\a + 1, \b - 1.5) -- (\a + 1, \b - 2);
			\def\a{0}
			\def\b{0}
			\node[anchor=west] at (\a+3.8, \b+1) {$\cdots$};
			\draw[very thick, blue] (\a+5, \b) -- (\a+5, \b+2) node[anchor=south, black] {$\tau$};
			\node[anchor=north] at (\a + 1.75, \b) {$a$};
			\node[anchor=north] at (\a + 3, \b) {$b$};
			\node[anchor=north] at (\a + 4.25, \b) {$c$};
		}
	\end{align*}
	\caption{Action of $\sigma_i$ on the anyon fusion basis.}
	\label{fig:sigmaiactionfusion}
\end{figure}

   \item We then use the $F$- and $R$-moves repeatedly to unwind the braided anyonic lines and rewrite them in terms of the vectors of the fusion basis. 
    \item Repeat this procedure for each vector in the fusion basis to establish the braid group generator.
\end{enumerate}

\begin{figure}[h]
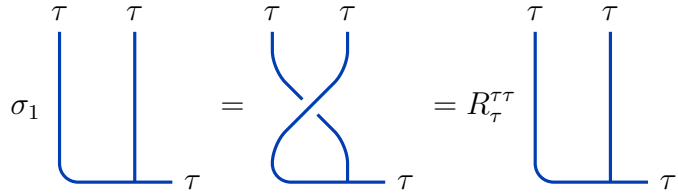

	\centering
	\begin{align*}
		\sigma_{1}\tikz[baseline=5ex, scale=1]{
			\def\a{0}
			\def\b{0}
			\draw[very thick, blue, rounded corners=7] (\a, \b+2) node[anchor=south, black] {$\tau$} -- (\a, \b) -- (\a + 1.5, \b) node[anchor=west, black] {$\tau$};
			\draw[very thick, blue] (\a+1, \b) -- (\a+1, \b+2) node[anchor=south, black] {$\tau$};
		}=
		\tikz[baseline=5ex, scale=1]{
			\def\a{0}
			\def\b{2}
			\draw[very thick, blue, rounded corners=7] (\a + 1, \b) node[anchor=south, black] {$\tau$} -- (\a + 1, \b - 0.5) -- (\a + 1.4, \b - 0.9);
			\draw[very thick, blue, rounded corners=7] (\a + 1.6, \b - 1.1) -- (\a + 2, \b - 1.5) -- (\a + 2, \b - 2);
			\draw[very thick, blue, rounded corners=7] (\a + 2, \b) node[anchor=south, black] {$\tau$} -- (\a + 2, \b - 0.5) -- (\a + 1, \b - 1.5) -- (\a + 1, \b - 2) --  (\a + 2.5, \b - 2) node[anchor=west, black] {$\tau$};
		}=R_{\tau}^{\tau\tau}
		\tikz[baseline=5ex, scale=1]{
			\def\a{0}
			\def\b{0}
			\draw[very thick, blue, rounded corners=7] (\a, \b+2) node[anchor=south, black] {$\tau$} -- (\a, \b) -- (\a + 1.5, \b) node[anchor=west, black] {$\tau$};
			\draw[very thick, blue] (\a+1, \b) -- (\a+1, \b+2) node[anchor=south, black] {$\tau$};
		}
	\end{align*}
	\caption{Construction of $\sigma_1$, the lone generator of $\mathcal{B}_2$. Here $R^{\tau\tau}_\tau=\lambda$.}
	\label{fig:B2rep}
\end{figure}

We illustrate this first for the 2-strand case or $\mathcal{B}_2$ in Fig. \ref{fig:B2rep}.

The lone generator $\sigma_1$ is just the scalar $\lambda$ as the 2-strand fusion space is one-dimensional. Note that for Fibonacci anyons $R^{\tau\tau}_\tau$ takes a specific value, but we do not use that here to keep the representation as general as possible. We will see that this is only possible for the 2-strand and 3-strand braid groups where we obtain a one-parameter family of representations. This is because the far commutativity relations play no role in these cases eliminating a constraint on the representations. We will lose this freedom when the number of strands is more than three, fixing the value of the lone parameter.

\begin{figure}[h]
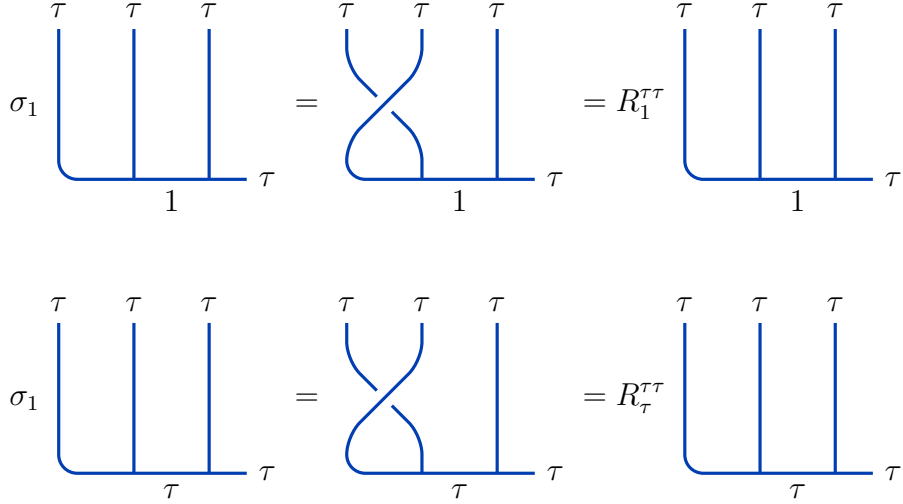

	\centering
	\begin{align*}
		\sigma_{1}\tikz[baseline=5ex, scale=1]{
			\def\a{0}
			\def\b{0}
			\draw[very thick, blue, rounded corners=7] (\a, \b+2) node[anchor=south, black] {$\tau$} -- (\a, \b) -- (\a + 2.5, \b) node[anchor=west, black] {$\tau$};
			\node[anchor=north] at (\a + 1.5, \b) {$1$};
			\draw[very thick, blue] (\a+1, \b) -- (\a+1, \b+2) node[anchor=south, black] {$\tau$};
			\draw[very thick, blue] (\a+2, \b) -- (\a+2, \b+2) node[anchor=south, black] {$\tau$};
		}=
		\tikz[baseline=5ex, scale=1]{
			\def\a{0}
			\def\b{2}
			\draw[very thick, blue, rounded corners=7] (\a + 1, \b) node[anchor=south, black] {$\tau$} -- (\a + 1, \b - 0.5) -- (\a + 1.4, \b - 0.9);
			\draw[very thick, blue, rounded corners=7] (\a + 1.6, \b - 1.1) -- (\a + 2, \b - 1.5) -- (\a + 2, \b - 2);
			\draw[very thick, blue, rounded corners=7] (\a + 3, \b -2) -- (\a + 3, \b) node[anchor=south, black] {$\tau$};
			\draw[very thick, blue, rounded corners=7] (\a + 2, \b) node[anchor=south, black] {$\tau$} -- (\a + 2, \b - 0.5) -- (\a + 1, \b - 1.5) -- (\a + 1, \b - 2) --  (\a + 3.5, \b - 2) node[anchor=west, black] {$\tau$};
			\node[anchor=north] at (\a + 2.5, \b - 2) {$1$};
		}=R_{1}^{\tau\tau}
		\tikz[baseline=5ex, scale=1]{
			\def\a{0}
			\def\b{0}
			\draw[very thick, blue, rounded corners=7] (\a, \b+2) node[anchor=south, black] {$\tau$} -- (\a, \b) -- (\a + 2.5, \b) node[anchor=west, black] {$\tau$};
			\node[anchor=north] at (\a + 1.5, \b) {$1$};
			\draw[very thick, blue] (\a+1, \b) -- (\a+1, \b+2) node[anchor=south, black] {$\tau$};
			\draw[very thick, blue] (\a+2, \b) -- (\a+2, \b+2) node[anchor=south, black] {$\tau$};
		}
	\end{align*}
	\begin{align*}
		\sigma_{1}\tikz[baseline=5ex, scale=1]{
			\def\a{0}
			\def\b{0}
			\draw[very thick, blue, rounded corners=7] (\a, \b+2) node[anchor=south, black] {$\tau$} -- (\a, \b) -- (\a + 2.5, \b) node[anchor=west, black] {$\tau$};
			\node[anchor=north] at (\a + 1.5, \b) {$\tau$};
			\draw[very thick, blue] (\a+1, \b) -- (\a+1, \b+2) node[anchor=south, black] {$\tau$};
			\draw[very thick, blue] (\a+2, \b) -- (\a+2, \b+2) node[anchor=south, black] {$\tau$};
		}=
		\tikz[baseline=5ex, scale=1]{
			\def\a{0}
			\def\b{2}
			\draw[very thick, blue, rounded corners=7] (\a + 1, \b) node[anchor=south, black] {$\tau$} -- (\a + 1, \b - 0.5) -- (\a + 1.4, \b - 0.9);
			\draw[very thick, blue, rounded corners=7] (\a + 1.6, \b - 1.1) -- (\a + 2, \b - 1.5) -- (\a + 2, \b - 2);
			\draw[very thick, blue, rounded corners=7] (\a + 3, \b -2) -- (\a + 3, \b) node[anchor=south, black] {$\tau$};
			\draw[very thick, blue, rounded corners=7] (\a + 2, \b) node[anchor=south, black] {$\tau$} -- (\a + 2, \b - 0.5) -- (\a + 1, \b - 1.5) -- (\a + 1, \b - 2) --  (\a + 3.5, \b - 2) node[anchor=west, black] {$\tau$};
			\node[anchor=north] at (\a + 2.5, \b - 2) {$\tau$};
		}=R_{\tau}^{\tau\tau}
		\tikz[baseline=5ex, scale=1]{
			\def\a{0}
			\def\b{0}
			\draw[very thick, blue, rounded corners=7] (\a, \b+2) node[anchor=south, black] {$\tau$} -- (\a, \b) -- (\a + 2.5, \b) node[anchor=west, black] {$\tau$};
			\node[anchor=north] at (\a + 1.5, \b) {$\tau$};
			\draw[very thick, blue] (\a+1, \b) -- (\a+1, \b+2) node[anchor=south, black] {$\tau$};
			\draw[very thick, blue] (\a+2, \b) -- (\a+2, \b+2) node[anchor=south, black] {$\tau$};
		}
	\end{align*}
	\caption{Construction of $\sigma_1$, a generator of $\mathcal{B}_3$. Here $R^{\tau\tau}_1=\kappa$ and $R^{\tau\tau}_\tau=\lambda$.}
	\label{fig:B3sigma1}
\end{figure}

Moving on to the 3-strand case we need to find two generators, $\sigma_1$ and $\sigma_2$ that are two-dimensional, as this fusion space contains just two basis vectors. 
The generator $\sigma_1$ is obtained by just using the $R$-moves as shown in Fig. \ref{fig:B3sigma1}.


This results in 
\begin{equation}
\sigma_1 = \left(\begin{array}{cc} \kappa & 0 \\ 0 & \lambda\end{array}\right), 
\end{equation}
or
\begin{equation}\label{eq:B3sigma1}
    \sigma_1=R.
\end{equation}
Again we have left $\lambda$ and $\kappa$ arbitrary to keep the representation as general as possible. 

For the generator $\sigma_2$ we need to use both the $F$- and the $R$-moves to fix this representation. Furthermore we take $F$ to be unitary or, $F^{-1}=F$ and the form of $F$ as 
\begin{equation}\label{eq:FFib}
F = \left(\begin{array}{cc} a & b \\ b & -a\end{array}\right),
\end{equation}
with $a^2+b^2=1$ as $F^2=1_{2\times 2}$. The action of $\sigma_2$ on the two basis vectors of the 3-strand fusion space is shown in Figs. \ref{fig:B3sigma2-0}-\ref{fig:B3sigma2-1}.

\begin{figure}[h]
	\centering
	\begin{align*}
		\tikz[baseline=-5ex, scale=1]{
			\def\a{0}
			\def\b{0}
			\draw[very thick, blue, rounded corners=7] (\a, \b) node[anchor=south, black] {$\tau$} -- (\a, \b -2) -- (\a + 2.5, \b - 2) node[anchor=west, black] {$\tau$};
			\draw[very thick, blue, rounded corners=7] (\a + 1, \b) node[anchor=south, black] {$\tau$} -- (\a + 1, \b - 0.5) -- (\a + 1.4, \b - 0.9);
			\draw[very thick, blue, rounded corners=7] (\a + 1.6, \b - 1.1) -- (\a + 2, \b - 1.5) -- (\a + 2, \b - 2);
			\draw[very thick, blue, rounded corners=7] (\a + 2, \b) node[anchor=south, black] {$\tau$} -- (\a + 2, \b - 0.5) -- (\a + 1, \b - 1.5) -- (\a + 1, \b - 2);
			\node[anchor=north] at (\a + 1.5, \b - 2) {$1$};
		}&=\left(F_{\tau}^{\tau\tau\tau}\right)_{11}
		\tikz[baseline=-5ex, scale=1]{
			\def\a{0}
			\def\b{0}
			\draw[very thick, blue, rounded corners=7] (\a, \b) node[anchor=south, black] {$\tau$} -- (\a, \b -2) -- (\a + 2.5, \b - 2) node[anchor=west, black] {$\tau$};
			\draw[very thick, blue, rounded corners=7] (\a + 1, \b) node[anchor=south, black] {$\tau$} -- (\a + 1, \b - 0.25) -- (\a + 1.45, \b - 0.67);
			\draw[very thick, blue, rounded corners=7] (\a + 1.6, \b - 0.79) -- (\a + 2, \b - 1.25) -- (\a + 1, \b - 2);
			\draw[very thick, blue, rounded corners=7] (\a + 2, \b) node[anchor=south, black] {$\tau$} -- (\a + 2, \b - 0.25) -- (\a + 1, \b - 1.25) -- (\a + 1.5, \b - 1.6);
			\node[anchor=north] at (\a + 1.5, \b - 1.6) {$1$};
		}
		+
		\left(F_{\tau}^{\tau\tau\tau}\right)_{1\tau}
		\tikz[baseline=-5ex, scale=1]{
			\def\a{0}
			\def\b{0}
			\draw[very thick, blue, rounded corners=7] (\a, \b) node[anchor=south, black] {$\tau$} -- (\a, \b -2) -- (\a + 2.5, \b - 2) node[anchor=west, black] {$\tau$};
			\draw[very thick, blue, rounded corners=7] (\a + 1, \b) node[anchor=south, black] {$\tau$} -- (\a + 1, \b - 0.25) -- (\a + 1.45, \b - 0.67);
			\draw[very thick, blue, rounded corners=7] (\a + 1.6, \b - 0.79) -- (\a + 2, \b - 1.25) -- (\a + 1, \b - 2);
			\draw[very thick, blue, rounded corners=7] (\a + 2, \b) node[anchor=south, black] {$\tau$} -- (\a + 2, \b - 0.25) -- (\a + 1, \b - 1.25) -- (\a + 1.5, \b - 1.6);
			\node[anchor=north] at (\a + 1.5, \b - 1.6) {$\tau$};
		}\\
	&=\left(F_{\tau}^{\tau\tau\tau}\right)_{11}R_{1}^{\tau\tau}
	\tikz[baseline=-5ex, scale=1]{
		\def\a{0}
		\def\b{0}
		\draw[very thick, blue, rounded corners=7] (\a, \b) node[anchor=south, black] {$\tau$} -- (\a, \b -2) -- (\a + 2.5, \b - 2) node[anchor=west, black] {$\tau$};
		\draw[very thick, blue, rounded corners=7] (\a + 1, \b) node[anchor=south, black] {$\tau$} -- (\a + 1, \b - 0.5) -- (\a + 1.5, \b - 1);
		\draw[very thick, blue, rounded corners=7] (\a + 2, \b) node[anchor=south, black] {$\tau$} -- (\a + 2, \b - 0.5) -- (\a + 1, \b - 1.5) -- (\a + 1, \b - 2);
		\node[anchor=north] at (\a + 1.4, \b - 1.2) {$1$};
	}
	+
	\left(F_{\tau}^{\tau\tau\tau}\right)_{1\tau}R_{\tau}^{\tau\tau}
	\tikz[baseline=-5ex, scale=1]{
		\def\a{0}
		\def\b{0}
		\draw[very thick, blue, rounded corners=7] (\a, \b) node[anchor=south, black] {$\tau$} -- (\a, \b -2) -- (\a + 2.5, \b - 2) node[anchor=west, black] {$\tau$};
		\draw[very thick, blue, rounded corners=7] (\a + 1, \b) node[anchor=south, black] {$\tau$} -- (\a + 1, \b - 0.5) -- (\a + 1.5, \b - 1);
		\draw[very thick, blue, rounded corners=7] (\a + 2, \b) node[anchor=south, black] {$\tau$} -- (\a + 2, \b - 0.5) -- (\a + 1, \b - 1.5) -- (\a + 1, \b - 2);
		\node[anchor=north] at (\a + 1.4, \b - 1.2) {$\tau$};
	}\\
	&=\left(F_{\tau}^{\tau\tau\tau}\right)_{11}R_{1}^{\tau\tau}\left[\left(F_{\tau}^{\tau\tau\tau}\right)_{11}^{-1}
	\tikz[baseline=-5ex, scale=0.8]{
		\def\a{0}
		\def\b{0}
		\draw[very thick, blue, rounded corners=7] (\a, \b) node[anchor=south, black] {$\tau$} -- (\a, \b -2) -- (\a + 2.5, \b - 2) node[anchor=west, black] {$\tau$};
		\draw[very thick, blue, rounded corners=7] (\a + 1, \b) node[anchor=south, black] {$\tau$} -- (\a + 1, \b -2);
		\draw[very thick, blue, rounded corners=7] (\a + 2, \b) node[anchor=south, black] {$\tau$} -- (\a + 2, \b - 2);
		\node[anchor=north] at (\a + 1.5, \b - 2) {$1$};
	}
	+
	\left(F_{\tau}^{\tau\tau\tau}\right)_{1\tau}^{-1}
	\tikz[baseline=-5ex, scale=0.8]{
		\def\a{0}
		\def\b{0}
		\draw[very thick, blue, rounded corners=7] (\a, \b) node[anchor=south, black] {$\tau$} -- (\a, \b -2) -- (\a + 2.5, \b - 2) node[anchor=west, black] {$\tau$};
		\draw[very thick, blue, rounded corners=7] (\a + 1, \b) node[anchor=south, black] {$\tau$} -- (\a + 1, \b -2);
		\draw[very thick, blue, rounded corners=7] (\a + 2, \b) node[anchor=south, black] {$\tau$} -- (\a + 2, \b - 2);
		\node[anchor=north] at (\a + 1.5, \b - 2) {$\tau$};
	}\right]\\
	&+\left(F_{\tau}^{\tau\tau\tau}\right)_{1\tau}R_{\tau}^{\tau\tau}\left[\left(F_{\tau}^{\tau\tau\tau}\right)_{\tau 1}^{-1}
	\tikz[baseline=-5ex, scale=0.8]{
		\def\a{0}
		\def\b{0}
		\draw[very thick, blue, rounded corners=7] (\a, \b) node[anchor=south, black] {$\tau$} -- (\a, \b -2) -- (\a + 2.5, \b - 2) node[anchor=west, black] {$\tau$};
		\draw[very thick, blue, rounded corners=7] (\a + 1, \b) node[anchor=south, black] {$\tau$} -- (\a + 1, \b -2);
		\draw[very thick, blue, rounded corners=7] (\a + 2, \b) node[anchor=south, black] {$\tau$} -- (\a + 2, \b - 2);
		\node[anchor=north] at (\a + 1.5, \b - 2) {$1$};
	}
	+
	\left(F_{\tau}^{\tau\tau\tau}\right)_{\tau\tau}^{-1}
	\tikz[baseline=-5ex, scale=0.8]{
		\def\a{0}
		\def\b{0}
		\draw[very thick, blue, rounded corners=7] (\a, \b) node[anchor=south, black] {$\tau$} -- (\a, \b -2) -- (\a + 2.5, \b - 2) node[anchor=west, black] {$\tau$};
		\draw[very thick, blue, rounded corners=7] (\a + 1, \b) node[anchor=south, black] {$\tau$} -- (\a + 1, \b -2);
		\draw[very thick, blue, rounded corners=7] (\a + 2, \b) node[anchor=south, black] {$\tau$} -- (\a + 2, \b - 2);
		\node[anchor=north] at (\a + 1.5, \b - 2) {$\tau$};
	}\right]
	\end{align*}
	\caption{Action of $\sigma_2\in \mathcal{B}_3$ on the fusion basis vector $|1\rangle$. The elements of the $F$-matrix are in \eqref{eq:FFib}.}
	\label{fig:B3sigma2-0}
\end{figure}


\begin{figure}[h]
	\centering
	\begin{align*}
		\tikz[baseline=-5ex, scale=1]{
			\def\a{0}
			\def\b{0}
			\draw[very thick, blue, rounded corners=7] (\a, \b) node[anchor=south, black] {$\tau$} -- (\a, \b -2) -- (\a + 2.5, \b - 2) node[anchor=west, black] {$\tau$};
			\draw[very thick, blue, rounded corners=7] (\a + 1, \b) node[anchor=south, black] {$\tau$} -- (\a + 1, \b - 0.5) -- (\a + 1.4, \b - 0.9);
			\draw[very thick, blue, rounded corners=7] (\a + 1.6, \b - 1.1) -- (\a + 2, \b - 1.5) -- (\a + 2, \b - 2);
			\draw[very thick, blue, rounded corners=7] (\a + 2, \b) node[anchor=south, black] {$\tau$} -- (\a + 2, \b - 0.5) -- (\a + 1, \b - 1.5) -- (\a + 1, \b - 2);
			\node[anchor=north] at (\a + 1.5, \b - 2) {$1$};
		}&=\left(F_{\tau}^{\tau\tau\tau}\right)_{\tau 1}
		\tikz[baseline=-5ex, scale=1]{
			\def\a{0}
			\def\b{0}
			\draw[very thick, blue, rounded corners=7] (\a, \b) node[anchor=south, black] {$\tau$} -- (\a, \b -2) -- (\a + 2.5, \b - 2) node[anchor=west, black] {$\tau$};
			\draw[very thick, blue, rounded corners=7] (\a + 1, \b) node[anchor=south, black] {$\tau$} -- (\a + 1, \b - 0.25) -- (\a + 1.45, \b - 0.67);
			\draw[very thick, blue, rounded corners=7] (\a + 1.6, \b - 0.79) -- (\a + 2, \b - 1.25) -- (\a + 1, \b - 2);
			\draw[very thick, blue, rounded corners=7] (\a + 2, \b) node[anchor=south, black] {$\tau$} -- (\a + 2, \b - 0.25) -- (\a + 1, \b - 1.25) -- (\a + 1.5, \b - 1.6);
			\node[anchor=north] at (\a + 1.5, \b - 1.6) {$1$};
		}
		+
		\left(F_{\tau}^{\tau\tau\tau}\right)_{\tau\tau}
		\tikz[baseline=-5ex, scale=1]{
			\def\a{0}
			\def\b{0}
			\draw[very thick, blue, rounded corners=7] (\a, \b) node[anchor=south, black] {$\tau$} -- (\a, \b -2) -- (\a + 2.5, \b - 2) node[anchor=west, black] {$\tau$};
			\draw[very thick, blue, rounded corners=7] (\a + 1, \b) node[anchor=south, black] {$\tau$} -- (\a + 1, \b - 0.25) -- (\a + 1.45, \b - 0.67);
			\draw[very thick, blue, rounded corners=7] (\a + 1.6, \b - 0.79) -- (\a + 2, \b - 1.25) -- (\a + 1, \b - 2);
			\draw[very thick, blue, rounded corners=7] (\a + 2, \b) node[anchor=south, black] {$\tau$} -- (\a + 2, \b - 0.25) -- (\a + 1, \b - 1.25) -- (\a + 1.5, \b - 1.6);
			\node[anchor=north] at (\a + 1.5, \b - 1.6) {$\tau$};
		}\\
		&=\left(F_{\tau}^{\tau\tau\tau}\right)_{\tau 1}R_{1}^{\tau\tau}
		\tikz[baseline=-5ex, scale=1]{
			\def\a{0}
			\def\b{0}
			\draw[very thick, blue, rounded corners=7] (\a, \b) node[anchor=south, black] {$\tau$} -- (\a, \b -2) -- (\a + 2.5, \b - 2) node[anchor=west, black] {$\tau$};
			\draw[very thick, blue, rounded corners=7] (\a + 1, \b) node[anchor=south, black] {$\tau$} -- (\a + 1, \b - 0.5) -- (\a + 1.5, \b - 1);
			\draw[very thick, blue, rounded corners=7] (\a + 2, \b) node[anchor=south, black] {$\tau$} -- (\a + 2, \b - 0.5) -- (\a + 1, \b - 1.5) -- (\a + 1, \b - 2);
			\node[anchor=north] at (\a + 1.4, \b - 1.2) {$1$};
		}
		+
		\left(F_{\tau}^{\tau\tau\tau}\right)_{\tau\tau}R_{\tau}^{\tau\tau}
		\tikz[baseline=-5ex, scale=1]{
			\def\a{0}
			\def\b{0}
			\draw[very thick, blue, rounded corners=7] (\a, \b) node[anchor=south, black] {$\tau$} -- (\a, \b -2) -- (\a + 2.5, \b - 2) node[anchor=west, black] {$\tau$};
			\draw[very thick, blue, rounded corners=7] (\a + 1, \b) node[anchor=south, black] {$\tau$} -- (\a + 1, \b - 0.5) -- (\a + 1.5, \b - 1);
			\draw[very thick, blue, rounded corners=7] (\a + 2, \b) node[anchor=south, black] {$\tau$} -- (\a + 2, \b - 0.5) -- (\a + 1, \b - 1.5) -- (\a + 1, \b - 2);
			\node[anchor=north] at (\a + 1.4, \b - 1.2) {$\tau$};
		}\\
		&=\left(F_{\tau}^{\tau\tau\tau}\right)_{\tau 1}R_{1}^{\tau\tau}\left[\left(F_{\tau}^{\tau\tau\tau}\right)_{11}^{-1}
		\tikz[baseline=-5ex, scale=0.8]{
			\def\a{0}
			\def\b{0}
			\draw[very thick, blue, rounded corners=7] (\a, \b) node[anchor=south, black] {$\tau$} -- (\a, \b -2) -- (\a + 2.5, \b - 2) node[anchor=west, black] {$\tau$};
			\draw[very thick, blue, rounded corners=7] (\a + 1, \b) node[anchor=south, black] {$\tau$} -- (\a + 1, \b -2);
			\draw[very thick, blue, rounded corners=7] (\a + 2, \b) node[anchor=south, black] {$\tau$} -- (\a + 2, \b - 2);
			\node[anchor=north] at (\a + 1.5, \b - 2) {$1$};
		}
		+
		\left(F_{\tau}^{\tau\tau\tau}\right)_{1\tau}^{-1}
		\tikz[baseline=-5ex, scale=0.8]{
			\def\a{0}
			\def\b{0}
			\draw[very thick, blue, rounded corners=7] (\a, \b) node[anchor=south, black] {$\tau$} -- (\a, \b -2) -- (\a + 2.5, \b - 2) node[anchor=west, black] {$\tau$};
			\draw[very thick, blue, rounded corners=7] (\a + 1, \b) node[anchor=south, black] {$\tau$} -- (\a + 1, \b -2);
			\draw[very thick, blue, rounded corners=7] (\a + 2, \b) node[anchor=south, black] {$\tau$} -- (\a + 2, \b - 2);
			\node[anchor=north] at (\a + 1.5, \b - 2) {$\tau$};
		}\right]\\
		&+\left(F_{\tau}^{\tau\tau\tau}\right)_{\tau\tau}R_{\tau}^{\tau\tau}\left[\left(F_{\tau}^{\tau\tau\tau}\right)_{\tau 1}^{-1}
		\tikz[baseline=-5ex, scale=0.8]{
			\def\a{0}
			\def\b{0}
			\draw[very thick, blue, rounded corners=7] (\a, \b) node[anchor=south, black] {$\tau$} -- (\a, \b -2) -- (\a + 2.5, \b - 2) node[anchor=west, black] {$\tau$};
			\draw[very thick, blue, rounded corners=7] (\a + 1, \b) node[anchor=south, black] {$\tau$} -- (\a + 1, \b -2);
			\draw[very thick, blue, rounded corners=7] (\a + 2, \b) node[anchor=south, black] {$\tau$} -- (\a + 2, \b - 2);
			\node[anchor=north] at (\a + 1.5, \b - 2) {$1$};
		}
		+
		\left(F_{\tau}^{\tau\tau\tau}\right)_{\tau\tau}^{-1}
		\tikz[baseline=-5ex, scale=0.8]{
			\def\a{0}
			\def\b{0}
			\draw[very thick, blue, rounded corners=7] (\a, \b) node[anchor=south, black] {$\tau$} -- (\a, \b -2) -- (\a + 2.5, \b - 2) node[anchor=west, black] {$\tau$};
			\draw[very thick, blue, rounded corners=7] (\a + 1, \b) node[anchor=south, black] {$\tau$} -- (\a + 1, \b -2);
			\draw[very thick, blue, rounded corners=7] (\a + 2, \b) node[anchor=south, black] {$\tau$} -- (\a + 2, \b - 2);
			\node[anchor=north] at (\a + 1.5, \b - 2) {$\tau$};
		}\right]
	\end{align*}
	\caption{Action of $\sigma_2\in \mathcal{B}_3$ on the fusion basis vector $|\tau\rangle$. The elements of the $F$-matrix are in \eqref{eq:FFib}.}
	\label{fig:B3sigma2-1}
\end{figure}

We end up with the generator
\begin{equation}
\sigma_2 = \left(\begin{array}{cc} a^2\kappa + b^2\lambda & ab\left(-\lambda+\kappa\right) \\ ab\left(-\lambda+\kappa\right) & b^2\kappa + a^2\lambda\end{array}\right) = \left(\begin{array}{cc} a & b \\ b & -a\end{array}\right)\left(\begin{array}{cc} \kappa & 0 \\ 0 & \lambda\end{array}\right)\left(\begin{array}{cc} a & b \\ b & -a\end{array}\right),
\end{equation}
or 
\begin{equation}\label{eq:B3sigma2}
\sigma_2 = FRF.
\end{equation}
Thus from \eqref{eq:B3sigma1} and \eqref{eq:B3sigma2}, we see that
\begin{equation}
    \sigma_1\sigma_2\sigma_1 = \sigma_2\sigma_1\sigma_2,
\end{equation}
provided $F$ and $R$ satisfy the hexagon identity. Note that there are no far commutativity relations to be satisfied in $\mathcal{B}_3$. However to prove the braid relations for $\sigma_1$ and $\sigma_2$ and to obtain the representations of $\mathcal{B}_N$, it is convenient to use the Jones representation as this greatly simplifies the ensuing algebra. As a consequence, we also obtain in the process non-local representations of the Temperley-Lieb algebra on the anyonic fusion space. 

Before going to the Jones representation we note that we could have also braided the anyon strands in the opposite way in Figs. \ref{fig:B3sigma1}, \ref{fig:B3sigma2-0} and \ref{fig:B3sigma2-1} resulting in the $\mathcal{B}_3$ generators, 
\begin{equation}\label{eq:B3repalt}
    \sigma_1 = R^{-1},~~\sigma_2=FR^{-1}F.
\end{equation}
These generators satisfy the braid relation, $\sigma_1\sigma_2\sigma_1=\sigma_2\sigma_1\sigma_2$ provided the alternate hexagon identity, Eq. \ref{hexagonidalt} is satisfied. However we will use the braiding conventions used in Figs. \ref{fig:B3sigma1}, \ref{fig:B3sigma2-0}, \ref{fig:B3sigma2-1} and will not pursue the opposite convention any more. 


\subsection{Temperley-Lieb algebra and the Jones representation}
\label{subsec:TLJrepresentation}
\begin{figure}[h]
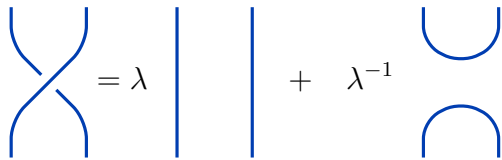

	\centering
	\begin{align*}
	\tikz[baseline=-6ex, scale=1]{
		\def\a{0}
		\def\b{0}
		\draw[very thick, blue, rounded corners=7] (\a + 1, \b) -- (\a + 1, \b - 0.5) -- (\a + 1.4, \b - 0.9);
		\draw[very thick, blue, rounded corners=7] (\a + 1.6, \b - 1.1) -- (\a + 2, \b - 1.5) -- (\a + 2, \b - 2);
		\draw[very thick, blue, rounded corners=7] (\a + 2, \b) -- (\a + 2, \b - 0.5) -- (\a + 1, \b - 1.5) -- (\a + 1, \b - 2);
	} = \lambda
	\tikz[baseline=-6ex, scale=1]{
		\def\a{0}
		\def\b{0}
		\node at (\a - 0.25, 0) {};
		\draw[very thick, blue, rounded corners=7] (\a, \b) -- (\a, \b - 2);
		\draw[very thick, blue, rounded corners=7] (\a + 1, \b) -- (\a + 1, \b - 2);
		\node at (\a + 1.25, 0) {};
	}
	+\;\;\;
	\lambda^{-1}
	\tikz[baseline=-6ex, scale=1]{
		\def\a{0}
		\def\b{0}
		\node at (\a - 0.25, 0) {};
		\draw[very thick, blue, rounded corners=7] (\a, \b) -- (\a, \b - 0.5) -- (\a + 0.5, \b - 0.75) -- (\a + 1, \b - 0.5) -- (\a + 1, \b);
		\draw[very thick, blue, rounded corners=7] (\a, \b - 2) -- (\a, \b - 1.5) -- (\a + 0.5, \b - 1.25) -- (\a + 1, \b - 1.5)-- (\a + 1, \b - 2);
	}
	\end{align*}
	\caption{Jones representation of the braid group in pictures. Here $A=\lambda$.}
	\label{fig:Jonesrep}
\end{figure}

Consider the $N$-strand Temperley-Lieb algebra $TL_N(\delta)$, generated by $U_i$ (with $i=1,\ldots, N-1$) satisfying the relations
\begin{eqnarray}\label{eq:TLrelations}
    U_i^2 & = & \delta U_i, \nonumber \\
    U_iU_{i\pm 1}U_i & = & U_i, \nonumber \\
    U_iU_j & = & U_jU_i, \qquad  |i-j|>1.
\end{eqnarray}
This algebra is usually defined on a ring $\mathbb{Z}\left[A, A^{-1}\right]$ and $\delta = -A^2-A^{-2}$. Define now the $\mathcal{B}_N$ generator using the TL algebra as
\begin{equation}\label{eq:Jonesrep}
    \sigma_i = A~I + A^{-1}~U_i.
\end{equation}
This is seen to satisfy the braid relations from
\begin{equation}
    \sigma_i\sigma_{i+1}\sigma_i = A^3I + A\left(U_i + U_{i+1}\right) + A^{-1}\left[U_iU_{i+1}+U_{i+1}U_i\right],
\end{equation}
which is symmetric under the interchange, $i\leftrightarrow i+1$. Far commutativity follows from the TL relations \eqref{eq:TLrelations}. This is known as the {\it Jones representation} and is pictorially depicted in Fig. \ref{fig:Jonesrep}.


This expression leads to the construction of the {\it bracket polynomial} as a state sum model and is related to the Jones polynomial. We will use this representation to obtain $\mathcal{B}_N$ representations on $N$-strands.


We start with $\mathcal{B}_3$. We will closely follow the notation of \cite{kauffmanlomonaco}.
The generator $\sigma_1$ in \eqref{eq:B3sigma1} can be decomposed as
\begin{equation}
    \sigma_1 = \lambda I + \lambda^{-1} \left(\begin{array}{cc} \delta & 0 \\ 0 & 0\end{array}\right),
\end{equation}
with $\delta=\lambda\left(\kappa-\lambda\right)$. We now identify $\lambda$ with $A$ in \eqref{eq:Jonesrep} and equate $-\lambda^2-\lambda^{-2} = \lambda\left(\kappa-\lambda\right)$ to find $\kappa = -\lambda^{-3}$. Then
\begin{equation}
    U_1 = \left(\begin{array}{cc} \delta & 0 \\ 0 & 0\end{array}\right), \qquad  U_1^2 = \delta U_1.
\end{equation}
Thus imposing the Jones representation leads to a relation between $\lambda$ and $\kappa$. Next we look at $\sigma_2$ in \eqref{eq:B3sigma2}. We find 
\begin{equation}
    \sigma_2 = \lambda I + \lambda^{-1}U_2 = F\left(\lambda I + \lambda^{-1}U_1\right)F,
\end{equation}
which implies $U_2=FU_1F$ and
\begin{equation}
    U_2^2 = FU_1FFU_1F = FU_1^2F = \delta FU_1F = \delta U_2,
\end{equation}
as desired. More compactly, we can write
\begin{equation}
    U_1 = \delta \ket{w}\bra{w},\qquad U_2 = \delta \ket{v}\bra{v},
\end{equation}
with $\ket{w} = \left(\begin{array}{c} 1 \\ 0 \end{array}\right)$ and $\ket{v}=F\ket{w}$. From this we see that 
\begin{equation}
    U_1U_2U_1 = \delta^2a^2 U_1,
\end{equation}
and so we have $a = \delta^{-1}$. This gives us a $\mathcal{B}_3$ representation, 
\begin{equation}
    \sigma_1 = \left(\begin{array}{cc}
    -\lambda^{-3} & 0 \\
       0 & \lambda 
    \end{array}\right),~~\sigma_2 = FRF= \frac{1}{1+\lambda^4}\left(\begin{array}{cc}
    \lambda^5 & \frac{\sqrt{1+\lambda^4+\lambda^8}}{\lambda} \\
       \frac{\sqrt{1+\lambda^4+\lambda^8}}{\lambda} & -\lambda^{-3} 
    \end{array}\right),
\end{equation}
in terms of the parameter $\lambda$, a 1-parameter family. In this process we have also obtained a representation of $TL_3(\delta)$ on the anyonic fusion space. Note that we have still not used the specific form of the parameters for Fibonacci anyons. 

Next we proceed to obtain the representations of $\mathcal{B}_N$ using the Jones form.
We divide the generators into those acting on the left end, $\left[\sigma_1, \sigma_2\right]$, those acting in the bulk, $\sigma_i$ (with $i=3,\ldots, N-2$), and the generator acting on the right end, $\sigma_{N-1}$. 

\begin{figure}[h!]
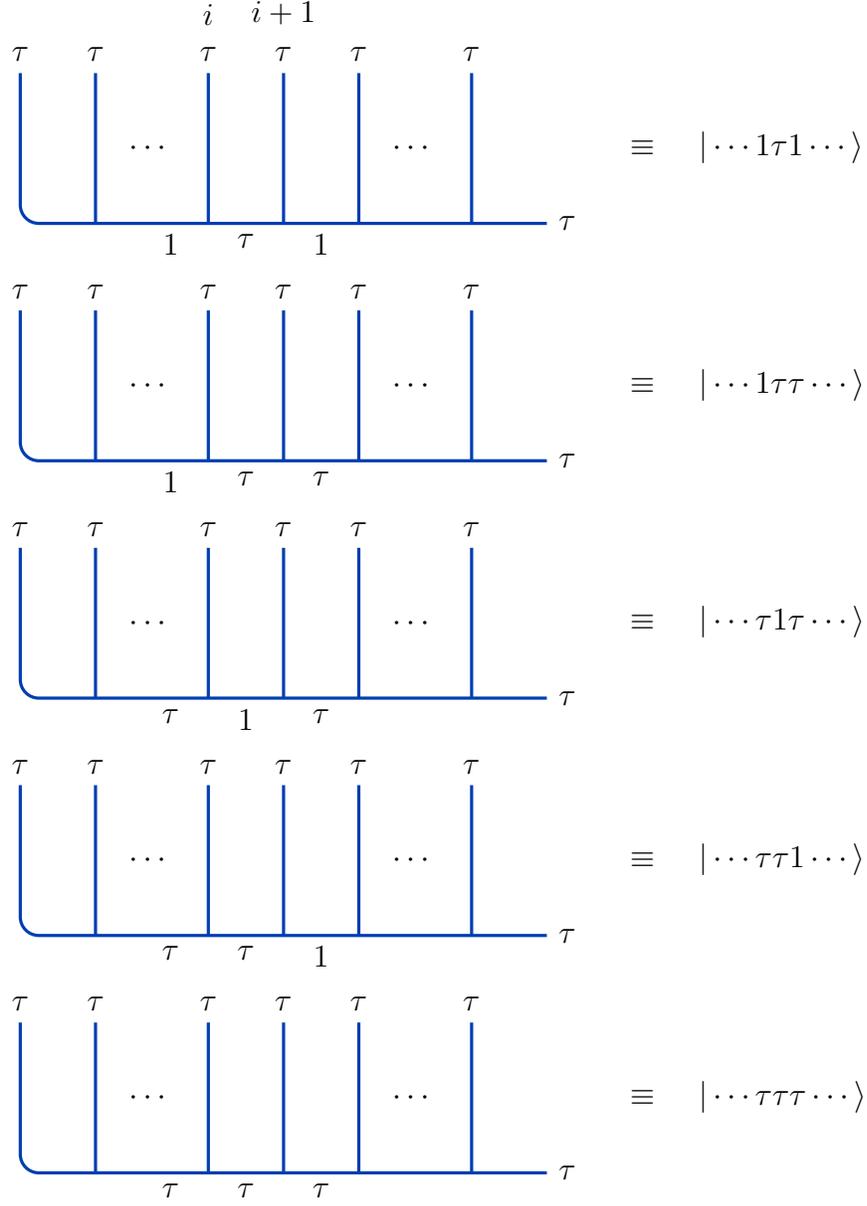

	\centering
	\begin{align*}
		\tikz[baseline=5ex, scale=1]{
			\def\a{0}
			\def\b{0}
			\node[anchor=south] at (\a + 2.5, \b + 2.5) {$i$};
			\node[anchor=south] at (\a + 3.5, \b + 2.5) {$i+1$};
			\draw[very thick, blue, rounded corners=7] (\a, \b+2) node[anchor=south, black] {$\tau$} -- (\a, \b) -- (\a + 7, \b) node[anchor=west, black] {$\tau$};
			\draw[very thick, blue] (\a+1, \b) -- (\a+1, \b+2) node[anchor=south, black] {$\tau$};
			\node[anchor=west] at (\a+1.3, \b+1) {$\cdots$};
			\draw[very thick, blue] (\a+2.5, \b) -- (\a+2.5, \b+2) node[anchor=south, black] {$\tau$};
			\draw[very thick, blue] (\a+3.5, \b) -- (\a+3.5, \b+2) node[anchor=south, black] {$\tau$};
			\draw[very thick, blue] (\a+4.5, \b) -- (\a+4.5, \b+2) node[anchor=south, black] {$\tau$};
			\node[anchor=west] at (\a+4.8, \b+1) {$\cdots$};
			\draw[very thick, blue] (\a+6, \b) -- (\a+6, \b+2) node[anchor=south, black] {$\tau$};
			\node[anchor=north] at (\a + 2, \b) {$1$};
			\node[anchor=north] at (\a + 3, \b) {$\tau$};
			\node[anchor=north] at (\a + 4, \b) {$1$};
		}\;\;\;\;&\equiv\;\;\;\; |\cdots 1\tau 1\cdots \rangle\\
	\tikz[baseline=5ex, scale=1]{
		\def\a{0}
		\def\b{0}
		\draw[very thick, blue, rounded corners=7] (\a, \b+2) node[anchor=south, black] {$\tau$} -- (\a, \b) -- (\a + 7, \b) node[anchor=west, black] {$\tau$};
		\draw[very thick, blue] (\a+1, \b) -- (\a+1, \b+2) node[anchor=south, black] {$\tau$};
		\node[anchor=west] at (\a+1.3, \b+1) {$\cdots$};
		\draw[very thick, blue] (\a+2.5, \b) -- (\a+2.5, \b+2) node[anchor=south, black] {$\tau$};
		\draw[very thick, blue] (\a+3.5, \b) -- (\a+3.5, \b+2) node[anchor=south, black] {$\tau$};
		\draw[very thick, blue] (\a+4.5, \b) -- (\a+4.5, \b+2) node[anchor=south, black] {$\tau$};
		\node[anchor=west] at (\a+4.8, \b+1) {$\cdots$};
		\draw[very thick, blue] (\a+6, \b) -- (\a+6, \b+2) node[anchor=south, black] {$\tau$};
		\node[anchor=north] at (\a + 2, \b) {$1$};
		\node[anchor=north] at (\a + 3, \b) {$\tau$};
		\node[anchor=north] at (\a + 4, \b) {$\tau$};
	}\;\;\;\;&\equiv\;\;\;\; |\cdots 1\tau \tau\cdots \rangle\\
	\tikz[baseline=5ex, scale=1]{
		\def\a{0}
		\def\b{0}
		\draw[very thick, blue, rounded corners=7] (\a, \b+2) node[anchor=south, black] {$\tau$} -- (\a, \b) -- (\a + 7, \b) node[anchor=west, black] {$\tau$};
		\draw[very thick, blue] (\a+1, \b) -- (\a+1, \b+2) node[anchor=south, black] {$\tau$};
		\node[anchor=west] at (\a+1.3, \b+1) {$\cdots$};
		\draw[very thick, blue] (\a+2.5, \b) -- (\a+2.5, \b+2) node[anchor=south, black] {$\tau$};
		\draw[very thick, blue] (\a+3.5, \b) -- (\a+3.5, \b+2) node[anchor=south, black] {$\tau$};
		\draw[very thick, blue] (\a+4.5, \b) -- (\a+4.5, \b+2) node[anchor=south, black] {$\tau$};
		\node[anchor=west] at (\a+4.8, \b+1) {$\cdots$};
		\draw[very thick, blue] (\a+6, \b) -- (\a+6, \b+2) node[anchor=south, black] {$\tau$};
		\node[anchor=north] at (\a + 2, \b) {$\tau$};
		\node[anchor=north] at (\a + 3, \b) {$1$};
		\node[anchor=north] at (\a + 4, \b) {$\tau$};
	}\;\;\;\;&\equiv\;\;\;\; |\cdots \tau 1\tau\cdots \rangle\\
	\tikz[baseline=5ex, scale=1]{
		\def\a{0}
		\def\b{0}
		\draw[very thick, blue, rounded corners=7] (\a, \b+2) node[anchor=south, black] {$\tau$} -- (\a, \b) -- (\a + 7, \b) node[anchor=west, black] {$\tau$};
		\draw[very thick, blue] (\a+1, \b) -- (\a+1, \b+2) node[anchor=south, black] {$\tau$};
		\node[anchor=west] at (\a+1.3, \b+1) {$\cdots$};
		\draw[very thick, blue] (\a+2.5, \b) -- (\a+2.5, \b+2) node[anchor=south, black] {$\tau$};
		\draw[very thick, blue] (\a+3.5, \b) -- (\a+3.5, \b+2) node[anchor=south, black] {$\tau$};
		\draw[very thick, blue] (\a+4.5, \b) -- (\a+4.5, \b+2) node[anchor=south, black] {$\tau$};
		\node[anchor=west] at (\a+4.8, \b+1) {$\cdots$};
		\draw[very thick, blue] (\a+6, \b) -- (\a+6, \b+2) node[anchor=south, black] {$\tau$};
		\node[anchor=north] at (\a + 2, \b) {$\tau$};
		\node[anchor=north] at (\a + 3, \b) {$\tau$};
		\node[anchor=north] at (\a + 4, \b) {$1$};
	}\;\;\;\;&\equiv\;\;\;\; |\cdots \tau \tau 1\cdots \rangle\\
	\tikz[baseline=5ex, scale=1]{
		\def\a{0}
		\def\b{0}
		\draw[very thick, blue, rounded corners=7] (\a, \b+2) node[anchor=south, black] {$\tau$} -- (\a, \b) -- (\a + 7, \b) node[anchor=west, black] {$\tau$};
		\draw[very thick, blue] (\a+1, \b) -- (\a+1, \b+2) node[anchor=south, black] {$\tau$};
		\node[anchor=west] at (\a+1.3, \b+1) {$\cdots$};
		\draw[very thick, blue] (\a+2.5, \b) -- (\a+2.5, \b+2) node[anchor=south, black] {$\tau$};
		\draw[very thick, blue] (\a+3.5, \b) -- (\a+3.5, \b+2) node[anchor=south, black] {$\tau$};
		\draw[very thick, blue] (\a+4.5, \b) -- (\a+4.5, \b+2) node[anchor=south, black] {$\tau$};
		\node[anchor=west] at (\a+4.8, \b+1) {$\cdots$};
		\draw[very thick, blue] (\a+6, \b) -- (\a+6, \b+2) node[anchor=south, black] {$\tau$};
		\node[anchor=north] at (\a + 2, \b) {$\tau$};
		\node[anchor=north] at (\a + 3, \b) {$\tau$};
		\node[anchor=north] at (\a + 4, \b) {$\tau$};
	}\;\;\;\;&\equiv\;\;\;\; |\cdots \tau \tau \tau\cdots \rangle\\
	\end{align*}
	\caption{The basis vectors in the bulk on which $\sigma_i$ has a non-trivial action.}
	\label{fig:bulkbasisstates}
\end{figure}

The bulk generators $\sigma_{i, i+1}$ act on the basis states shown in Fig. \ref{fig:bulkbasisstates}.

\begin{figure}[h]
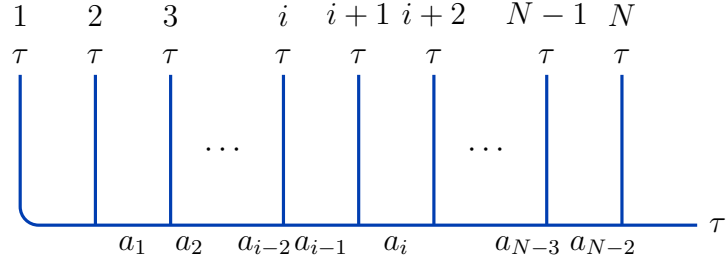

	\centering
	\tikz[baseline=5ex, scale=1]{
		\def\a{0}
		\def\b{0}
		\node[anchor=south] at (\a - 1, \b + 2.5) {$1$};
		\node[anchor=south] at (\a, \b + 2.5) {$2$};
		\node[anchor=south] at (\a + 1, \b + 2.5) {$3$};
		\node[anchor=south] at (\a + 2.5, \b + 2.5) {$i$};
		\node[anchor=south] at (\a + 3.5, \b + 2.5) {$i+1$};
		\node[anchor=south] at (\a + 4.5, \b + 2.5) {$i+2$};
		\node[anchor=south] at (\a + 6, \b + 2.5) {$N-1$};
		\node[anchor=south] at (\a + 7, \b + 2.5) {$N$};
		\draw[very thick, blue, rounded corners=7] (\a-1, \b+2) node[anchor=south, black] {$\tau$} -- (\a-1, \b) -- (\a + 8, \b) node[anchor=west, black] {$\tau$};
		\draw[very thick, blue] (\a, \b) -- (\a, \b+2) node[anchor=south, black] {$\tau$};
		\draw[very thick, blue] (\a+1, \b) -- (\a+1, \b+2) node[anchor=south, black] {$\tau$};
		\node[anchor=west] at (\a+1.3, \b+1) {$\cdots$};
		\draw[very thick, blue] (\a+2.5, \b) -- (\a+2.5, \b+2) node[anchor=south, black] {$\tau$};
		\draw[very thick, blue] (\a+3.5, \b) -- (\a+3.5, \b+2) node[anchor=south, black] {$\tau$};
		\draw[very thick, blue] (\a+4.5, \b) -- (\a+4.5, \b+2) node[anchor=south, black] {$\tau$};
		\node[anchor=west] at (\a+4.8, \b+1) {$\cdots$};
		\draw[very thick, blue] (\a+6, \b) -- (\a+6, \b+2) node[anchor=south, black] {$\tau$};
		\draw[very thick, blue] (\a+7, \b) -- (\a+7, \b+2) node[anchor=south, black] {$\tau$};
		\node[anchor=north] at (\a + 0.5, \b) {$a_{1}$};
		\node[anchor=north] at (\a + 1.25, \b) {$a_{2}$};
		\node[anchor=north] at (\a + 2.25, \b) {$a_{i-2}$};
		\node[anchor=north] at (\a + 3, \b) {$a_{i-1}$};
		\node[anchor=north] at (\a + 4, \b) {$a_{i}$};
		\node[anchor=north] at (\a + 5.75, \b) {$a_{N-3}$};
		\node[anchor=north] at (\a + 6.75, \b) {$a_{N-2}$};
	}
	\caption{Indexing of the anyons in the basis vectors with respect to the position of the fusing anyons. This indexing is important to index the local boson and fermion states of the supersymmetric Hilbert space. The anyon labels $a$ are either $1$ or $\tau$. The anyon $a_i$ corresponds to $c_ic_{i+1}$ in the supersymmetric Hilbert space. Here $c$ is either a local boson or a fermion. }
	\label{fig:basisindices}
\end{figure}

The braiding of the anyonic lines indexed by $i$ and $i+1$ correspond to the basis states with indices $i-2$, $i-1$ and $i$ as illustrated in Fig. \ref{fig:basisindices}.

Thus we obtain five states
$$\left\{ \ket{\cdots 1_{i-2}\tau_{i-1}1_i\cdots}, \ket{\cdots 1_{i-2}\tau_{i-1}\tau_i\cdots}, \ket{\cdots \tau_{i-2}1_{i-1}\tau_i\cdots}, \ket{\cdots \tau_{i-2}\tau_{i-1}1_i\cdots}, \ket{\cdots \tau_{i-2}\tau_{i-1}\tau_i\cdots} \right\},$$
on which the bulk generators act non-trivially. The action of $\sigma_i$ on these states can be obtained by using the $F$- and $R$-moves. The final expressions for the action of the braid and TL generators are
\begin{eqnarray}
    \sigma_i~\ket{\cdots 1_{i-2}\tau_{i-1}1_i\cdots} & = & \kappa~\ket{\cdots 1_{i-2}\tau_{i-1}1_i\cdots}, \nonumber \\
    \sigma_i~\ket{\cdots 1_{i-2}\tau_{i-1}\tau_i\cdots} & = & \lambda~\ket{\cdots 1_{i-2}\tau_{i-1}\tau_i\cdots}, \nonumber \\
    \sigma_i~\ket{\cdots \tau_{i-2}\tau_{i-1}1_i\cdots} & = & \lambda~\ket{\cdots \tau_{i-2}\tau_{i-1}1_i\cdots}, \nonumber \\
    \sigma_i~ \ket{\cdots \tau_{i-2}1_{i-1}\tau_i\cdots} & = & \left(a^2\kappa + b^2\lambda\right)~\ket{\cdots \tau_{i-2}1_{i-1}\tau_i\cdots} + ab\left(\kappa-\lambda\right)~\ket{\cdots \tau_{i-2}\tau_{i-1}\tau_i\cdots}, \nonumber \\
    \sigma_i~ \ket{\cdots \tau_{i-2}\tau_{i-1}\tau_i\cdots} & = & ab\left(\kappa-\lambda\right)~\ket{\cdots \tau_{i-2}1_{i-1}\tau_i\cdots} + \left(b^2\kappa + a^2\lambda\right)~\ket{\cdots \tau_{i-2}\tau_{i-1}\tau_i\cdots}, \nonumber \\
\end{eqnarray}
and 
\begin{eqnarray}
    U_i~\ket{\cdots 1_{i-2}\tau_{i-1}1_i\cdots} & = & \delta~\ket{\cdots 1_{i-2}\tau_{i-1}1_i\cdots}, \nonumber \\
    U_i~\ket{\cdots 1_{i-2}\tau_{i-1}\tau_i\cdots} & = & 0, \nonumber \\
    U_i~\ket{\cdots \tau_{i-2}\tau_{i-1}1_i\cdots} & = & 0, \nonumber \\
    U_i~ \ket{\cdots \tau_{i-2}1_{i-1}\tau_i\cdots} & = & a~\ket{\cdots \tau_{i-2}1_{i-1}\tau_i\cdots} + b~\ket{\cdots \tau_{i-2}\tau_{i-1}\tau_i\cdots}, \nonumber \\
    U_i~ \ket{\cdots \tau_{i-2}\tau_{i-1}\tau_i\cdots} & = & b~\ket{\cdots \tau_{i-2}1_{i-1}\tau_i\cdots} + \delta b^2~\ket{\cdots \tau_{i-2}\tau_{i-1}\tau_i\cdots}, \nonumber \\
\end{eqnarray}
respectively. Next we find the condition for the TL relations to hold for the bulk TL generators, $U_i$. It is easily seen that
\begin{equation}
    U_iU_{i+1}U_i~\ket{\tau_{i-2}\tau_{i-1}\tau_i\tau_{i+1}} = \delta^2b^4~ U_i~\ket{\tau_{i-2}\tau_{i-1}\tau_i\tau_{i+1}},
\end{equation}
which implies that $\delta^2b^4 =1$ or
\begin{equation}
    \delta - \frac{1}{\delta} = \pm 1, \qquad  \delta = \phi \equiv  \frac{1+\sqrt{5}}{2},
\end{equation}
which reduces the representation to that of the Fibonacci anyons. The solution for $\phi$ corresponds to the +1 value of $\delta-\delta^{-1}$. The other root, $-\frac{1}{\phi}=\frac{1-\sqrt{5}}{2}$ corresponds to the {\it Galois conjugate} of the Fibonacci anyons, known as the {\it Yang-Lee anyons}. The $F$-matrices, solved using the pentagon equation of (\ref{pentid}), is unitary for the Fibonacci case and non-unitary for the Yang-Lee case \cite{Ardonne2010MicroscopicMO}.  

Now we move on to the representations of $\sigma_1$ and $\sigma_2$, the generators on the left end of the anyonic fusion space. The two basis states on which $\sigma_1$ acts are $\left\{\ket{1_1\cdots}, \ket{\tau_1\cdots}\right\}$ and the three basis states on which $\sigma_2$ acts are given by $\left\{\ket{1_1\tau_2\cdots}, \ket{\tau_1\tau_2\cdots}, \ket{\tau_1 1_2\cdots}\right\}$. The action of the braid and TL generators on these states are given by
\begin{eqnarray}
    \sigma_1~\ket{1_1\cdots} & = & \kappa~\ket{1_1\cdots}, \nonumber \\
    \sigma_1~\ket{\tau_1\cdots} & = & \lambda~\ket{\tau_1\cdots}, \nonumber \\
    \sigma_2~\ket{1_1\tau_2\cdots} & = & \left(a^2\kappa + b^2\lambda\right)~\ket{1_1\tau_2\cdots} + ab\left(\kappa-\lambda\right)~\ket{\tau_1\tau_2\cdots}, \nonumber \\
    \sigma_2~\ket{\tau_1\tau_2\cdots} & = & ab\left(\kappa-\lambda\right)~\ket{1_1\tau_2\cdots} + \left(b^2\kappa + a^2\lambda\right)~\ket{\tau_1\tau_2\cdots}, \nonumber \\
    \sigma_2~\ket{\tau_11_2\cdots} & = & \lambda~\ket{\tau_11_2\cdots},
\end{eqnarray}
and 
\begin{eqnarray}
    U_1~\ket{1_1\cdots} & = & \delta~\ket{1_1\cdots}, \nonumber \\
    U_1~\ket{\tau_1\cdots} & = & 0, \nonumber \\
    U_2~\ket{1_1\tau_2\cdots} & = & a~\ket{1_1\tau_2\cdots} + b~\ket{\tau_1\tau_2\cdots}, \nonumber \\
    U_2~\ket{\tau_1\tau_2\cdots} & = & b~\ket{1_1\tau_2\cdots} + \delta b^2~\ket{\tau_1\tau_2\cdots}, \nonumber \\
    U_2~\ket{\tau_11_2\cdots} & = & 0,
\end{eqnarray}
respectively. 

The generator on the right end of the fusion space $\sigma_{N-1}$ acts on the basis states $\ket{\cdots 1_{N-3}\tau_{N-2}}$, $\ket{\cdots \tau_{N-3}1_{N-2}}$, and $\ket{\cdots \tau_{N-3}\tau_{N-2}}$. The action of $\sigma_{N-1}$ and $U_{N-1}$ on these states are summarized as
\begin{eqnarray}
    \sigma_{N-1}~\ket{\cdots 1_{N-3}\tau_{N-2}} & = & \lambda~\ket{\cdots 1_{N-3}\tau_{N-2}}, \nonumber \\
    \sigma_{N-1}~\ket{\cdots \tau_{N-3}1_{N-2}} & = & \left(a^2\kappa + b^2\lambda\right)~\ket{\cdots \tau_{N-3}1_{N-2}} + ab\left(\kappa-\lambda\right)~\ket{\cdots \tau_{N-3}\tau_{N-2}}, \nonumber \\
    \sigma_{N-1}~\ket{\cdots \tau_{N-3}\tau_{N-2}} & = & ab\left(\kappa-\lambda\right)~\ket{\cdots \tau_{N-3}1_{N-2}} + \left(b^2\kappa + a^2\lambda\right)~\ket{\cdots \tau_{N-3}\tau_{N-2}},
\end{eqnarray}
and 
\begin{eqnarray}
    U_{N-1}~\ket{\cdots 1_{N-3}\tau_{N-2}} & = & 0, \nonumber \\
    U_{N-1}~\ket{\cdots \tau_{N-3}1_{N-2}} & = & a~\ket{\cdots \tau_{N-3}1_{N-2}} + b~\ket{\cdots \tau_{N-3}\tau_{N-2}}, \nonumber \\
    U_{N-1}~\ket{\cdots \tau_{N-3}\tau_{N-2}} & = & b~\ket{\cdots \tau_{N-3}1_{N-2}} + \delta b^2~\ket{\cdots \tau_{N-3}\tau_{N-2}},
\end{eqnarray}
respectively.

The construction of the generators of $\mathcal{B}_N$ outlined in this section carries over for other anyon systems with the changes occurring in the number of fusion basis states and the $F$- and $R$-matrices. For example in the case of $SU(2)_4$ containing the Jones-Kauffman anyon system, $\{1, \tau, \mu\}$ we can follow the steps outlined above to construct the braid group generators. In particular we find that the $\mathcal{B}_2$ representation is one-dimensional and is generated by $\sigma_1=R^{\tau\tau}_\tau$ similar to the Fibonacci case. This is also true for $\mathcal{B}_2$ in other anyon systems with no multiplicity in the fusion channels. The fusion basis for three $\tau$ Jones-Kauffman anyons is three dimensional, (Fig. \ref{fig:JKfusionbasis}) and hence the three-strand braid group, $\mathcal{B}_3$ is also three dimensional. We find the braid generators to be $\sigma_1=R$ and $\sigma_2=FRF$ where the $R$- and $F$-matrices are that of the Jones-Kauffman anyons \cite{Bonderson2007NonAbelianAA}. We can continue this procedure for higher-strand braid groups to determine all the generators. The generators will have a larger dimension than that of the braid groups supported on the Fibonacci anyon fusion basis as there are more fusion channels in the case of the Jones-Kauffman anyons. Clearly this procedure can be applied to other $SU(2)_k$ anyons.

We remind the reader that the Jones-Kauffman and Ising anyons cannot realize universal computation through braiding alone \cite{PhysRevA.92.012301, Bravyi_2006}, in contrast to the Fibonacci anyons and so we do not go into the braid generators of these systems any further.

\section{Braid group on supersymmetric zero modes}
\label{sec:Zeromodesbraidgroup}
Using the correspondence between the anyonic fusion basis and the product zero modes of the supersymmetric systems above we can write down the $\mathcal{B}_N$ generators in terms of operators acting on the Hilbert spaces of the spin chain. We will write down the expressions for the bulk and boundary generators in what follows.

We begin with the action of the bulk operator $\sigma_i$ on $\ket{\cdots 1_{i-2}\tau_{i-1}1_i\cdots}$ which corresponds to
\begin{equation}
    \sigma_i~\ket{\cdots b_{i-2}b_{i-1}f_if_{i+1}\cdots} = \kappa~\ket{\cdots b_{i-2}b_{i-1}f_if_{i+1}\cdots}
\end{equation}
in the supersymmetric Hilbert space. The ellipses in the state above correspond to a Fibonacci sequence. For each of those Fibonacci sequences we have a similar relation as given above. This implies that on this state $\sigma_i$ acts as $\sum\kappa\left(\cdots B_{i-2}B_{i-1}G_iG_{i+1}\cdots\right)$, where the $\sum$ represents a sum over projectors to the appropriate Fibonacci sequence. 

Going further we see that the action on the state $\ket{\cdots \tau_{i-2}1_{i-1}\tau_i\cdots}$ corresponds to
\begin{equation}
    \sigma_i~ \ket{\cdots b_{i-2}f_{i-1}f_ib_{i+1}\cdots}  =  \left(a^2\kappa + b^2\lambda\right)\ket{\cdots b_{i-2}f_{i-1}f_ib_{i+1}\cdots} + ab\left(\kappa-\lambda\right)\ket{\cdots b_{i-2}f_{i-1}b_if_{i+1}\cdots}
\end{equation}
and 
\begin{equation}
    \sigma_i~ \ket{\cdots f_{i-2}b_{i-1}b_if_{i+1}\cdots}  =  \left(a^2\kappa + b^2\lambda\right)\ket{\cdots f_{i-2}b_{i-1}b_if_{i+1}\cdots} + ab\left(\kappa-\lambda\right)\ket{\cdots f_{i-2}b_{i-1}f_ib_{i+1}\cdots}.
\end{equation}
The ellipses in the states above represent a Fibonacci sequence as before. Thus the action of $\sigma_i$ on this state is represented by
\begin{eqnarray}
    & & \sum  ~\left(a^2\kappa + b^2\lambda\right)\left[\left(\cdots B_{i-2}G_{i-1}G_iB_{i+1}\cdots\right) + \left(\cdots G_{i-2}B_{i-1}B_iG_{i+1}\cdots\right)\right] \nonumber \\ & &+ \sum ~ab\left(\kappa-\lambda\right)\left[\left(\cdots B_{i-2}G_{i-1}q_iq_{i+1}^\dag\cdots\right) + \left(\cdots G_{i-2}B_{i-1}q_i^\dag q_{i+1}\cdots\right)\right].
\end{eqnarray}
Following this logic the full expression for the bulk and boundary  generators on the supersymmetric space are given by
\begin{eqnarray}
    \sigma_i & = & \sum~\Big[\kappa\left\{\cdots B_{i-2}B_{i-1}G_iG_{i+1}\cdots + \cdots G_{i-2}G_{i-1}B_iB_{i+1}\cdots\right\}  \nonumber \\
    & &+ \left. \lambda\left\{\cdots B_{i-2}B_{i-1}G_iB_{i+1}\cdots + \cdots G_{i-2}G_{i-1}B_iG_{i+1}\cdots\right. \right. \nonumber \\
    & &+ \left. \left. \cdots B_{i-2}G_{i-1}B_iB_{i+1}\cdots + \cdots G_{i-2}B_{i-1}G_iG_{i+1}\cdots \right\} \right. \nonumber \\
    & &+ \left. \left(a^2\kappa + b^2\lambda\right)\left\{\cdots B_{i-2}G_{i-1}G_iB_{i+1}\cdots + \cdots G_{i-2}B_{i-1}B_iG_{i+1}\cdots\right\} \right. \nonumber \\
    & &+ \left. \left(b^2\kappa + a^2\lambda\right)\left\{\cdots B_{i-2}G_{i-1}B_iG_{i+1}\cdots + \cdots G_{i-2}B_{i-1}G_iB_{i+1}\cdots\right\} \right. \nonumber \\
    & &+ \left. ab\left(\kappa-\lambda\right)\left\{\cdots B_{i-2}G_{i-1}q_iq_{i+1}^\dag\cdots + \cdots G_{i-2}B_{i-1}q_i^\dag q_{i+1}\cdots \right. \right. \nonumber \\
    & &+ \left. \left. \cdots G_{i-2}B_{i-1}q_iq_{i+1}^\dag\cdots + \cdots B_{i-2}G_{i-1}q_i^\dag q_{i+1}\cdots \right\} \right], \qquad  i=3,\ldots, N-2,
    \end{eqnarray}
\begin{equation}
    \sigma_1 = \sum~\kappa\left(B_1B_2\cdots + G_1G_2\cdots\right) + \lambda\left(B_1G_2\cdots + G_1B_2\cdots\right),
\end{equation}
\begin{eqnarray}
    \sigma_2 & = & \sum~\left(a^2\kappa + b^2\lambda\right)\left[B_1B_2G_3\cdots + G_1G_2B_3\cdots \right] \nonumber \\
    & &+ \left(b^2\kappa + a^2\lambda\right)\left[B_1FG_2B_3\cdots + G_1B_2G_3\cdots \right] \nonumber \\
    & &+ \lambda \left[B_1G_2G_3\cdots + G_1B_2B_3\cdots \right] \nonumber \\
    & &+ ab\left(\kappa-\lambda\right)\left[1_1q_2^\dag q_3\cdots + 1_1q_2q_3^\dag\right]
\end{eqnarray}
and 
\begin{eqnarray}
    \sigma_{N-1}&=&   \sum~\left(a^2\kappa + b^2\lambda\right)\left[\cdots B_{N-3}G_{N-2}G_{N-1} + \cdots G_{N-3}B_{N-2}B_{N-1}\right] \nonumber \\
    & &+ \left(b^2\kappa + a^2\lambda\right)\left[\cdots B_{N-3}G_{N-2}B_{N-1} + \cdots G_{N-3}B_{N-2}G_{N-1}\right] \nonumber \\
    & &+ \lambda \left[\cdots B_{N-3}B_{N-2}G_{N-1} + \cdots G_{N-3}G_{N-2}B_{N-1}\right] \nonumber \\
    & &+ ab\left(\kappa-\lambda\right)\left[\cdots B_{N-3}G_{N-2}\left(q_{N-1}+q_{N-1}^\dag\right) + \cdots G_{N-3}B_{N-2}\left(q_{N-1}+q_{N-1}^\dag\right)\right],\cr &&
\end{eqnarray}
respectively. The $\sum$ is a sum over the projectors to the Fibonacci sequences. 
For example, consider a 4-site Fibonacci sequence $b_jb_{j+1}f_{j+2}b_{j+3}$. The projector to this sequence is given by $B_jB_{j+1}G_{j+2}B_{j+3}$. The ellipses in the expressions above for the braid generators represent the projectors to these kinds of sequences. This also shows that the braid generators act non-trivially only on a few sites, while it leaves the remaining sites unchanged.

As an explicit example we write down the generators of the three-strand braid group realized on the product zero modes of the SUSY spin chain as
\begin{eqnarray}
    \sigma_1  & = & -\lambda^{-3}\left[B_1B_2 + G_1G_2\right] + \lambda\left[B_1G_2 + G_1B_2\right] \nonumber \\
    \sigma_2 & = & \lambda\left[B_1B_2 + G_1G_2\right] -\lambda^{-3}\left[B_1G_2 + G_1B_2\right] -\frac{1}{\lambda\left(\lambda^{-2}+\lambda^2\right)}\left[B_1B_2 + G_1G_2 -B_1G_2 - G_1B_2\right] \nonumber \\ & + & \frac{\sqrt{1+\lambda^{-4}+\lambda^4}}{\lambda\left(\lambda^{-2}+\lambda^2\right)}\left[q_2 + q_2^\dag \right].
    \end{eqnarray}
As matrices on this four dimensional space spanned by $\{\ket{bb},\ket{bf}, \ket{ff}, \ket{fb}\}$ we see that
\begin{equation}
    \sigma_1 = \left(\begin{array}{cc}
        R & 0 \\
        0 & R
    \end{array}\right),~~  \sigma_2 = \left(\begin{array}{cc}
        FRF & 0 \\
        0 & FRF
    \end{array}\right),
\end{equation}
which is easily seen to satisfy the braid relations.

\subsection{Supersymmetric quantum circuits}
\label{subsec:SUSYqc}

Some comments about the relevance of these results for topological quantum computing are now in order. The important observation is that the braid generators defined on the supersymmetric Hilbert space act non-trivially only on the product zero modes of the Nicolai-like system. All the other states of the spectrum contain at least one set of three consecutive $f$'s or $b$'s. Such states are killed by the braid generators. Because of this, one can show that the braid generators constructed on the zero modes of the spin chain commute with the supercharge $Q$. This can be proved in general as follows. The Hilbert space of the supersymmetric system can be split into two orthogonal subspaces, the space of the zero modes $\mathcal{H}_0$ and the space of positive energy modes $\mathcal{H}_+$, which occur as supersymmetric doublets. Furthermore, the space of zero modes $\mathcal{H}_0$ consists of product zero modes $\mathcal{H}_{0}^P$ and entangled zero modes $\mathcal{H}_{0}^E$, which are orthogonal to each other. Thus the total supersymmetric Hilbert space is a union of these three orthogonal spaces. Let us verify that the braid generators commute with the supercharges on each of these three spaces. We will use the fact that $\sigma_i\ket{pz} \in \mathcal{H}_{p0}$, where $\ket{pz}$ is a product zero mode. 
\begin{enumerate}
    \item Commutativity on $\mathcal{H}_{0}^P$:
    \begin{equation}
        Q\sigma_i~\ket{pz} = 0,\qquad \sigma_iQ~\ket{pz} = 0,
    \end{equation}
    by the definition of the zero modes.
    \item Commutativity on $\mathcal{H}_{0}^E$:
    \begin{equation}
        Q\sigma_i~\ket{ez} = 0,\qquad \sigma_iQ~\ket{ez} = 0.
    \end{equation}
    This follows from the fact that $\sigma_i$ kills all the states that are orthogonal to the product zero modes. Also $Q\ket{ez}\not\in\mathcal{H}_{0}^P$, as by definition of zero modes they belong to the cohomology of the supercharge $Q$, that is those states that cannot be written as $Q$ of some other state.
    \item Commutativity on $\mathcal{H}_+$:
    \begin{equation}
        Q\sigma_i~\ket{+} = 0,~~\sigma_iQ~\ket{+} = 0.
    \end{equation}
    This again follows from the fact that the braid generators annihilate all states orthogonal to those in $\mathcal{H}_{0}^P$ and that $Q\ket{+}\not\in\mathcal{H}_{0}^P$.
\end{enumerate}
Thus the braid generators commute with the supercharges and hence are supersymmetric. As a consequence of this, any quantum gate realized as an element of the braid group $\mathcal{B}_N$ will continue to commute with the supercharges and hence the quantum circuits built out of such gates are supersymmetric. Additionally, these braid generators also commute with the supersymmetric Hamiltonian. 

Furthermore, if we encode information in the product zero modes and act on them with quantum gates built out of the braid group realized on this space, we are ensured that the states cannot move into orthogonal subspaces which can be viewed as errors. Thus we can suppress such errors during information processing. We see this as a potential advantage offered by the supersymmetric system, as the product zero modes precisely emulate the basis states of the fusion basis. 

We would like to emphasize here that the product zero modes do not realize non-Abelian anyons. By realizing the fusion space braid groups on the product zero modes of these supersymmetric systems we are merely emulating the effects that non-Abelian anyons would offer via braiding and fusing. Thus this circumvents the actual need to find non-Abelian anyons in a physical system to eventually build a quantum computer. We could in principle equivalently realize a topological quantum computer on a supersymmetric spin chain by simply working with the product zero modes (classical states) of such a system.

\subsection{Robustness of topological quantum computing}
\label{subsec:stability}
Braid groups are inherently topological as their properties are insensitive to the length and shape of the strands. This is also expected to hold for the physical systems carrying braid group representations. We can ensure this for the SUSY systems if deforming them does not mix the fusion basis with the non-fusion basis. To this effect we can think of two possibilities of deformed SUSY Hamiltonians that continue to support the braid group and hence topological quantum computing. 
\begin{enumerate}
    \item In the first case the deformed Hamiltonians leave each of the product zero modes unchanged. In such a situation the braid group and consequently the quantum gates built using them are unchanged as well. This is the most ideal scenario with regard to quantum computing.
    \item For the second case we consider perturbations that can shuffle the product zero modes among themselves hence lifting the degeneracy but nevertheless they do not connect the product zero modes to positive energy SUSY doublets or to entangled zero modes. Such systems continue to support the braid group which is a rotated version of the braid group computed in the previous case.
\end{enumerate}

In what follows we will consider two ways of deforming the SUSY Hamiltonian, either by directly perturbing them by adding terms to the SUSY Hamiltonian or by deforming the supercharges generating the SUSY Hamiltonian. We shall analyze this for all the SUSY systems considered so far.

\subsection*{SUSY preserving perturbations}
An arbitrary perturbation to the SUSY Hamiltonian can lift the degeneracy of the product zero modes. This is also true for generic quantum platforms for quantum computing. On the other hand we expect SUSY preserving deformations to keep the topological quantum computation stable by not mixing the fusion and non-fusion bases. We will now check this in the different cases.

\paragraph{Fibonacci anyons} - The only terms preserving the supercharges in (\ref{eq:Q1}) and its conjugate is the identity operator and hence these are trivial perturbations. They represent a constant shift in the energy levels and hence the deformed system continues to support the braid group and hence topological quantum computing.

\paragraph{Jones-Kauffman anyons} - We can find more non-trivial local perturbations that can be added to this SUSY system. The local terms of the form 
\begin{equation}\nonumber
\left(\begin{array}{cc}
     I & 0  \\
     0 & 0
\end{array}\right)_i,~~ \left(\begin{array}{cc}
     0 & 0  \\
     0 & I
\end{array}\right)_i,    
\end{equation}
with $I$ being the identity matrix on every site $i$ leaves the space of product zero modes for the supercharge built using (\ref{eq:Qjk}) unchanged. A more non-trivial choice corresponds to the local perturbation,
\begin{equation}\nonumber
   P_i= \left(\begin{array}{cc}
     0 & I  \\
     I & 0
\end{array}\right)_i,
\end{equation}
on every site $i$. This operator interchanges $\left(\begin{array}{c}
     b  \\ 0\end{array}\right)$, $\left(\begin{array}{c}
     f  \\ 0\end{array}\right)$ with $\left(\begin{array}{c}
     0  \\ b\end{array}\right)$, $\left(\begin{array}{c}
     0  \\ f\end{array}\right)$. In terms of the Jones-Kauffman anyons , $\{1, \tau, \mu\}$ and under the correspondence (\ref{eq:JKzIdentification}), we see that this local perturbation interchanges the anyons $1$ and $\mu$ and leaves $\tau$ unchanged. This is true when the local perturbation is of the form $\sum\limits_{i=1}^N~P_i$ or $\sum\limits_{i=1}^{N-1}~P_iP_{i+1}$. Thus such local perturbations shuffle some of the product zero modes and lift the degeneracy. Note that though the perturbations change the eigenvalues of the product zero modes, they do not change the structure of them. That is, the perturbed states continue to be sequences of $1$, $\tau$ and $\mu$ such that they do not contain $11$, $1\mu$, $\mu 1$ and $\mu\mu$.  For example when $N=3$, the space of product zero modes in the fusion basis of four $\tau$ anyons is spanned by, $\{\ket{1\tau}, \ket{\tau 1}, \ket{\tau\mu}, \ket{\mu\tau}, \ket{\tau\tau}\}$. The eigenstates and the corresponding eigenvalues of the perturbed Hamiltonian, 
     \begin{equation}
         \tilde{H} = H_{SUSY} + P_1 + P_2 + P_3,
     \end{equation}
     are given by $\{\ket{1\tau} + \ket{\mu\tau}, \ket{\tau 1} + \ket{\tau\mu} \}$ with eigenvalue 3, $\{-\ket{1\tau} + \ket{\mu\tau}, -\ket{\tau 1} + \ket{\tau\mu} \}$ with eigenvalue -1 and $\ket{\tau\tau}$ with eigenvalue 0. Clearly these states, though they are entangled, are still sequences built out of the configurations, $\tau\tau$, $1\tau$, $\tau 1$, $\tau\mu$ and $\mu\tau$ and they do not contain the forbidden configurations $11$, $1\mu$, $\mu 1$ and $\mu\mu$ (See Table \ref{tab:afJK}). This rotated basis spans the same space as the product zero modes and they continue to support the braid group. The difference is that these states have different energy eigenvalues. We illustrate the change in the rotated braid generator by looking at $\sigma_1$ generating $\mathcal{B}_4$. In the fusion basis $\{\ket{1\tau}, \ket{\tau 1}, \ket{\tau\mu}, \ket{\mu\tau}, \ket{\tau\tau}\}$,
     \begin{equation}
         \sigma_1 = \left(\begin{array}{ccccc}
             R^{\tau\tau}_1 &  &  &  & \\
              & R^{\tau\tau}_\tau & & & \\
              & & R^{\tau\tau}_\mu & & \\
              & & & R^{\tau\tau}_\tau & \\
              & & & & R^{\tau\tau}_\tau
         \end{array}\right)
     \end{equation}
     with $R^{\tau\tau}_1$, $R^{\tau\tau}_\tau$ and $R^{\tau\tau}_\mu$ being the $R$-matrix elements in the Jones-Kauffman case, obtained by solving the pentagon and hexagon equations \cite{Bonderson2007NonAbelianAA}. For the perturbed Hamiltonian, $\tilde{H}$, this operator becomes,
     \begin{equation}
         \tilde{\sigma_1} = \left(\begin{array}{ccccc}
             \frac{R^{\tau\tau}_1+R^{\tau\tau}_\mu}{2} &  & -\frac{R^{\tau\tau}_1-R^{\tau\tau}_\mu}{2} &  & \\
              & R^{\tau\tau}_\tau & & & \\
             -\frac{R^{\tau\tau}_1-R^{\tau\tau}_\mu}{2} & & \frac{R^{\tau\tau}_1+R^{\tau\tau}_\mu}{2} & & \\
              & & & R^{\tau\tau}_\tau & \\
              & & & & R^{\tau\tau}_\tau
         \end{array}\right).
     \end{equation}
     We can write down the other deformed braid generators in a similar manner. The arguments of Sec. \ref{subsec:SUSYqc} continue to hold for the deformed braid generators $\tilde{\sigma_i}$ and the braid group continues to be supersymmetric and as a consequence the quantum circuit built out of the deformed braid generators will also be supersymmetric.

\paragraph{Ising anyons} - The perturbations that commute with the supercharge generating the Jones-Kauffman fusion basis also commute with the supercharges generating the Ising anyon fusion basis. As in the Jones-Kauffman case, the perturbations, $\left(\begin{array}{cc}
     I & 0  \\
     0 & 0
\end{array}\right)_i,~~ \left(\begin{array}{cc}
     0 & 0  \\
     0 & I
\end{array}\right)_i,$ leave the product zero modes invariant shifting the total spectrum by a constant. Thus the braid group and the resulting quantum gates is unchanged in this case.

On the other hand the perturbations $P_i=\left(\begin{array}{cc}
     0 & I  \\
     I & 0
\end{array}\right)_i$, having the same effect on the local bosons and fermions as in the Jones-Kauffman case, shuffle the product zero modes lifting the degeneracy. However they do not mix the fusion basis with the non-fusion basis and hence as seen before the perturbed version of this SUSY system continues to support topological quantum computation due to Ising anyons. 

\subsection*{Deformed SUSY algebras}
We can find deformed SUSY Hamiltonians by deforming the supercharges generating them. The resulting Hamiltonians are part of the SUSY algebra unlike the deformed SUSY Hamiltonians considered earlier. We consider two possible ways to deform the supercharges.

\paragraph{Disordered supercharges} - In the first case introduce a site-dependent parameter for every local supercharge, that is use $\tilde{q}_i=\left(\begin{array}{cc}
    0 & \alpha_i \\
    0 & 0
\end{array}\right).$ Then the SUSY Hamiltonian corresponding to the supercharge (\ref{eq:Q1}),
\begin{eqnarray}
    H & = & \sum\limits_{i=1}^{N-2}~\left[B_iB_{i+1}B_{i+2} + G_iG_{i+1}G_{i+2} \right] \nonumber \\
    & + & \sum\limits_{i=1}^{N-3}~\left[\theta_i\left(B_{i+1}B_{i+2} + G_{i+1}G_{i+2}\right)\theta_{i+3}^\dag + h.c. \right] \nonumber \\
    & + & \sum\limits_{i=1}^{N-4}~\left[\theta_i\theta_{i+1}\theta_{i+3}^\dag\theta_{i+4}^\dag + h.c.   \right],
\end{eqnarray}
appears with site-dependent couplings breaking bulk translational invariance. Clearly this does not alter the structure of the product zero modes as they have eigenvalues zero. Hence deforming the local supercharges on each site does not change the space of product zero modes. Such deformations are possible for all the three SUSY systems considered in this paper as the local supercharge $q$ is the fundamental building block for all three of them.

\paragraph{Modified supercharges} -
Next we consider adding the class of supersymmetric terms
\begin{equation}
    \sum\limits_{i=1}^{N-2k}~\theta_i\theta_{i+1}\theta_{i+2}\theta_{i+3}\cdots\theta_{i+2k},~~k\in\{2,3,\cdots\},
\end{equation}
to the supercharge in (\ref{eq:Q1}) and this generates a new SUSY algebra. Every product zero mode of the supercharge in (\ref{eq:Q1}) is also a zero mode of this new term and they exhaust the possible product zero modes and hence the new SUSY system will continue to have the same number of product zero modes in their ground state spectrum. Note that we cannot add the linear term $\sum\limits_{i=1}^N~\theta_i$ to the supercharge in (\ref{eq:Q1}) as this term does not have any zero mode. This follows from the form of the local supercharges, (\ref{eq:localq1}). Thus the space of product zero modes is stable to these deformations as well. Such terms, when appropriately modified, can also be added to the supercharges generating the Jones-Kauffman anyon system, (\ref{eq:Qjk}).

\subsection{Braid group on the zero modes of qudit SUSY systems}
\label{sec:SISbraidgroup}

Up to this point we have considered supersymmetric systems with on each site a copy of $\mathbb{C}^2$, namely we have considered qubits. To address cases where the local Hilbert space on the chain is not $\mathbb{C}^2$ but is instead $\mathbb{C}^3$ for example, then we need to construct the appropriate supercharges. The supersymmetric systems constructed out of {\it symmetric inverse semigroups (SIS's)} precisely provide such examples \cite{Padmanabhan:2017ekk}. The reader is referred to that reference for details.

Consider the system built out of the SIS $\mathcal{S}^3_1$, namely the partial bijections on a set of three elements. On a single site we have the Hilbert space $\mathbb{C}^3$ (a qutrit) and the supercharge is given by
\begin{equation}\label{eq:S31q}
    q = \frac{1}{\sqrt{2}}\left(\begin{array}{ccc} 0 & 1 & 1 \\ 0 & 0 & 0 \\ 0 & 0 & 0\end{array}\right),\qquad q^\dag = \frac{1}{\sqrt{2}}\left(\begin{array}{ccc} 0 & 0 & 0 \\ 1 & 0 & 0 \\ 1 & 0 & 0\end{array}\right),
\end{equation}
which are easily to seen satisfy the supersymmetry relations $q^2=\left(q^\dag\right)^2=0$.
The local $\mathbb{C}^3$ is spanned by three states,
$$ \left\{\ket{b} = \left(\begin{array}{c} 1 \\0 \\0\end{array}\right),~ \ket{f} = \frac{1}{\sqrt{2}}\left(\begin{array}{c} 0 \\1 \\ 1\end{array}\right),~\ket{z} = \frac{1}{\sqrt{2}}\left(\begin{array}{c} 0 \\1 \\-1\end{array}\right) \right\},$$
which satisfy
\begin{equation}
    q\ket{f} = \ket{b},~q^\dag\ket{b} = \ket{f}, \qquad q\ket{z}=q^\dag\ket{z}=0.
\end{equation}
Consider the global supercharges in \eqref{eq:Q1} constructed out the local supercharges in \eqref{eq:S31q}. These systems continue to have both product and entangled zero modes. The product zero modes are obtained by filling each site with $f$'s and $b$'s like the previous cases, but now additionally we can also fill a site with the local zero mode, $z$. This drastically enlarges the number of product zero modes for this system. Nevertheless these states continue to support the Fibonacci braid group. The local zero modes can be accommodated anywhere in a zero sequence. Adjacent to these local zero modes we fill up the sites with a Fibonacci sequence out of the $f$'s and $b$'s. Then we can classify the product zero modes of the ${\cal S}^3_1$ system according to the number of local zero modes and their respective positions in the zero sequences. Thus a general zero sequence with a fixed number of local zero modes looks like 
$$ \ket{\underbrace{\cdots}_{\textrm{Fib sequence}} z_{i_1}\underbrace{\cdots}_{\textrm{Fib sequence}} z_{i_2}\underbrace{\cdots}_{\textrm{group of 0's}} z_{i_k} \underbrace{\cdots}_{\textrm{Fib sequence}} }.$$
This state includes an isolated local zero mode at site $i_1$ and a group of local zero modes at sites $i_2$ to $i_k$. Fibonacci sequences made out of $f$'s and $b$'s fill up the remaining sites. The dimension of this subspace is precisely $F(d_1)F(d_2)F(d_3)$, where $d_1$, $d_2$ and $d_3$ are the number of sites with Fibonacci sequences. We can realize the direct product group $\mathcal{B}_{d_1+1}\times \mathcal{B}_{d_2+1}\times \mathcal{B}_{d_3+1}$ on this subspace. In a similar manner, we can construct a direct product of Fibonacci braid groups on other such subspaces, which are determined by the number and precise location of the local zero modes. Thus in general we can construct the $\mathcal{B}_{d_1}\times \mathcal{B}_{d_2}\times\cdots\times \mathcal{B}_{d_p}$ on the product zero modes of the $\mathcal{S}^3_1$ system. 

More generally, a ${\cal S}^p_1$ system on a single site is supported on $\mathbb{C}^p$. The supercharges are,
\begin{equation}
    q = \frac{1}{\sqrt{p-1}}\left(\begin{array}{cccc} 0 & 1 & \cdots & 1 \\ 0 & 0 & \cdots & 0 \\ \vdots & \vdots & & \vdots \\  0 & 0 & \cdots & 0 \end{array}\right), \qquad  q^\dag = \frac{1}{\sqrt{p-1}}\left(\begin{array}{cccc} 0 & 0 & \cdots & 0 \\ 1 & 0 & \cdots & 0 \\ \vdots & \vdots & & \vdots \\  1 & 0 & \cdots & 0 \end{array}\right).
\end{equation}
The system has a single local boson given by $\left(\begin{array}{cccc} 1 & 0 & \cdots & 0\end{array}\right)^T$ and a single local fermion given by $\left(\begin{array}{cccc} 0 & 1 & \cdots & 1\end{array}\right)^T$. The remaining $p-2$ states are local zero modes. The system built out of the global supercharge has product zero modes which are very similar to the $\mathcal{S}^3_1$ system with the difference that now we have $p-2$ choices for the local zero modes instead of just a single choice. Thus the $\mathcal{S}^p_1$ system also realizes a direct product of the Fibonacci braid group. 

We can go further and consider the local supersymmetric system built on $\mathbb{C}^5$, using $\mathcal{S}^5_2$. In this case the supercharges are
\begin{equation}
    q \propto \left(\begin{array}{ccccc} 0 & 0 & 1 & 1 & 1 \\ 0 & 0 & 1 & 1 & 1 \\ 0 & 0 & 0 & 0 & 0 \\ 0 & 0 & 0 & 0 & 0 \\ 0 & 0 & 0 & 0 & 0\end{array}\right), \qquad  q^\dag \propto \left(\begin{array}{ccccc} 0 & 0 & 0 & 0 & 0 \\ 0 & 0 & 0 & 0 & 0 \\ 1 & 1 & 0 & 0 & 0 \\ 1 & 1 & 0 & 0 & 0 \\ 1 & 1 & 0 & 0 & 0\end{array}\right).
\end{equation}
There are three zero modes for this supercharge, plus a boson and a fermion. These are given by
$$ \left\{\ket{z_1}=\left(\begin{array}{c} 0 \\ 0 \\ 1 \\ \omega \\ \omega^2\end{array}\right), \ket{z_2}=\left(\begin{array}{c} 0 \\ 0 \\ 1 \\ \omega^2 \\ \omega\end{array}\right), \ket{z_3}=\left(\begin{array}{c} 1 \\ -1 \\ 0 \\ 0 \\ 0\end{array}\right), \ket{f}=\left(\begin{array}{c} 0 \\ 0 \\ 1 \\ 1 \\ 1\end{array}\right), \ket{b}=\left(\begin{array}{c} 1 \\ 1 \\ 0 \\ 0 \\ 0\end{array}\right)\right\},$$
with $\omega$ being a cube root of unity. Clearly, the structure of the product zero states of a global supersymmetric system built out of the above supercharge is similar to the $\mathcal{S}^p_1$ case and hence we can realize a direct product of the Fibonacci braid groups on these product zero modes.

We can now conclude that this property continues to hold for an arbitrary supersymmetric system built out of the inverse semigroup $\mathcal{S}^n_r$ with $r<n$. In this case the supercharge is given by,
\begin{equation}
    q \propto \sum\limits_{j=r+1}^n\sum\limits_{k=1}^r~ e_{kj},\qquad  q^\dag \propto \sum\limits_{j=r+1}^n\sum\limits_{k=1}^r~ e_{jk},
\end{equation}
where the $e_{jk}$ denote matrices with the only non-zero entry at the $(j, k)$-th location. These systems are supported on $\mathbb{C}^n$ which is spanned by $n-2$ zero modes plus a fermion and a boson. The global supersymmetric systems built out of these supercharges continue to host a direct product of the Fibonacci braid groups as in the previous examples. We also note that when $n$ is divisible by $r$ there are no local zero modes. In such a case we have more number of bosons/fermions to fill each site on the chain. The resulting product states continue to host the Fibonacci braid group. 

\section{Outlook}
\label{sec:Outlook}

Supersymmetric spin chains, like the ones we have considered in this paper, are interesting arenas with an underlying rich structure of ground states which can be mapped to the fusion space of different anyonic systems, like the Fibonacci anyons, Jones-Kauffman anyons and the Ising anyons. We have limited our attention to just a few choices of supercharges. Of course there are many more possible choices, also to accommodate systems more general than qubits, as we have seen in Sec. \ref{sec:SISbraidgroup}. 

Exploring more possibilities is then a very natural development of this analysis. For example, one could consider a supercharge built out of the local term
\bea
\left(\begin{array}{cc} \theta & 0 \\ 0 & 0 \end{array}\right)_j\left(\begin{array}{cc} \theta & 0 \\ 0 & 0 \end{array}\right)_{j+1}\left(\begin{array}{cc} \theta & 0 \\ 0 & 0 \end{array}\right)_{j+2} + \left(\begin{array}{cc} 0 & 0 \\ 0 & \theta \end{array}\right)_j\left(\begin{array}{cc} 0 & 0 \\ 0 & \theta \end{array}\right)_{j+1}\left(\begin{array}{cc} 0 & 0 \\ 0 & \theta \end{array}\right)_{j+2},
\eea
with $\theta$ being the anticommuting variable constructed out of the local supercharge $q$. This model has local states in $\mathbb{C}^4$ which are spanned by states similar to the one found for the Jones-Kauffman or Ising case. If these states are placed in correspondence with the Jones-Kauffman labels, $\{1, \mu, \tau \}$ as in Sec. \ref{subsec:JK2Q2}, then we see that the sequences $\mu\mu$, $1\mu$ and $\mu 1$, see Table \ref{tab:afJK}, are no longer forbidden. It would be interesting to find a representation of the braid group on such spaces as well. We suspect that such a system can be mapped to the Fibonacci anyons by a linear transformation among the anyon labels. The resulting model may have a different multiplicity for the fusion channels when compared to the original Fibonacci model, which has multiplicity 1 for the different fusion channels. 

Also looking for other supersymmetric spin chains corresponding to other anyonic models would be a worthwhile attempt. Finding more examples could be a basis for establishing a larger and deeper connection between anyons and supersymmetric spin chains.

Quantum information is stored in long range entangled states in error correcting codes exhibiting topological order, like Kitaev's toric code. These are fault-tolerant ways of realizing topological quantum computation. On the other hand, it is worth noting that here the logical qubits need to be encoded in product zero modes of supersymmetric systems. This could imply a lower cost in encoding quantum information. 

It is well-known that the Fibonacci model provides universal quantum computing through braiding alone \cite{bonesteel2005braid,field2018introduction,simon2006topological,https://doi.org/10.48550/arxiv.1604.06429}. This is not true for the level $k=4$ systems, where one needs to introduce non-topological gates to achieve universality \cite{https://doi.org/10.48550/arxiv.1501.02841,PhysRevA.92.012301}. While we constructed the braid group corresponding to the Fibonacci model on the supersymmetric zero mode space, it is possible to construct the braid groups corresponding to the level $k=4$ and $k=2$ systems as well. We then need to consider the non-topological gates in the supersymmetric Hilbert space. This would be an important check to see if the supersymmetric system continues to support quantum computation in these cases.

On a more technical note, it is well known that the diagrammatics of the Temperley-Lieb recoupling theory and the Jones representation lead to the bracket polynomials. We could adapt these computations to the supersymmetric Hilbert space, find the associated topological invariants and see if they are related to the Witten index of these theories.

Finally, the supersymmetric systems considered here include both product and entangled zero modes. While the product zero modes support the appropriate braid group, it is natural to check if a similar representation is possible on the space of entangled zero modes as well. We hope to come back to this question in the future.

\section*{Acknowledgements}
We thank A. P. Balachandran, Hosho Katsura and Hajime Moriya for useful discussions. We also thank the anonymous referee of JHEP for useful comments on the manuscript. DT is supported in part by the INFN grant {\it Gauge and String Theory (GAST)} and would like to acknowledge FAPESP’s partial support through the grants 2016/01343-7 and 2019/21281-4. IJ is partially supported by DST's INSPIRE Faculty Fellowship through the grant \\ DST/INSPIRE/04/2019/000015. 

\appendix
\section{Entangled ground states}

In Section \ref{sec:SUSYzeromodes} we have counted the number of product zero modes of our Nicolai-like supersymmetric spin chain with the choice of supercharges \eqref{eq:localq1}. For completeness, we report here also the counting of the entangled ground states. Before counting them we look at their structure. This can be contrasted with that of the product zero modes which are made of sequences of $f$'s and $b$'s such that three consecutive $f$'s or $b$'s do not appear. The entangled states are superpositions of sequences containing at least one set of three consecutive $f$'s or three consecutive $b$'s. For example when $N=4$, the supercharge in (\ref{eq:Q1}) and its conjugate annihilate the states,
\begin{equation}
    \frac{1}{\sqrt{2}}\left[\ket{b_1b_2b_3f_4}+\ket{f_1b_2b_3b_4}\right],~~ 
    \frac{1}{\sqrt{2}}\left[\ket{f_1f_2f_3b_4}-\ket{b_1f_2f_3f_4}\right],
\end{equation}
  and when $N=6$ they annihilate,
  \begin{equation}
      \frac{1}{\sqrt{6}}\left[\ket{b_1b_2b_3f_4f_5f_6} + 2\ket{f_1b_2b_3b_4f_5f_6} - \ket{f_1f_2f_3b_4b_5b_6} \right].
  \end{equation}
  The general structure of the entangled ground states for arbitrary $N$ is a harder problem. Nevertheless we can count the number of such states for each $N$ as shown below.

The action of the Hamiltonian preserves the number of $|b\rangle$'s, and so we can decompose the whole space (having dimension $2^{N}$) into eigen subspaces with the number of $|b\rangle$'s being kept fixed. Therefore one may analyze the other states of the Hamiltonian on each of those subspaces. 
Let $f_{E}(N)$ be the number of entangled ground states on $N$ sites. From numerical calculations, $f_{E}(N)$ on each subspace are as in Table \ref{tab:entangled_ground}.
\begin{table}
    \centering
    \begin{tabular}{|c|cccccccccc|c|}
        \hline
		$\#b=$ & 1 & 2 & 3 & 4 & 5 & 6 & 7 & 8 & 9 & 10 & $f_{E}(N)$\\\hline
		$N=3$ & &&&&&&&&&&0\\
		$N=4$ & 1& &1&&&&&&&&2\\
		$N=5$ &  &2&2&&&&&&&&4\\
		$N=6$ &  &3&4&3&&&&&&&10\\
		$N=7$ &  &2&9&9&2&&&&&&22\\
		$N=8$ &  & &12&20&12&&&&&&44\\
		$N=9$ &  & &9&36&36&9&&&&&90\\
		$N=10$ & & &3&45&78&45&3&&&&174\\
		$N=11$ & & & &37&131&131&37&&&&336\\\hline
	\end{tabular}
    \caption{Counting of the entangled ground states.}
    \label{tab:entangled_ground}
\end{table}
Combining $f_{E}(N)$ with $f_{P}(N)$ in \eqref{eqn:product_ground_states_fibo}, the total number of ground states $f_{G}(N)$ is as in Table \ref{tab:all_ground_states}, namely
\begin{align}\label{eqn:entangled_ground_states}
	f_{E}(N)=2(f_{E}(N-2)+f_{E}(N-3))+f_{P}(N-3).
\end{align}
\begin{table}
    \centering
    \begin{tabular}{|c|cc|c|}
        \hline
		$N$ & $f_{E}(N)$ & $f_{P}(N)$ & $f_{G}(N)$\\\hline
		3 & 0&6&6\\
		4 & 2&10&12\\
		5 & 4&16&20\\
		6 & 10&26&36\\
		7 & 22&42&64\\
		8 & 44&68&112\\
		9 & 90&110&200\\
		10 & 174&178&352\\
		11 & 336&288&624\\\hline
	\end{tabular}
    \caption{Counting of all the ground states. This table coincides with table II of \cite{sannomiya2017supersymmetry}.}
    \label{tab:all_ground_states}
\end{table}
From this table, one may predict that $f_{G}(N)$ satisfies the recursive relation
\begin{align}\label{eqn:all_ground_states_formula}
    f_{G}(N)=2(f_{G}(N-2)+f_{G}(N-3)).
\end{align}

Moreover, we may find the generating function of $f_{E}(N)$ as follows. Let us define $e(t)=\sum_{N=1}^{\infty}f_{E}(N)t^{N}$, $g(t)=\sum_{N=1}^{\infty}f_{G}(t)t^{N}$. Since $f_{G}(N)=2(f_{G}(N-2)+f_{G}(N-3))$, using the same technique as in \eqref{eqn: general_generating_function} along with the initial conditions $f_{G}(1)=2, f_{G}(2)=4, f_{G}(3)=6$, we may obtain that
\begin{align*}
	2t^{2}g(t)+2t^{3}g(t)&=2f_{G}(1)t^{3}+g(t)-f_{G}(1)t-f_{G}(2)t^{2}-f_{G}(3)t^{3}\\
	&=g(t)-2t-4t^{2}-2t^{3}.
\end{align*}
Consequently,
\begin{align*}
	g(t)=\frac{2t+4t^{2}+2t^{3}}{1-2t^{2}-2t^{3}}.
\end{align*}
Now using the expression of $g_{F}(t)$ and the fact that $g(t)=e(t)+g_{F}(t)$, we obtain
\begin{align*}
	e(t)=\frac{2t+4t^{2}+2t^{3}}{1-2t^{2}-2t^{3}} - \frac{2(t+t^{2})}{1-t-t^{2}}.
\end{align*}

\section{Generating functions}
Here we discuss the generating functions for the counting formulas appearing in the main text. Let $h:\mathbb{N}\to \mathbb{N}$ be a function satisfying
\begin{align}\label{eqn: general_recursion_relation_for_generating_function}
	h(N)=a[h(N-1)+bh(N-2)],
\end{align}
where $a, b\in \mathbb{N}$. Define the formal power series (generating function)
\begin{align*}
	z(t)=\sum_{N=1}^{\infty}h(N)t^{N}.
\end{align*}
Using \eqref{eqn: general_recursion_relation_for_generating_function}, one sees that
\begin{align*}
	atz(t)+abt^{2}z(t)
	&=z(t)-h(1)t-(h(2)-ah(1))t^{2},
\end{align*}
which solved for $z(t)$ yields
\begin{align}\label{eqn: general_generating_function}
	z(t)=\frac{h(1)t+(h(2)-ah(1))t^{2}}{1-at-abt^{2}}
\end{align}
Using the above formula, we calculate the generating functions of \eqref{eqn:product_ground_states_fibo}, \eqref{eqn:fibonacci anyon formula}, \eqref{eqn: recursion relation of j(N) zero modes}, \eqref{eqn:JK_anyon_counting_equation}, as follows. 

\subsection{Fibonacci case}

Let us define $g_{F}(t)=\sum_{N=1}^{\infty}f_{P}(N)t^{N}$, $a_{F}(t)=\sum_{N=1}^{\infty}f(N)t^{N}$, where $f_{P}$ and $f$ are same as in \eqref{eqn:product_ground_states_fibo} and \eqref{eqn:fibonacci anyon formula}, respectively.
Using the initial conditions $f_{P}(1)=2, f_{P}(2)=4, f(1)=1, f(2)=1$, we obtain
\begin{align}
	g_{F}(t)=\frac{2(t+t^{2})}{1-t-t^{2}},\qquad 
    a_{F}(t)=\frac{t}{1-t-t^{2}}.\label{eqn: generating_function_of_fibo_anyon}
\end{align}
The above expressions are obtained from \eqref{eqn: general_generating_function} by using $a=1=b$.

\subsection{Jones-Kauffman case}

Let us define $g_{JK}(t)=\sum_{N=1}^{\infty}j_{P}(N)t^{N}$, $a_{JK}(t)=\sum_{N=1}^{\infty}j(N)t^{N}$, where $j_{P}$ and $j$ are same as in \eqref{eqn: recursion relation of j(N) zero modes} and \eqref{eqn:JK_anyon_counting_equation}, respectively. We observe that by using $a=2=b$ in \eqref{eqn: general_recursion_relation_for_generating_function} we obtain \eqref{eqn: recursion relation of j(N) zero modes}, and using $a=1, b=2$ we obtain \eqref{eqn:JK_anyon_counting_equation}. Therefore applying \eqref{eqn: general_generating_function} along with the initial conditions $j_{P}(1)=4, j_{P}(2)=16, j(1)=1, j(2)=1$ we obtain
\begin{align}
	g_{JK}(t)=\frac{4t+8t^{2}}{1-2t-4t^{2}}\qquad 
    a_{JK}(t)=\frac{t}{1-t-2t^{2}}.\label{eqn: generating_function_of_JK_anyon}
\end{align}


\bibliographystyle{unsrt}
\normalem
\bibliography{refs}
\end{document}